\documentclass[%
aps,
twocolumn,
superscriptaddress,
amsmath,amssymb,
pra,
floatfix,
longbibliography,
10pt
]{revtex4-2}
\usepackage{url}
\usepackage{graphicx}% Include figure files
\usepackage{bm}% bold math
\usepackage{amsmath,amsfonts,amssymb}
\usepackage{mathtools}
\usepackage{xcolor}

\usepackage{silence}
\usepackage{physics}
\usepackage{dsfont}
\makeatletter
\def\pre@bibdata{MTv02Notes}
\newcommand{\thickhline}{%
    \noalign {\ifnum 0=`}\fi \hrule height 1pt
    \futurelet \reserved@a \@xhline
}
\makeatother

\makeatletter

\newcommand{\AppendixTocStart}{%
  \let\l@part\AppendixTocSavedPart
  \let\l@section\AppendixTocSavedSection
  \let\l@subsection\AppendixTocSavedSubsection
  \let\l@subsubsection\AppendixTocSavedSubsubsection
}

\newcommand{\appendixtableofcontents}{%
  \begingroup

  \setcounter{tocdepth}{2}%

  \let\AppendixTocSavedPart\l@part
  \let\AppendixTocSavedSection\l@section
  \let\AppendixTocSavedSubsection\l@subsection
  \let\AppendixTocSavedSubsubsection\l@subsubsection

  \let\l@part\@gobbletwo
  \let\l@section\@gobbletwo
  \let\l@subsection\@gobbletwo
  \let\l@subsubsection\@gobbletwo

  \let\appendix\relax

  \begin{center}
    {\small\bfseries CONTENTS}
  \end{center}

  \vspace{0.5cm}

  \@starttoc{toc}%

  \endgroup
}

\makeatother

\usepackage{multirow}
\usepackage{tabu}
\usepackage{textcomp,booktabs}
\usepackage{colortbl}
\usepackage[T1]{fontenc}
\definecolor{mygray}{gray}{0.9}
\definecolor{mypink}{rgb}{0.99,0.91,0.95}
\definecolor{mycyan}{cmyk}{0.3,0,0,0}

\usepackage[colorlinks, linkcolor=red, anchorcolor=green, citecolor=blue]{hyperref}

\usepackage[T1]{fontenc}
\newcommand{\T}{\mathbf{T}}
\newcommand{\1}{\mathds{1}}

\usepackage{amsthm}
\usepackage{tcolorbox}
\tcbuselibrary{theorems}
\newtheorem{thm}{Theorem}
\newtheorem{prop}[thm]{Proposition}

\newtcbtheorem[number within=section, use counter*=thm]{mythm}{Theorem}%
{colback=red!5,colframe=red!35!red,fonttitle=\bfseries}{thm}

\newtcbtheorem[number within=section, use counter*=thm]{mylem}{Lemma}%
{colback=green!5,colframe=green!35!black,fonttitle=\bfseries}{lem}

\newtcbtheorem[number within=section, use counter*=thm]{mydef}{Definition}%
{colback=yellow!5,colframe=yellow!70!black,fonttitle=\bfseries}{def}

\newtcbtheorem[number within=section, use counter*=thm]{mycor}{Corollary}%
{colback=blue!5,colframe=blue!70!black,fonttitle=\bfseries}{cor}

\allowdisplaybreaks[2]
\providecommand{\ii}{\mathrm i}
\providecommand{\smin}{\sigma_{\min}}
\providecommand{\Mfun}{\mathcal M}
\providecommand{\comp}[1]{#1^{[2]}}

\definecolor{mtrevisionorange}{RGB}{183,83,0}

\begin{document}
\title{Hybrid Uncertainty in the Real Forms of \texorpdfstring{$\mathfrak{sl}(2,\mathbb C)$}{sl(2,C)}}

\author{Xiaoli Hu}
\email{xiaolihumath@163.com}
\affiliation{School of Artificial Intelligence, Jianghan University, Wuhan, Hubei 430056, China}

\author{Xiaoyong Yang}
\email{yang_math123@163.com}
\affiliation{School of Mathematics and Statistics, Central China Normal University, Wuhan, Hubei 430079, China}

\author{Naihuan Jing}
\email{jing@ncsu.edu}
\affiliation{Department of Mathematics, North Carolina State University, Raleigh, NC 27695, USA}

%%%%%%%%%%%%%%%%%%%%%%%%%%%%%%%%%%%%%%%%%%%%%%%%%%%%%%%%%%%%%%%%%%%%%%%%%%%%%%%%%%%%%%%%%%%%%%%%%%%%%%%%%%%%%%%%%%%%%%

%\date{\today}

%%%%%%%%%%%%%%%%%%%%%%%%%%%%%%%%%%%%%%%%%%%%%%%%%%%%%%%%%%%%%%%%%%%%%%%%%%%%%%%%%%%%%%%%%%%%%%%%%%%%%%%%%%%%%%%%%%%%%%

\begin{abstract}
An optimized commutator bound can lie strictly below the quantum infimum even when its minimizing mean is realized by a quantum state. 
We study hybrid uncertainty, defined as the sum of pairwise products of standard deviations.  
For any spanning list of nonzero real linear combinations of the generators in a positive discrete series representation of $\mathfrak{su}(1,1)$, the commutator bound has a unique minimizing mean, realized by a Perelomov coherent state. 
We prove that the optimized bound equals the infimum over states with finite second moments if and only if the symmetrized covariance matrix of the prescribed observables is diagonal in this state. 
When this condition fails, the quantum infimum is strictly greater than the bound; for a nonorthogonal list, we obtain an explicit positive lower bound on this gap.
For irreducible spin-$j$ representations of $\mathfrak{su}(2)$, we derive the exact minimum of the spectral hybrid uncertainty over states for any spanning list of spin observables.
For a fixed state, we compare separate measurements made before a random target is disclosed with direct measurement of the target known in advance.
For these protocols, the excess mean squared error coefficient equals twice the listed hybrid uncertainty, and its minimum over orthogonal combinations of the observables equals twice the spectral hybrid uncertainty.
\end{abstract}

%%%%%%%%%%%%%%%%%%%%%%%%%%%%%%%%%%%%%%%%%%%%%%%%%%%%%%%%%%%%%%%%%%%%%%%%%%%%%%%%%%%%%%%%%%%%%%%%%%%%%%%%%%%%%%%%%%%%%%

\maketitle

%%%%%%%%%%%%%%%%%%%%%%%%%%%%%%%%%%%%%%%%%%%%%%%%%%%%%%%%%%%%%%%%%%%%%%%%%%%%%%%%%%%%%%%%%%%%%%%%%%%%%%%%%%%%%%%%%%%%%%

\section{Introduction}
\label{sec:main-introduction}
Uncertainty relations constrain the outcome statistics of incompatible quantum measurements~\cite{Kennard1927,Robertson1929}.
Relations based on variances are used to certify entanglement and to study precision limits in quantum sensing and metrology~\cite{HofmannTakeuchi2003,Guhne2004,Schwonnek2017,Pezze2018,ChiewGessner2022}.
Entropic uncertainty relations are used in security proofs for quantum key distribution~\cite{Berta2010,Coles2017}, while uncertainty relations for quantum processes constrain the outcome statistics of incompatible process measurements~\cite{XiaoProcess2021,XiaoInterventions2023}.
For a fixed list of observables, finding the optimal state-independent bound requires minimizing the uncertainty over all admissible states.
For Lie algebra generators, commutator expectations depend on the first moments of the generators.
Minimizing a commutator bound over the physically allowed means in a fixed representation therefore yields a state-independent lower bound.
However, a state that realizes the minimizing mean need not saturate the commutator bound.
We ask when this optimized bound equals the quantum infimum and when a strictly positive gap remains despite the existence of a state realizing the minimizing mean.

The Robertson uncertainty relation bounds the product of standard deviations using the commutator expectation, while the Schr\"odinger refinement also includes the covariance~\cite{Robertson1929,Schrodinger1930}.
For several observables, their variances and covariances must arise from a common quantum state.
The covariance matrix is positive semidefinite, and its principal minors and spectral invariants enter multivariate uncertainty relations~\cite{Robertson1934,TrifonovDonev1998,Trifonov2000}.
Sum and product uncertainty relations provide complementary bounds for multiple observables, including angular momentum components~\cite{Maccone2014,KechrimparisWeigert2014,MaEtAl2017,ChenFei2015}.
The operator algebra and its representation further restrict which moments a quantum state can realize~\cite{MoroderKeylLutkenhaus2008}.
Refinements involving quantum Fisher information also strengthen variance-based uncertainty bounds~\cite{Huang2012,Schwonnek2017,Giorda2019,ChiewGessner2022}.
For irreducible representations of compact semisimple Lie algebras, an invariant sum of generator variances admits an exact minimum determined by the highest weight and root system~\cite{deGuise2018,HuJingZhangXiao2026}.
The covariance spectrum gives the principal variances of angular momentum independently of the original choice of measurement axes~\cite{RivasLuis2008}.
Graph methods and state polynomial optimization use algebraic relations among observables to derive uncertainty bounds and certify their tightness in particular cases~\cite{deGois2023,MoranHuber2024}.
Unlike an invariant sum of variances, the sum of pairwise products studied here depends on the prescribed observables.
Attaining its optimized commutator bound requires one state to realize the minimizing mean and saturate every pairwise commutator inequality.
If no state attains the bound, a sequence of states could still approach it.

We establish an exactness criterion for any spanning list of nonzero real linear combinations of the generators in a positive discrete series representation of $\mathfrak{su}(1,1)$.
The commutator bound has a unique minimizing mean on the boundary of the physically allowed region, which is realized by a Perelomov coherent state~\cite{Perelomov1986}.
In this coherent state, the pairwise Robertson-Schr\"odinger inequalities are saturated~\cite{Trifonov2000}, but the real covariances between the prescribed observables can be nonzero.
The optimized commutator bound equals the quantum infimum if and only if the covariance matrix of this coherent state is diagonal for the prescribed observable list.
If the matrix is not diagonal, we prove a strictly positive gap by excluding sequences of states whose uncertainty approaches the optimized commutator bound.
The proof uses the unique coherent state realizing the minimizing mean, the spectral gap of an associated generator, and lower semicontinuity of second moments.
For the nonorthogonal list $\bm A_*=(K_x,K_y,K_x+K_z)^{\T}$, we establish an explicit positive gap and show that a pure state achieves lower uncertainty than any mixture of Perelomov coherent states.
For weighted observables related by unitarily implementable Lorentz transformations, a coherent state attains the optimized commutator bound.
For irreducible spin-$j$ representations of $\mathfrak{su}(2)$, we also obtain the exact spectral minimum over states and orthogonal combinations of any spanning list of spin observables.

We consider a randomly signed target observable and compare measuring it directly when its signs are known in advance with measuring the prescribed observables separately before the signs are disclosed.
At a fixed state, the excess mean squared error coefficient of separately measuring the prescribed observables, relative to directly measuring the target when it is known in advance, equals twice the listed hybrid uncertainty after optimal copy allocation~\cite{McCleanEtAl2016,RubinBabbushMcClean2018,CrawfordEtAl2021,GonthierEtAl2022}.
Allowing orthogonal combinations of the observables to be measured reduces the minimum excess coefficient to twice the spectral hybrid uncertainty.
The comparison assumes known component variances and, when optimizing over orthogonal combinations, the covariance matrix; their estimation costs are excluded.
These identities concern the specified protocols using independent copies, not arbitrary collective measurements.
Section~\ref{sec:main-functionals} defines the functionals and covariance bounds; Secs.~\ref{sec:main-criterion} and~\ref{sec:main-gap} give the exactness criterion and strict gaps; Secs.~\ref{sec:main-spin} and~\ref{sec:main-sampling} give the spin and sampling results.
The Appendices contain the proofs and operator domain conditions.

%%%%%%%%%%%%%%%%%%%%%%%%%%%%%%%%%%%%%%%%%%%%%%%%%%%%%%%%%%%%%%%%%%%%%%%%%%%%%%%%%%%%%%%%%%%%%%%%%%%%%%%%%%%%%%%%%%%%%%

\section{Hybrid Uncertainty and Covariance Structure}
\label{sec:main-functionals}

The sum of variances is invariant under orthogonal mixing of observables, whereas the sum of pairwise
products of standard deviations generally depends on the measured combinations.
Let $\bm A=(A_1,\ldots,A_N)^{\T}$ be a finite list of self-adjoint observables, not necessarily linearly independent, and let $\rho$ be a quantum state with
finite second moments.
Writing
$\overline A_{r,\rho}
=A_r-\langle A_r\rangle_\rho\1$,
we define the real covariance matrix by
$[\Gamma_\rho(\bm A)]_{rs}
=\operatorname{Re}
\langle\overline A_{r,\rho}
\overline A_{s,\rho}\rangle_\rho$,
whose diagonal entries are $(\Delta_\rho A_r)^2$.
The listed hybrid uncertainty is
\begin{align}
\Mfun_\rho(\bm A)
&\coloneqq
\sum_{r<s}
\Delta_\rho A_r\Delta_\rho A_s.
\label{eq:main-listed}
\end{align}

To minimize over orthogonal combinations of the prescribed observables, let
$\lambda_1,\ldots,\lambda_N\geqslant0$
be the eigenvalues of $\Gamma_\rho(\bm A)$.
We introduce the spectral hybrid uncertainty by
\begin{align}
\Mfun_\rho^{\mathrm{sp}}(\bm A)
&\coloneqq
\sum_{i<j}\sqrt{\lambda_i\lambda_j}
=\min_{O\in\mathrm O(N)}
\Mfun_\rho(O\bm A).
\label{eq:main-spectral}
\end{align}
The minimum is attained precisely when
$O\Gamma_\rho(\bm A)O^{\T}$ is diagonal.
Thus, for a fixed quantum state, the spectral uncertainty
is the minimum listed hybrid uncertainty obtainable by measuring
orthogonal linear combinations of the prescribed observables.
It depends only on the covariance eigenvalues,
whereas the listed hybrid uncertainty also depends on the
choice of measured observables~\cite{RivasLuis2008}.
For unbounded generators, we impose the domain and
finite-moment conditions specified in
Subsec.~\ref{sec:su11-weight-states}.

The second covariance invariant is
$C_2^{(N)}(\Gamma_{\rho}(\bm A))=\sum_{i<j}\lambda_i\lambda_j$,
the sum of the $2\times2$ principal minors of $\Gamma_{\rho}(\bm A)$.
The covariance inequalities established in
Appendix~\ref{Covariance_based_inequalities} give
\begin{align}
\Mfun_\rho(\bm A)
&\geqslant
\Mfun_\rho^{\mathrm{sp}}(\bm A)
\geqslant
\sqrt{C_2^{(N)}(\Gamma_\rho(\bm A))}\,.
\label{eq:main-covariance-chain}
\end{align}
For a reference list
$\bm X=(X_1,\ldots,X_r)^{\T}$
and $\bm A=M\bm X$, the covariance transforms as
$\Gamma_\rho(\bm A)
=M\Gamma_\rho(\bm X)M^{\T}$.
If $M\in\mathbb R^{N\times r}$ has full column rank
with $r\geqslant2$, this yields
\begin{align}
\Mfun_\rho^{\mathrm{sp}}(M\bm X)
&\geqslant
\sigma_1(M)\sigma_2(M)
\sqrt{C_2^{(r)}(\Gamma_\rho(\bm X))}\, ,
\label{eq:main-covariance-transfer}
\end{align}
where $\sigma_1(M)\leqslant\sigma_2(M)$
are its two smallest singular values.
The transfer inequality therefore gives bounds for spanning lists $M\bm X$ from the covariance constraints on $\bm X$.

A complementary bound applies to the prescribed observable list without orthogonal mixing.
The pairwise Robertson uncertainty relations give
\begin{align}
\Mfun_\rho(\bm A)
&\geqslant
\frac12\sum_{r<s}
\left|\langle[A_r,A_s]\rangle_\rho\right|.
\label{eq:main-commutator-bound}
\end{align}
For Lie algebra generators, the commutator bound depends only on first moments and can lie strictly below the quantum minimum.
Consider the qubit spin triple $\bm J=(J_x,J_y,J_z)^{\T}$, with $J_a=\sigma_a/2$.
For the maximally mixed state $\rho=\1_2/2$, the commutator expectations vanish and $\Mfun_\rho(\bm J)=3/4$, so the optimized commutator bound is zero but is not attained by this state.
The quantum minimum is $\min_\rho\Mfun_\rho(\bm J)=1/4$, attained by a spin eigenstate.
The minimizing first moments are therefore realized by a quantum state, although the optimized commutator bound $0$ is strictly below the quantum minimum $1/4$.
We next ask whether this gap can persist for a positive optimized commutator bound.
We establish an exactness criterion for the positive discrete series in Sec.~\ref{sec:main-criterion} and demonstrate a quantitative positive gap in Sec.~\ref{sec:main-gap}.

%%%%%%%%%%%%%%%%%%%%%%%%%%%%%%%%%%%%%%%%%%%%%%%%%%%%%%%%%%%%%%%%%%%%%%%%%%%%%%%%%%%%%%%%%%%%%%%%%%%%%%%%%%%%%%%%%%%%%%
\section{Exactness of the Optimized Commutator Bound}
\label{sec:main-criterion}

\begin{figure}[t]
\centering
\includegraphics[
    width=0.9\columnwidth
]{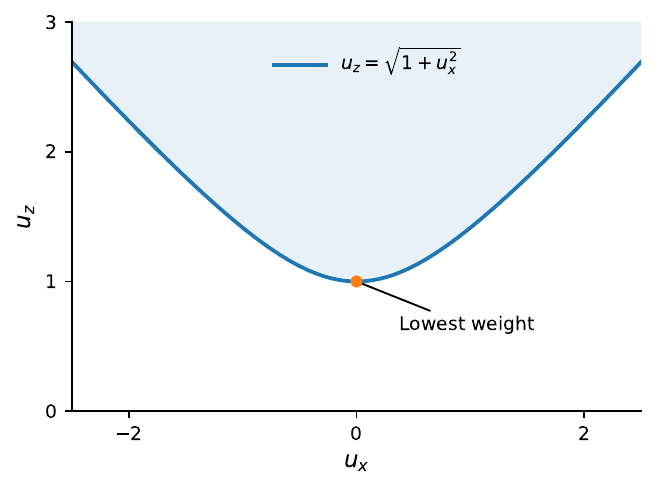}
\caption{\textbf{Allowed first-moment geometry.}
The normalized generator means in the positive
discrete series satisfy
$u_z\geqslant\sqrt{1+u_x^2+u_y^2}$.
The shaded region shows the section $u_y=0$,
whose boundary is realized by Perelomov coherent
states.
The point $(u_x,u_z)=(0,1)$ corresponds to
the lowest-weight state.
The commutator bound is optimized over this
representation-constrained region.}
\label{fig:first-moment-geometry}
\end{figure}

We now consider the positive discrete series
$\mathcal D_k^+$ of $\mathfrak{su}(1,1)$,
where $k>0$ is the lowest eigenvalue of $K_z$.
Let $\bm K=(K_x,K_y,K_z)^{\T}$ and denote by
$\mathcal S_2(\mathcal D_k^+)$ the states with finite
second moments of these generators.
The representation constrains their normalized
expectation values $\bm u=\langle\bm K\rangle_\rho/k$
to the region
\begin{align}
\mathcal C_1
&\coloneqq
\left\{
\bm u\in\mathbb R^3:
u_z\geqslant\sqrt{1+u_x^2+u_y^2}
\right\}.
\label{eq:main-mean-region}
\end{align}
The commutator bound is optimized over this allowed mean region, illustrated in Fig.~\ref{fig:first-moment-geometry}.

%%%%%%%%%%%%%%%%%%%%%%%%%%%%%%%%%%%%%%%%%%%%%%%%%%%%%%%%%%%%%%%%%%%%%%%%%%%%%%%%%%%%%%%%%%%%%%%%%%%%%%%%%%%%%%%%%%%%%%%%%%%%%%%%%

\subsection{The optimized commutator bound}

Consider a prescribed list of real linear combinations of the generators
$\bm A=M\bm K$, where
$M\in\mathbb R^{N\times3}$ has rank three and no zero row.
Writing the $r$th row of $M$ as $\bm m_r^{\T}$,
the $\mathfrak{su}(1,1)$ commutation relations give
\begin{align}
[A_r,A_s]
&=\ii\,\bm c_{rs}\cdot\bm K,
\quad
\bm c_{rs}
=\eta_3(\bm m_r\times\bm m_s),
\label{eq:main-frame-commutators}
\end{align}
where $\eta_3=\operatorname{diag}(1,1,-1)$.
Consequently,
$|\langle[A_r,A_s]\rangle_\rho|
=k|\bm c_{rs}\cdot\bm u|$.
The commutator bound therefore depends on the state
only through its first moments, reducing its optimization
to the allowed region $\mathcal C_1$.

We define the optimized commutator coefficient by
\begin{align}
\gamma_N(M)
&\coloneqq
\min_{\bm u\in\mathcal C_1}
\sum_{r<s}
|\bm c_{rs}\cdot\bm u|.
\label{eq:main-conic-coefficient}
\end{align}
The absolute value objective and the constraint defining $\mathcal C_1$ admit a second-order cone formulation, given explicitly in Appendix~\ref{app:su11-proofs}.
Unlike the singular value estimate, this optimization uses the commutator vectors $\bm c_{rs}$ and the constraint $\bm u\in\mathcal C_1$.

By Thm.~\ref{thm:su11_conic_hybrid_bound},
every $\rho\in\mathcal S_2(\mathcal D_k^+)$ satisfies
\begin{align}
\Mfun_\rho(M\bm K)
&\geqslant
\frac{k}{2}\gamma_N(M)
\geqslant
\frac{k}{2}\sigma_1(M)\sigma_2(M),
\label{eq:main-conic-bound}
\end{align}
where $\sigma_1(M)$ and $\sigma_2(M)$ are the two smallest singular values of $M$.
The first inequality follows by minimizing the commutator expectations, but it need not be tight.
Equality requires one state to realize the minimizing mean and saturate every pairwise commutator bound.
We next give a necessary and sufficient covariance condition for this equality.
%%%%%%%%%%%%%%%%%%%%%%%%%%%%%%%%%%%%%%%%%%%%%%%%%%%%%%%%%%%%%%%%%%%%%%%%%%%%%%%%%%%%%%%%%%%%%%%%%%%%%%%%%%%%%%%%%%%%%%%%%%%%%%%%%
\subsection{The covariance criterion}

The commutator objective has a unique minimizing mean $\bm u_M$, which lies on the boundary $u_{M,z}^2-u_{M,x}^2-u_{M,y}^2=1$.
This mean determines a unique admissible state $\rho_M$, which is a Perelomov coherent state.
Define
\begin{align}
T_M
&=M(\eta_3+\bm u_M\bm u_M^{\T})M^{\T},
\end{align}
so that $\Gamma_{\rho_M}(M\bm K)=kT_M/2$.

\begin{thm}
\label{thm:su11_conic_criterion}
Let $M\in\mathbb R^{N\times3}$ have rank three and no zero row.
For every $k>0$, the equality
\begin{align}
\inf_{\rho\in\mathcal S_2(\mathcal D_k^+)}
\Mfun_\rho(M\bm K)
&=\frac{k}{2}\gamma_N(M)
\label{eq:main-conic-criterion}
\end{align}
holds if and only if $T_M$ is diagonal.
When this condition holds, the infimum is attained uniquely by $\rho_M$.
Otherwise, the quantum infimum is strictly greater than the optimized commutator bound.
\end{thm}

For fixed $M$, checking the covariance of $\rho_M$ determines whether the optimized commutator bound equals the infimum over all admissible states.
Since $T_M$ is independent of $k$, the condition is the same for every $k>0$, while the bound scales linearly with $k$.

The covariance condition follows from the pairwise uncertainty relations.
In the state $\rho_M$, every Robertson-Schr\"odinger inequality is saturated~\cite{TrifonovDonev1998,Trifonov2000}, as shown in Appendix~\ref{sec:su11-conic-criterion}.
A pairwise commutator bound is then saturated if and only if the corresponding real covariance vanishes.
Thus, all off-diagonal covariances must vanish for the optimized bound to be exact; orthogonal mixing would change the prescribed list and the bound being tested.

%%%%%%%%%%%%%%%%%%%%%%%%%%%%%%%%%%%%%%%%%%%%%%%%%%%%%%%%%%%%%%%%%%%%%%%%%%%%%%%%%%%%%%%%%%%%%%%%%%%%%%%%%%%%%%%%%%%%%%%%%%%%%%%%%
\subsection{Why the gap is strictly positive}

In an infinite-dimensional Hilbert space, failure to attain a lower bound does not imply that the quantum infimum is strictly greater than that bound.
To exclude approximate saturation, suppose that a sequence of admissible states $\rho_n$ satisfies $\Mfun_{\rho_n}(M\bm K)\to k\gamma_N(M)/2$.
Since the commutator objective diverges as $\|\bm u\|_2\to\infty$ and has a unique minimizer, the normalized means $\langle\bm K\rangle_{\rho_n}/k$ converge to $\bm u_M$.
The operator $H_{\bm u_M}=u_{M,z}K_z-u_{M,x}K_x-u_{M,y}K_y$ is unitarily equivalent to $K_z$, with a unique ground state projector $\rho_M$ and a unit spectral gap.
Convergence of its expectation to $k$ therefore forces $\rho_n$ to converge to $\rho_M$ in trace norm, as established in Appendix~\ref{sec:su11-conic-criterion}.

Trace norm convergence alone does not guarantee convergence of the second moments of unbounded observables.
Applying trace norm convergence to bounded spectral truncations shows that the second moments are lower semicontinuous.
Together with convergence of the first moments, this gives
\begin{align}
\Mfun_{\rho_M}(M\bm K)
\leqslant
\liminf_{n\to\infty}
\Mfun_{\rho_n}(M\bm K)
=
\frac{k}{2}\gamma_N(M).
\end{align}
If $T_M$ is not diagonal, however, $\Mfun_{\rho_M}(M\bm K)$ strictly exceeds this bound, contradicting the preceding inequality.
No sequence of admissible states can therefore approach the optimized commutator bound, so the quantum infimum is strictly greater than $k\gamma_N(M)/2$.
The full argument, including the treatment of unbounded second moments, is given in Appendix~\ref{sec:su11-conic-criterion}.

%%%%%%%%%%%%%%%%%%%%%%%%%%%%%%%%%%%%%%%%%%%%%%%%%%%%%%%%%%%%%%%%%%%%%%%%%%%%%%%%%%%%%%%%%%%%%%%%%%%%%%%%%%%%%%%%%%%%%%%%%%%%%%%%%

\section{Exact Bounds and Strict Quantum Gaps}
\label{sec:main-gap}
%The covariance criterion established in Sec.~\ref{sec:main-criterion} distinguishes when the optimized commutator bound is exact and when it lies strictly below the quantum infimum.
We apply the exactness criterion first to observables related by Lorentz transformations, for which the bound is attained, and then to a nonorthogonal list with a strictly positive gap.

\subsection{Exact Minima under Lorentz Transformations}

Consider the prescribed observables $\bm A=DL\bm K$, where $D=\operatorname{diag}(a,b,c)$ with $a,b,c>0$, and $L$ is a Lorentz transformation preserving the allowed mean region $\mathcal C_1$.
We require $L$ to admit the unitary implementation specified in Eq.~\eqref{eq:su11-lorentz-unitary} of Appendix~\ref{app:su11-proofs}.

For this family, the commutator objective has a unique minimizer $\bm u_M=L^{-1}(0,0,1)^{\T}$, with $\gamma_3(DL)=ab$.
The covariance matrix of the corresponding coherent state satisfies $T_{DL}=\operatorname{diag}(a^2,b^2,0)$ and is therefore diagonal.
The exactness criterion in Eq.~\eqref{eq:main-conic-criterion} then gives
\begin{align}
\min_{\rho}\Mfun_\rho(DL\bm K)
&=
\min_{\rho}\Mfun_\rho^{\mathrm{sp}}(DL\bm K)
=
\frac{abk}{2}.
\label{eq:main-lorentz-exact}
\end{align}
Both minima are attained by the same coherent state, as established in Cor.~\ref{cor:su11_lorentz_frame_minima} of Appendix~\ref{app:su11-proofs}.

\begin{figure}[t]
\centering
\includegraphics[width=0.9\columnwidth]{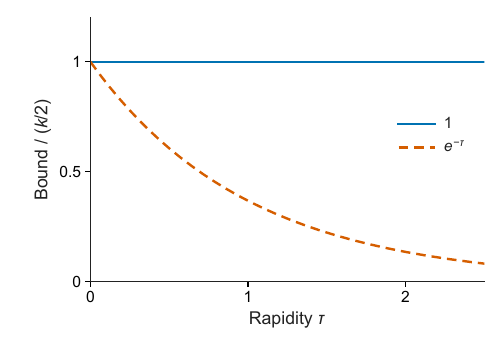}
\caption{\textbf{Exact minima under Lorentz boosts.}
For an $xz$ Lorentz boost $L_\tau$, the singular-value estimate decreases as $(k/2)e^{-\tau}$, whereas the optimized commutator bound and the exact quantum minimum remain $k/2$ for all $\tau\geqslant0$.
The ratio between the exact minimum and the singular value estimate grows as $e^\tau$.}
\label{fig:lorentz-benchmark}
\end{figure}

For an $xz$ Lorentz boost $L_\tau$ with rapidity $\tau\geqslant0$, the singular values are $e^{-\tau}$, $1$, and $e^\tau$.
Consequently, the singular-value estimate decreases as $(k/2)e^{-\tau}$, whereas the optimized commutator bound and the exact quantum minimum both remain equal to $k/2$.
Figure~\ref{fig:lorentz-benchmark} shows that the ratio of the exact minimum to the singular value estimate grows as $e^\tau$.
As $\tau\to\infty$, the singular value bound approaches zero, while the exact minimum remains $k/2$.

%%%%%%%%%%%%%%%%%%%%%%%%%%%%%%%%%%%%%%%%%%%%%%%%%%%%%%%%%%%%%%%%%%%%%%%%%%%%%%%%%%%%%%%%%%%%%%%%%%%%%%%%%%%%%%%%%%%%%%
\subsection{A Strict Gap above the Optimized Commutator Bound}

We now consider a nonorthogonal observable list for which the optimized commutator bound is not tight.
Let $\bm A_*=(K_x,K_y,K_x+K_z)^{\T}$ in the positive discrete series.
Its optimized commutator coefficient is $\gamma_3(M_*)=\sqrt3$, with unique minimizing mean $\bm u_*=(-1,0,2)^{\T}/\sqrt3$.
The coherent state $\rho_{\mathrm m}$ realizing this mean satisfies
$\operatorname{Cov}_{\rho_{\mathrm m}}(K_x,K_x+K_z)=k/3$.
By Thm.~\ref{thm:su11_conic_criterion}, the quantum infimum is therefore strictly above the optimized commutator bound $k\sqrt3/2$.
We bound the gap explicitly and compare the quantum minimum with the minimum over mixtures of Perelomov coherent states.

% Let $m_Q(k)$ denote the minimum of
% $\Mfun_\rho(\bm A_*)$ over
% $\mathcal S_2(\mathcal D_k^+)$, and let
% $m_{\mathrm{coh}}(k)$ denote its minimum over
% Perelomov coherent-state mixtures with finite second moments.

\begin{figure}[t]
\centering
\includegraphics[width=0.9\columnwidth]{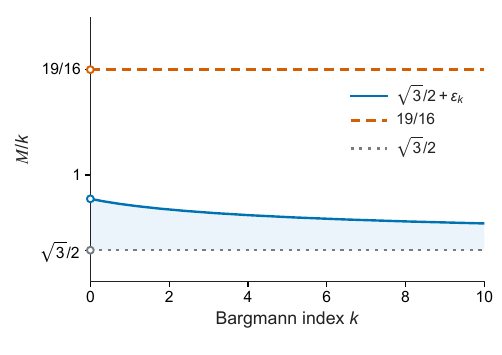}
\caption{\textbf{Gap above the optimized commutator bound.}
For $\bm A_*=(K_x,K_y,K_x+K_z)^{\T}$, the gray dotted line denotes the optimized commutator bound $\sqrt3/2$ in units of $k$.
The blue curve gives the certified lower bound $\sqrt3/2+\epsilon_k$ on $\frac1k\min_{\rho\in \mathcal S_2(\mathcal D_k^+)}\Mfun_\rho(\bm A_*)$, while the orange dashed line is the exact coherent mixture minimum $19/16$.
The shaded region between the gray and blue curves certifies a strictly positive quantum gap.
The exact quantum minimum is not plotted.}
\label{fig:quantum-gap}
\end{figure}

\begin{thm}
\label{thm:su11_quantitative_quantum_gap}
For every Bargmann index $k>0$, define
$\epsilon_k=[4(1+\sqrt{(3k+6)/2})]^{-1}>0$.
Then
\begin{align}
\frac{\sqrt3}{2}+\epsilon_k
&\leqslant
%\frac{m_Q(k)}{k}
\frac{1}{k}\min_{\rho\in \mathcal S_2(\mathcal D_k^+)}\Mfun_\rho(\bm A_*)
<
\frac{19}{16}
%=\frac{m_{\mathrm{coh}}(k)}{k}.
\label{eq:main-quantum-gap}
\end{align}
The quantum minimum is attained by a pure state.
\end{thm}
Thus, no mixture of Perelomov coherent states minimizes the listed hybrid uncertainty.
The exact value of $\min_{\rho\in \mathcal S_2(\mathcal D_k^+)}\Mfun_\rho(\bm A_*)$ remains unknown.
Figure~\ref{fig:quantum-gap} displays the proved bounds without plotting the unknown minimum.

The coherent state realizing the minimizing mean differs from the coherent state minimizing the listed hybrid uncertainty.
The latter has normalized mean $(-3/4,0,5/4)^{\T}$, rather than $\bm u_*=(-1,0,2)^{\T}/\sqrt3$.
Moreover, the total variance formula and the Cauchy-Schwarz inequality show that mixtures of Perelomov coherent states cannot reduce the uncertainty below $19k/16$.
A pure state obtained by perturbing the optimal coherent state has strictly smaller uncertainty, as demonstrated by Eq.~\eqref{eq:su11-lowering-perturbation} and Cor.~\ref{cor:su11_coherent_state_limitation}.
In this example, the mean minimizing the commutator bound differs from that of the optimal coherent state, and the best coherent state does not minimize the uncertainty over all quantum states.

A single bosonic mode provides a physical realization of the $\mathfrak{su}(1,1)$ generators through quadratic field quadratures~\cite{BrifMann1997}.
For a single bosonic mode, let
$X=(a+a^\dagger)/\sqrt2$ and
$P=(a-a^\dagger)/(i\sqrt2)$
be the canonical quadratures, where
$[a,a^\dagger]=1$ and $[X,P]=\ii$.
The $\mathfrak{su}(1,1)$ generators are
$K_x=\frac{X^2-P^2}{4},
K_y=-\frac{XP+PX}{4},
K_z=\frac{X^2+P^2}{4}.$
The nonorthogonal list $\bm A_*$ includes
$A_3=K_x+K_z=X^2/2$.
The subspaces spanned by the even photon number states $\{|0\rangle,|2\rangle,\ldots\}$ and the odd
photon number states $\{|1\rangle,|3\rangle,\ldots\}$
carry positive discrete series representations with $k=1/4$ and $k=3/4$, respectively.
Since the observables in $\bm A_*$ are quadratic in $X$ and $P$, their variances involve fourth order moments.
For either representation, Eq.~\eqref{eq:main-quantum-gap}
gives a lower bound on the hybrid uncertainty strictly above the optimized commutator bound for every density operator $\rho$ supported on the corresponding subspace and satisfying
$\langle K_a^2\rangle_\rho<\infty$
for $a=x,y,z$.

%%%%%%%%%%%%%%%%%%%%%%%%%%%%%%%%%%%%%%%%%%%%%%%%%%%%%%%%%%%%%%%%%%%%%%%%%%%%%%%%%%%%%%%%%%%%%%%%%%%%%%%%%%%%%%%%%%%%%%

\section{Spectral Uncertainty in Spin Systems}
\label{sec:main-spin}

For spin observables, we minimize the spectral hybrid uncertainty over quantum states and orthogonal combinations of the prescribed observables.

\subsection{Exact Spectral Bounds for Spin Observables}

For the irreducible spin-$j$ representation with $j>0$, let $\bm J=(J_x,J_y,J_z)^{\T}$ denote the self-adjoint angular momentum generators.
Their covariance matrix satisfies the tight state-independent inequality~\cite{ShabbirBjork2016}
\begin{align}
C_2^{(3)}\bigl(\Gamma_\rho(\bm J)\bigr)
&\geqslant \frac{j^2}{4}.
\label{eq:main-spin-covariance-bound}
\end{align}
For $j\geqslant1$, Thm.~\ref{thm:Spin_covariance_refinement} in Appendix~\ref{app:su2-proofs} further refines this bound by retaining the mean spin length.
With the covariance transfer relation of Sec.~\ref{sec:main-functionals}, this bound gives the exact spectral minimum for any spanning list of spin observables.

\begin{thm}
\label{thm:spin-frame-minimum}
Let $\bm A=M\bm J$, where $M\in\mathbb R^{N\times3}$ has rank three, with singular values $0<\sigma_1(M)\leqslant\sigma_2(M)\leqslant\sigma_3(M)$.
For every irreducible spin-$j$ representation with $j>0$,
\begin{align}
\min_\rho\Mfun_\rho^{\mathrm{sp}}(M\bm J)
&=\frac{j}{2}\sigma_1(M)\sigma_2(M),
\label{eq:main-spin-spectral-minimum}
\end{align}
where the minimum is over all states in the representation.
It is attained by any spin coherent state polarized along a unit right singular vector of $M$ associated with $\sigma_3(M)$.
\end{thm}

At such a minimizing state, the transformed covariance has only two nonzero eigenvalues, $j\sigma_1(M)^2/2$ and $j\sigma_2(M)^2/2$.
A spin coherent state polarized along a largest singular direction is optimal; the minimum depends on the two smallest singular values.
For a weighted orthogonal triple, polarization along a coefficient row of maximal norm also diagonalizes the covariance for the prescribed observable list.
Consequently, the listed and spectral uncertainties attain the same minimum,
\begin{align}
\min_\rho\Mfun_\rho(M\bm J)
=
\min_\rho\Mfun_\rho^{\mathrm{sp}}(M\bm J)
=
\frac{j}{2}\sigma_1(M)\sigma_2(M).
\label{eq:main-spin-orthogonal-minima}
\end{align}
For nonorthogonal lists, the minima can differ, as the following qubit example demonstrates.
%%%%%%%%%%%%%%%%%%%%%%%%%%%%%%%%%%%%%%%%%%%%%%%%%%%%%%%%%%%%%%%%%%%%%%%%%%%%%%%%%%%%%%%%%%%%%%%%%%%%%%%%%%%%%%%%%%%%%%%%%%%%%%%%%
\subsection{Nonorthogonal Qubit Observables}

To distinguish the two minimizations, consider the qubit observables $\bm A=(J_x,\cos\theta J_x+\sin\theta J_y,J_z)^{\T}$, where $J_a=\sigma_a/2$ and $0<\theta\leqslant\pi/2$.
The first two observables become nonorthogonal for $0<\theta<\pi/2$, while the three directions continue to span the spin operator space.

\begin{figure}[t]
\centering
\includegraphics[width=0.9\columnwidth]{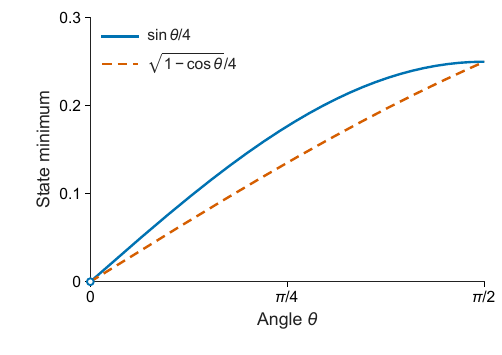}
\caption{\textbf{Listed and spectral minima for nonorthogonal qubit observables.}
The blue solid curve gives $\min_\rho\Mfun_\rho(\bm A)=\sin\theta/4$, while the orange dashed curve gives $\min_\rho\Mfun_\rho^{\mathrm{sp}}(\bm A)=\sqrt{1-\cos\theta}/4$.
The quantum state is optimized independently for each curve.}
%The minima differ for $0<\theta<\pi/2$ and coincide at $\theta=\pi/2$.
%The open points indicate the rank-deficient limit $\theta\to0^+$.}
\label{fig:qubit-minima}
\end{figure}

Theorem~\ref{thm:nonorthogonal_qubit_minima} in
Appendix~\ref{app:angular-proof} gives the exact
state-independent minima
\begin{align}
\min_\rho\Mfun_\rho(\bm A)
&=\frac{\sin\theta}{4},\\
\min_\rho\Mfun_\rho^{\mathrm{sp}}(\bm A)
&=\frac{\sqrt{1-\cos\theta}}{4}.
\label{eq:main-qubit-minima}
\end{align}
The spectral minimum is strictly smaller for
$0<\theta<\pi/2$, and the two minima coincide
at $\theta=\pi/2$, as shown in
Fig.~\ref{fig:qubit-minima}.

For $0<\theta<\pi/2$, the listed minimum is
attained precisely by eigenstates of $A_1$ or $A_2$,
whereas the spectral minimum is attained by
pure states with Bloch vectors
$\bm s_\pm=\pm(\cos(\theta/2),\sin(\theta/2),0)^{\T}$,
along the bisector of the two nonorthogonal directions.
At $\theta=\pi/2$, every pure qubit state
minimizes the spectral uncertainty, while
only eigenstates of the three Cartesian spin
components minimize the listed uncertainty.

The effect of measuring orthogonal combinations
of observables at a fixed state is examined
in Sec.~\ref{sec:main-sampling}.

%%%%%%%%%%%%%%%%%%%%%%%%%%%%%%%%%%%%%%%%%%%%%%%%%%%%%%%%%%%%%%%%%%%%%%%%%%%%%%%%%%%%%%%%%%%%%%%%%%%%%%%%%%%%%%%%%%%%%%%%%%%%%%%%%
\section{Hybrid Uncertainty as a Sampling Cost}
\label{sec:main-sampling}

We now compare measurement protocols in which the target observable is known before or only after the measurement data are collected.
For a fixed state, the difference between the leading mean squared error coefficients of these protocols equals twice the listed hybrid uncertainty after optimal copy allocation.
Fix a quantum state $\rho$ and a list of self-adjoint observables $\bm A=(A_1,\ldots,A_N)^{\T}$.
The task is to estimate the expectation value of a randomly selected target
\begin{align}
H_{\bm s}
&=\sum_{r=1}^{N}s_rA_r,
\quad s_r\in\{-1,+1\},
\label{eq:main-sign-target}
\end{align}
where the signs are independent and uniformly distributed. The two protocols differ in whether these signs are revealed before or after measurement.
We compare the statistical errors of measurement strategies with and without advance knowledge of the target, using the same state and total number of copies.

%%%%%%%%%%%%%%%%%%%%%%%%%%%%%%%%%%%%%%%%%%%%%%%%%%%%%%%%%%%%%%%%%%%%%%%%%%%%%%%%%%%%%%%%%%%%%%%%%%%%%%%%%%%%%%%%%%%%%%

\subsection{Direct and Separate Measurement Protocols}

\begin{figure*}[t]
\centering

\begin{minipage}[t]{0.60\textwidth}
    \vspace{0pt}
    \centering \includegraphics[width=\linewidth]    {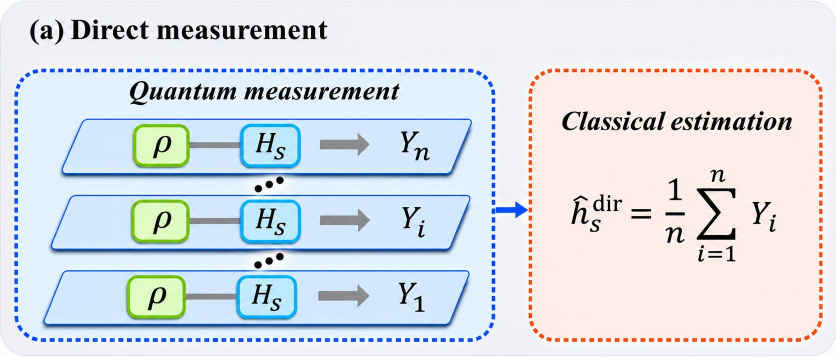}   \includegraphics[width=\linewidth]    {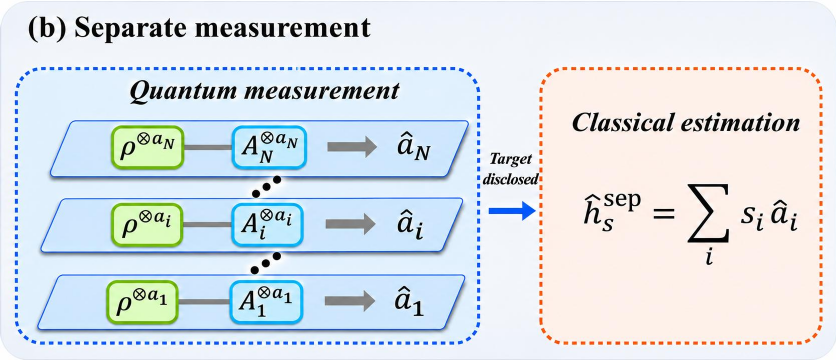}
\end{minipage}
\hfill
\begin{minipage}[t]{0.38\textwidth}
    \vspace{0pt}
    \centering    \includegraphics[width=\linewidth]
    {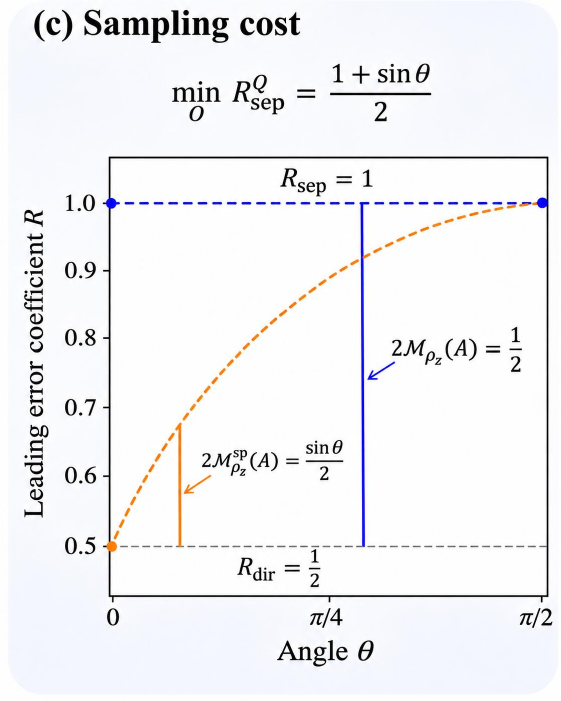}
\end{minipage}
\caption{\textbf{Direct and separate measurement protocols.}
The target is $H_{\bm s}=\sum_r s_rA_r$, with independent uniform signs $s_r=\pm1$ and $n$ independent copies of $\rho$.
\textbf{(a)} The direct protocol measures $H_{\bm s}$ after $\bm s$ is disclosed.
\textbf{(b)} The separate protocol measures each $A_r$ on its own batch before disclosure, then combines the sample means using the revealed signs.
\textbf{(c)} For $\rho_z=(\1_2+\sigma_z)/2$ and $\bm A=(J_x,\cos\theta J_x+\sin\theta J_y,J_z)^{\T}$, with $J_a=\sigma_a/2$, the curves show the leading target averaged error coefficients in $\operatorname{MSE}=R/n+o(n^{-1})$: $R_{\mathrm{dir}}=1/2$, $R_{\mathrm{sep}}=1$ after optimal batch allocation, and $\min_O R_{\mathrm{sep}}^O=(1+\sin\theta)/2$ after orthogonal mixing.
The arrows show the full excesses $2\Mfun_{\rho_z}(\bm A)=1/2$ and $2\Mfun_{\rho_z}^{\mathrm{sp}}(\bm A)=\sin\theta/2$ relative to direct measurement.}
\label{fig:sampling-cost-protocol}
\end{figure*}

Both protocols use $n$ independently prepared copies of $\rho$, as shown in Fig.~\ref{fig:sampling-cost-protocol}(a,b).
In the direct measurement protocol, the target label $\bm s$ is known before measurement, allowing $H_{\bm s}$ to be measured on every copy and its expectation estimated from the sample mean.
In the prescribed separate measurement protocol, the label is withheld until after measurement, so the copies are divided into batches of sizes $n_r$, with each batch used to measure one observable $A_r$.
Once $\bm s$ is revealed, the target expectation is estimated by combining the component sample means as $\widehat h_{\bm s}=\sum_r s_r\widehat a_r$.
The measurement settings and batch allocation are fixed before disclosure, with the component variances assumed available for optimal allocation.
Both protocols use measurements on individual copies, and the estimation budget excludes the cost of learning these variances.
The precise protocol and assumptions are specified in Def.~\ref{def:sign_target_estimation} of Appendix~\ref{app:hybrid-applications}.

Write the target averaged mean squared error as $\operatorname{MSE}=R/n+o(n^{-1})$, where $R$ is the leading coefficient.
Writing $d_r=\Delta_\rho A_r$, averaging over the independent random signs eliminates the cross covariances in the direct measurement protocol, giving $R_{\mathrm{dir}}=\sum_r d_r^2$.
For the separate measurement protocol, independent measurement batches contribute $\sum_r d_r^2/n_r$ to the estimation error.
The optimal allocation assigns copies in proportion to the component standard deviations, $n_r/n\to d_r/\sum_t d_t$, yielding $R_{\mathrm{sep}}=(\sum_r d_r)^2$.
Subtracting the direct coefficient from the optimized separate coefficient gives an excess of twice the listed hybrid uncertainty, as stated in Prop.~\ref{prop:main-sampling-identities} of subsec.~\ref{subsec:sampling_costs}.
The derivation, including zero-variance observables and integer sample allocations, is given in Thm.~\ref{thm:optimal_allocation_hybrid_excess} of Appendix~\ref{app:hybrid-applications}.

%%%%%%%%%%%%%%%%%%%%%%%%%%%%%%%%%%%%%%%%%%%%%%%%%%%%%%%%%%%%%%%%%%%%%%%%%%%%%%%%%%%%%%%%%%%%%%%%%%%%%%%%%%%%%%%%%%%%%%

\subsection{Orthogonal Mixing and Sampling Costs}\label{subsec:sampling_costs}

The preceding comparison assumes that the prescribed observables are measured separately.
We now allow their real orthogonal combinations to be selected before the target is disclosed.
For $O\in\mathrm O(N)$, let $\bm B=O\bm A$ and denote by $R_{\mathrm{sep}}^O$ the optimally allocated error coefficient for separate measurements of $\bm B$.
The original target is reconstructed with $\bm t=O\bm s$, since $H_{\bm s}=\bm t^{\T}\bm B$.
Although the transformed coefficients are generally neither independent nor restricted to signs, they satisfy $\mathbb E_{\bm s}[\bm t\bm t^{\T}]=\1_N$, leaving the target-averaged direct error unchanged.
The excess error coefficients are given by the listed and spectral hybrid uncertainties.

\begin{prop}
\label{prop:main-sampling-identities}
For a fixed state $\rho$ and a prescribed list $\bm A$ of self-adjoint observables with finite second moments, the delayed-target protocols satisfy
\begin{align}
R_{\mathrm{sep}}-R_{\mathrm{dir}}
&=2\Mfun_\rho(\bm A),
\label{eq:main-hybrid-cost}\\
\min_{O\in\mathrm O(N)}
\left[R_{\mathrm{sep}}^O-R_{\mathrm{dir}}\right]
&=2\Mfun_\rho^{\mathrm{sp}}(\bm A).
\label{eq:main-spectral-cost}
\end{align}
The second minimum is attained when $O$ diagonalizes $\Gamma_\rho(\bm A)$.
\end{prop}

Equation~\eqref{eq:main-hybrid-cost} gives the excess for the fixed list; Eq.~\eqref{eq:main-spectral-cost} gives its minimum under orthogonal mixing.
Their difference, $2[\Mfun_\rho(\bm A)-\Mfun_\rho^{\mathrm{sp}}(\bm A)]$, is the decrease in the separate measurement error coefficient obtained by optimal orthogonal mixing at fixed $\rho$.
Realizing this reduction generally requires changing the quantum measurements rather than merely processing the original measurement outcomes.
The two identities follow from Thms.~\ref{thm:optimal_allocation_hybrid_excess} and~\ref{thm:spectral_hybrid_sampling_excess} in Appendix~\ref{app:hybrid-applications}.

To illustrate this reduction, consider the qubit state $\rho_z=(\1_2+\sigma_z)/2$ and the nonorthogonal list $\bm A=(J_x,\cos\theta J_x+\sin\theta J_y,J_z)^{\T}$, where $J_a=\sigma_a/2$ and $0<\theta\leqslant\pi/2$.
For this fixed preparation, the three acquisition strategies yield
\begin{align}
R_{\mathrm{dir}}=\frac12,\,
R_{\mathrm{sep}}=1,\,
\min_{O\in\mathrm O(3)}R_{\mathrm{sep}}^O
=\frac{1+\sin\theta}{2}.
\label{eq:main-qubit-sampling}
\end{align}
Thus, orthogonal mixing reduces the sampling excess from $1/2$ to $\sin\theta/2$, without changing the quantum state or the target distribution.
As shown in Fig.~\ref{fig:sampling-cost-protocol}\textcolor{red}{(c)}, the reduction is largest as the two transverse measurement directions approach coincidence and vanishes when they become orthogonal.

The strict gap above the optimized commutator bound established in Sec.~\textcolor{red}{\ref{sec:main-gap}} has a direct consequence for the delayed-target estimation task.
For the nonorthogonal list $\bm A_*=(K_x,K_y,K_x+K_z)^{\T}$, combining Eqs.~\eqref{eq:main-hybrid-cost} and~\eqref{eq:main-quantum-gap} gives, for every admissible preparation $\rho\in\mathcal S_2(\mathcal D_k^+)$,
\begin{align}
R_{\mathrm{sep}}(\rho;\bm A_*)
-R_{\mathrm{dir}}(\rho;\bm A_*)
&\geqslant k(\sqrt3+2\epsilon_k),
\label{eq:main-sampling-gap}
\end{align}
where $\epsilon_k>0$ is defined in Eq.~\eqref{eq:main-quantum-gap}.
First moment optimization alone would yield only the weaker lower bound $k\sqrt3$.
Even after optimizing the state and batch allocation, the excess coefficient is at least $2k\epsilon_k$ larger than the estimate from the optimized commutator bound.
After orthogonal mixing, the excess is instead given by Eq.~\eqref{eq:main-spectral-cost}; the bound in Eq.~\eqref{eq:main-sampling-gap} concerns the original list.

For a fixed target error, the excess sampling coefficient determines the leading difference in the required numbers of copies.
Let $n_{\mathrm{sep}}^*(\varepsilon)$ and $n_{\mathrm{dir}}^*(\varepsilon)$ denote the minimum numbers of independently prepared copies required by the two protocols to achieve a target-averaged root mean square error at most $\varepsilon$.
Corollary~\ref{cor:finite_copy_requirements} in Appendix~\ref{app:hybrid-applications} gives
\begin{align}
n_{\mathrm{sep}}^*(\varepsilon)
-n_{\mathrm{dir}}^*(\varepsilon)
&=
\frac{2\Mfun_\rho(\bm A)}{\varepsilon^2}
+O(1),
\quad \varepsilon\to0.
\label{eq:main-finite-copy-cost}
\end{align}
The $O(1)$ correction arises from integer sample allocations and minimum batch sizes.
These copy requirements exclude the calibration cost of determining the component variances or, when choosing orthogonal combinations, the covariance matrix.
These copy bounds apply to the specified protocols using measurements on individual copies; they do not imply optimality over adaptive or collective measurements.

%%%%%%%%%%%%%%%%%%%%%%%%%%%%%%%%%%%%%%%%%%%%%%%%%%%%%%%%%%%%%%%%%%%%%%%%%%%%%%%%%%%%%%%%%%%%%%%%%%%%%%%%%%%%%%%%%%%%%%

\section{Discussion}
\label{sec:main-discussion}
A quantum state can realize the mean minimizing the commutator bound without attaining the bound itself.
For any spanning list in the positive discrete series of $\mathfrak{su}(1,1)$, this mean determines a unique Perelomov coherent state.
The optimized bound equals the quantum infimum precisely when the real covariances between the prescribed observables vanish in that state.
If these covariances do not vanish, no sequence of admissible states approaches the commutator bound, and the quantum infimum is strictly larger.
Coherent states realize boundary means, but need not minimize hybrid uncertainty for a fixed list of observables.
For the nonorthogonal list $\boldsymbol{A}_*=(K_x,K_y,K_x+K_z)^{\T}$, the best coherent state also minimizes the uncertainty over coherent state mixtures, but a pure state outside this family gives a strictly smaller value.

In the delayed target task, the listed hybrid uncertainty equals half the excess mean squared error coefficient of the prescribed separate protocol relative to direct target measurement; the spectral hybrid uncertainty gives half the minimum excess over orthogonal combinations.
For $\boldsymbol{A}_*$, the strict gap provides a stronger lower bound on the separate protocol's excess coefficient than optimizing the commutator expectations alone.
The protocols use independent copies and assume known component variances, as well as a known covariance matrix when orthogonal mixing is optimized.
The identities do not establish optimality over general quantum measurements.

Our criterion determines when the optimized commutator bound is exact, but it does not in general determine the size of a strict gap.
For $\boldsymbol{A}_*$, the exact value of $\min_{\rho\in \mathcal S_2(\mathcal D_k^+)}\Mfun_\rho(\bm A_*)$ and the form of the pure states attaining it remain to be determined.
For other representations, one must first establish whether a minimizing mean determines a unique state.
If it does not, proving a strict gap requires controlling sequences of states whose means approach the minimizing value.

%%%%%%%%%%%%%%%%%%%%%%%%%%%%%%%%%%%%%%%%%%%%%%%%%%%%%%%%%%%%%%%%%%%%%%%%%%%%%%%%%%%%%%%%%%%%%%%%%%%%%%%%%%%%%%%%%%%%%%
\section*{Acknowledgments}
X.H.
thanks Yunlong Xiao and Bing Yu for valuable discussions.
This research is supported by the China Scholarship Council and the Research Fund of Jianghan University (Grant No.~2023JCYJ08).
N. J. is funded by the Simons Foundation (Grant No.~MP-TSM-00002518).

%%%%%%%%%%%%%%%%%%%%%%%%%%%%%%%%%%%%%%%%%%%%%%%%%%%%%%%%%%%%%%%%%%%%%%%%%%%%%%%%%%%%%%%%%%%%%%%%%%%%%%%%%%%%%%%%%%%%%%
\bigskip
\section*{Author Contributions}

All authors contributed to the research and preparation of the manuscript.
ChatGPT-6 was used as an interactive aid for language editing, bibliographic research, figure-layout refinement, and LaTeX consistency checks.
The authors independently reviewed the results and take full responsibility for the content of the manuscript.

%%%%%%%%%%%%%%%%%%%%%%%%%%%%%%%%%%%%%%%%%%%%%%%%%%%%%%%%%%%%%%%%%%%%%%%%%%%%%%%%%%%%%%%%%%%%%%%%%%%%%%%%%%%%%%%%%%%%%%

\bibliography{Bib}

%=========================================================
% Supplemental Material
% =========================================================

\clearpage
\onecolumngrid
\appendix

%%%%%%%%%%%%%%%%%%%%%%%%%%%%%%%%%%%%%%%%%%%%%%%%%%%%%%%%%%%%%%%%%%%%%%%%%%%%%%%%%%%%%%%%%%%%%%%%%%%%%%%%%%%%%%%%%%

\section*{Appendices}

Throughout the appendices, observable lists may contain linearly dependent operators.
Appendix~\ref{Covariance_based_inequalities} introduces the covariance functionals and proves transfer bounds for real linear combinations.
Appendix~\ref{app:su2-proofs} treats spin covariance, gives the exact spectral minimum for spanning lists of spin observables, and analyzes a nonorthogonal qubit example.
Appendix~\ref{app:su11-proofs} treats the positive discrete series, including the required domain conditions, optimized commutator bounds, and attainability questions.
Appendix~\ref{app:hybrid-applications} connects the listed and spectral hybrid uncertainties with sampling costs for separate measurements and their orthogonal combinations.

\addtocontents{toc}{\protect\AppendixTocStart}
\appendixtableofcontents

%%%%%%%%%%%%%%%%%%%%%%%%%%%%%%%%%%%%%%%%%%%%%%%%%%%%%%%%%%%%%%%%%%%%%%%%%%%%%%%%%%%%%%%%%%%%%%%%%%%%%%%%%%%%%%%%%%%%%%%%
\clearpage
\section{Covariance Inequalities and Transfer Bounds}
\label{Covariance_based_inequalities}

We derive covariance bounds for finite lists of observables and show how they transform under real linear combinations.
Subsec.~\ref{app:covariance-functionals} defines the hybrid and spectral functionals, relates them to the second covariance invariant, and characterizes equality.
Subsec.~\ref{Cov_transfer_UR} derives singular value transfer bounds for total variance and the second covariance invariant, including a discussion of rank deficiency.

%%%%%%%%%%%%%%%%%%%%%%%%%%%%%%%%%%%%%%%%%%%%%%%%%%%%%%%%%%%%%%%%%%%%%%%%%%%%%%%%%%%%%%%%%%%%%%%%%%%%%%%%%%%%%%%%%%

\subsection{Covariance Functionals}
\label{app:covariance-functionals}

Let $\rho$ be a density operator on a $d$-dimensional Hilbert space $\mathcal H_d$, and let $\bm A=(A_1,\ldots,A_N)^{\T}$ be a finite list of self-adjoint observables.
For any observable $X$, we write $\langle X\rangle_\rho\coloneqq\Tr(\rho X)$ for its expectation value and $\overline X_\rho\coloneqq X-\langle X\rangle_\rho\1_d$ for its centered form, where $\1_d$ denotes the identity operator on $\mathcal H_d$ and the superscript $\T$ denotes transposition.
For $A_r,A_s\in\bm A$, their real covariance is defined by
\begin{align}
\operatorname{Cov}_\rho(A_r,A_s)
&\coloneqq
\frac12\left\langle
\overline A_{r,\rho} \overline A_{s,\rho}
+
\overline A_{s,\rho} \overline A_{r,\rho}
\right\rangle_\rho
\\
&=
\operatorname{Re}
\left\langle
\overline A_{r,\rho} \overline A_{s,\rho}
\right\rangle_\rho
\\
&=
\frac12\langle A_rA_s+A_sA_r\rangle_\rho
-
\langle A_r\rangle_\rho\langle A_s\rangle_\rho.
\label{eq:real-pairwise-covariance}
\end{align}

\begin{mydef}{Real Covariance Matrix}{real_covariance}
The real covariance matrix of $\bm A$ in the state $\rho$ is the $N\times N$ matrix $\Gamma_\rho(\bm A)$ defined entrywise by
\begin{align}
\bigl[\Gamma_\rho(\bm A)\bigr]_{rs}
\coloneqq
\operatorname{Cov}_\rho(A_r,A_s).
\end{align}
\end{mydef}

By construction, $\Gamma_\rho(\bm A)$ is real and symmetric, and its diagonal entries are the variances
\begin{align}
(\Delta_\rho A_r)^2
\coloneqq
\operatorname{Cov}_\rho(A_r,A_r)
=
\bigl\langle \overline{A}_{r,\rho}^{ 2}\bigr\rangle_\rho
=
\bigl[\Gamma_\rho(\bm A)\bigr]_{rr}.
\end{align}
Moreover, $\Gamma_\rho(\bm A)$ is positive semidefinite.
Indeed, for any $\bm v\in\mathbb R^N$, defining $B_{\bm v}\coloneqq\sum_r v_r\overline{A}_{r,\rho}$ gives
\begin{align}
\bm v^{\T}\Gamma_\rho(\bm A)\bm v
=
\bigl\langle B_{\bm v}^{ 2}\bigr\rangle_\rho
\geqslant 0.
\end{align}
Hence $\Delta_\rho A_r\geqslant0$ is the standard deviation of $A_r$ in the state $\rho$.

\begin{mydef}{Hybrid Uncertainty}{Hybrid_UR}
The hybrid uncertainty of the observable list $\bm A$ in the state $\rho$ is defined by
\begin{align}
\Mfun_\rho(\bm A)
\coloneqq
\sum_{r<s}\Delta_\rho A_r \Delta_\rho A_s.
\end{align}
\end{mydef}

Since the variances are the diagonal entries of the covariance matrix, the hybrid uncertainty can equivalently be written as
\begin{align}
\Mfun_\rho(\bm A)
&=
\sum_{r<s}
\sqrt{
\bigl[\Gamma_\rho(\bm A)\bigr]_{rr}
\bigl[\Gamma_\rho(\bm A)\bigr]_{ss}
}\\
&=
\frac12
\left[
\left(
\sum_{r=1}^{N}
\sqrt{\bigl[\Gamma_\rho(\bm A)\bigr]_{rr}}
\right)^2
-
\Tr\Gamma_\rho(\bm A)
\right].
\end{align}

The hybrid uncertainty depends on the variances of the prescribed observables and generally changes under orthogonal mixing.
To separate this coordinate dependence from the covariance spectrum, we introduce a spectral counterpart determined only by the covariance eigenvalues.
Let $\lambda_1,\ldots,\lambda_N$ denote the eigenvalues of $\Gamma_\rho(\bm A)$.
Since $\Gamma_\rho(\bm A)$ is positive semidefinite, its eigenvalues are nonnegative.

\begin{mydef}{Spectral Hybrid Uncertainty}{Spectral_hybrid_UR}
The spectral hybrid uncertainty of $\bm A$ in the state $\rho$ is defined by
\begin{align}
\Mfun_\rho^{\mathrm{sp}}(\bm A)
\coloneqq
\sum_{i<j}\sqrt{\lambda_i\lambda_j},
\end{align}
where $\lambda_1,\ldots,\lambda_N$ are the eigenvalues of $\Gamma_\rho(\bm A)$.
\end{mydef}

Equivalently, since $\Tr\sqrt{\Gamma_\rho(\bm A)}=\sum_i\sqrt{\lambda_i}$, the spectral hybrid uncertainty can be written as
\begin{align}
\Mfun_\rho^{\mathrm{sp}}(\bm A)
=
\frac12
\left[
\left(\Tr\sqrt{\Gamma_\rho(\bm A)}  \right)^2
-
\Tr\Gamma_\rho(\bm A)
\right].
\end{align}
The quantity $\Mfun_\rho^{\mathrm{sp}}(\bm A)$ depends only on the covariance eigenvalues and is therefore invariant under orthogonal mixing.

\begin{mylem}{Spectral Hybrid Minimum}{Spectral_min_UR}
For a fixed state $\rho$ and a finite list of self-adjoint observables $\bm A=(A_1,\ldots,A_N)^{\T}$,
\begin{align*}
\Mfun_\rho^{\mathrm{sp}}(\bm A)
=
\min_{O\in\mathrm O(N)}
\Mfun_\rho(O\bm A),
\end{align*}
where $\mathrm O(N)$ denotes the group of real $N\times N$ orthogonal matrices.
The minimum is attained if and only if $O\Gamma_\rho(\bm A)O^{\T}$ is diagonal.
\end{mylem}

\begin{proof}
Fix $O\in\mathrm O(N)$ and write $\Gamma\coloneqq\Gamma_\rho(\bm A)$ and $\Gamma_O\coloneqq\Gamma_\rho(O\bm A)$.
By covariance bilinearity,
\begin{align}
\Gamma_O=O\Gamma O^{\T},
\end{align}
so $\Tr\Gamma_O=\Tr \Gamma$.
Let $H\coloneqq\sqrt{\Gamma_O}$ be the unique positive semidefinite square root of $\Gamma_O$, so that $H^2=\Gamma_O$.
Since $H$ is real symmetric, each diagonal entry of $\Gamma_O$ satisfies
\begin{align}
(\Gamma_O)_{rr}
=
\sum_s H_{rs}^2.
\end{align}
Consequently,
\begin{align}
\sqrt{(\Gamma_O)_{rr}}
=
\left(H_{rr}^2+\sum_{s\neq r}H_{rs}^2\right)^{1/2}
\geqslant H_{rr},
\end{align}
where $H_{rr}\geqslant0$ follows from positivity of $H$.
Equality holds for every $r$ if and only if $H$ is diagonal.
Summing over $r$ gives
\begin{align}
\sum_r\sqrt{(\Gamma_O)_{rr}}
\geqslant
\Tr H
=
\Tr\sqrt{\Gamma_O}
=
\Tr\sqrt{\Gamma},
\end{align}
where the last equality follows because $\Gamma_O$ and $\Gamma$ have the same eigenvalues.
It follows that
\begin{align}
\Mfun_\rho(O\bm A)
&=
\frac12
\left[
\left(\sum_r\sqrt{(\Gamma_O)_{rr}}\right)^2
-\Tr\Gamma
\right]
\\
&\geqslant
\frac12
\left[
\left(\Tr\sqrt{\Gamma}\right)^2
-\Tr\Gamma
\right]
=
\Mfun_\rho^{\mathrm{sp}}(\bm A).
\end{align}
Equality holds if and only if $H_{rs}=0$ for every $r\neq s$, which is equivalent to $\Gamma_O$ being diagonal.
Since $\Gamma$ is real symmetric, an orthogonal transformation that diagonalizes it exists and attains the lower bound.
\end{proof}

A lower bound on the total variance $\Tr\Gamma$ alone does not imply a positive lower bound on the spectral hybrid uncertainty, which vanishes whenever $\Gamma$ has rank one.
We therefore consider the second characteristic coefficient of $\Gamma$~\cite{TrifonovDonev1998}.

\begin{mydef}{Second Covariance Invariant}{sec_cov_invariant}
For an $N\times N$ covariance matrix $\Gamma$ with eigenvalues $\lambda_1,\ldots,\lambda_N$, define
\begin{align}
\label{eq:functional_definitions}
C_2^{(N)}(\Gamma)
\coloneqq
\sum_{i<j}\lambda_i\lambda_j.
\end{align}
\end{mydef}

The superscript $(N)$ indicates the size of the covariance matrix and is omitted when no ambiguity arises.
The second covariance invariant can equivalently be written as
\begin{align}
C_2^{(N)}(\Gamma)
=
\frac12
\left[
\bigl(\Tr\Gamma\bigr)^2
-
\Tr(\Gamma^2)
\right].
\end{align}
For real symmetric $\Gamma$, it is also the sum of its $2\times2$ principal minors:
\begin{align}
C_2^{(N)}(\Gamma)
=
\sum_{r<s}
\left(
\Gamma_{rr}\Gamma_{ss}
-
\Gamma_{rs}^2
\right).
\label{eq:second-covariance-principal-minors}
\end{align}
The trace formula shows that $C_2^{(N)}(\Gamma)$ is invariant under real orthogonal changes of basis.
For $\Gamma\geqslant0$, it is nonnegative, with equality if and only if $\operatorname{rank}\Gamma\leqslant1$.

\begin{mycor}{Covariance Bound Hierarchy}{Cov_UR_chain}
For any state $\rho$ and any finite list $\bm A=(A_1,\ldots,A_N)^{\T}$ of self-adjoint observables with $N\geqslant2$, let $\Gamma\coloneqq\Gamma_\rho(\bm A)$.
Then
\begin{align}
\label{eq:covariance_chain}
\Mfun_\rho(\bm A)
\geqslant
\Mfun_\rho^{\mathrm{sp}}(\bm A)
\geqslant
\sqrt{C_2^{(N)}(\Gamma)}.
\end{align}
The first inequality is saturated if and only if $\Gamma$ is diagonal, whereas the second is saturated if and only if $\mathrm{rank} \Gamma\leqslant2$.
\end{mycor}

\begin{proof}
The first inequality and its equality condition follow directly from Eq.~\eqref{eq:main-spectral}.
Since $\lambda_i\geqslant0$, the definition of the spectral hybrid uncertainty gives
\begin{align}
\bigl[\Mfun_\rho^{\mathrm{sp}}(\bm A)\bigr]^2
=
\left(
\sum_{i<j}\sqrt{\lambda_i\lambda_j}
\right)^2
\geqslant
\sum_{i<j}\lambda_i\lambda_j
=
C_2^{(N)}(\Gamma),
\end{align}
which proves the second inequality.
Equality holds if and only if at most one product $\lambda_i\lambda_j$ is nonzero, equivalently if and only if $\mathrm{rank} \Gamma\leqslant2$.
\end{proof}

For a qubit, the equality conditions in Cor.~\ref{cor:Cov_UR_chain} admit a simple Bloch sphere interpretation.
Write $\rho=(\1_2+\bm r\cdot\bm\sigma)/2$, with $\|\bm r\|_2\leqslant1$, and let $\bm J=\bm\sigma/2$.
Then
\begin{align}
\Gamma_\rho(\bm J)
=
\frac14
\left(
\1_3-\bm r\bm r^{\T}
\right).
\end{align}
For a full-rank state, $\|\bm r\|_2<1$, so $\Gamma_\rho(\bm J)$ is positive definite.
Now consider a list of observables $\bm A=F\bm J$, where $F\in\mathbb R^{N\times3}$ has rank three and no zero rows.
Its covariance matrix satisfies
\begin{align}
\Gamma_\rho(\bm A)
=
F\Gamma_\rho(\bm J)F^{\T},
\end{align}
and therefore has rank three with strictly positive diagonal entries.
If $N>3$, such a matrix cannot be diagonal, so the first inequality in Eq.~\eqref{eq:covariance_chain} is necessarily strict.

%%%%%%%%%%%%%%%%%%%%%%%%%%%%%%%%%%%%%%%%%%%%%%%%%%%%%%%%%%%%%%%%%%%%%%%%%%%%%%%%%%%%%%%%%%%%%%%%%%%%%%%%%%%%%%%%%%

\subsection{Covariance Transfer of Uncertainty Bounds}
\label{Cov_transfer_UR}

Let $\bm X=(X_1,\ldots,X_r)^{\T}$ be a reference list of self-adjoint observables, and define $\bm A=M\bm X$ for a real coefficient matrix $M\in\mathbb R^{N\times r}$.
For linearly independent reference observables, the combinations $M\bm X$ span the same real linear space precisely when $\operatorname{rank}M=r$.
Bilinearity of the covariance gives~\cite[Eq.~(5)]{TrifonovDonev1998}
\begin{align}
\Gamma_\rho(\bm A)
=
M\Gamma_\rho(\bm X)M^{\T}.
\label{eq:covariance-linear-transfer}
\end{align}
We now derive lower bounds on the total variance and the second covariance invariant of $\bm A$ from those of $\bm X$.
For these transfer bounds, assume that $M$ has full column rank, so that $N\geqslant r$, and order its singular values as $0<\sigma_1(M)\leqslant\cdots\leqslant\sigma_r(M)$.
We write $\smin(M)\coloneqq\sigma_1(M)$.

Lemma~\ref{lem:Variance_sum_bound} transfers total variance bounds with a factor $\sigma_1(M)^2$.
For $r\geqslant2$, Lem.~\ref{lem:second-invariant-transfer} transfers bounds on the second covariance invariant with a factor $[\sigma_1(M)\sigma_2(M)]^2$.
The corresponding bounds on the hybrid and spectral hybrid uncertainties follow in Cor.~\ref{cor:Hybrid_U_transfer}.
These results require neither closure of the observables under commutation nor invertibility of $\Gamma_\rho(\bm X)$.

\begin{mylem}{Total Variance Transfer}{Variance_sum_bound}
Let $\bm X=(X_1,\ldots,X_r)^{\T}$ be a list of self-adjoint observables and let $\bm A=M\bm X$, where $M\in\mathbb R^{N\times r}$ has rank $r\geqslant1$.
Then, for every state $\rho$,
\begin{align}
\label{eq:app-variance-transfer}
\sum_{p=1}^{N}(\Delta_\rho A_p)^2
\geqslant
\smin(M)^2\Tr\Gamma_\rho(\bm X),
\end{align}
where $\smin(M)$ is the smallest singular value of $M$.
\end{mylem}

\begin{proof}
Let $\Gamma\coloneqq\Gamma_\rho(\bm X)$.
Using Eq.~\eqref{eq:covariance-linear-transfer} and cyclicity of the trace,
\begin{align}
\sum_{p=1}^{N}(\Delta_\rho A_p)^2
=
\Tr\Gamma_\rho(\bm A)
=
\Tr \left(M^{\T}M\Gamma\right).
\end{align}
Since the smallest eigenvalue of $M^{\T}M$ is $\smin(M)^2$, we have
\begin{align}
M^{\T}M-\smin(M)^2\1_r\geqslant0.
\end{align}
Together with $\Gamma\geqslant0$, this gives
\begin{align}
\Tr\left(M^{\T}M\Gamma\right)
-
\smin(M)^2\Tr\Gamma
&=
\Tr \left[
\Gamma^{1/2}
\left(
M^{\T}M-\smin(M)^2\1_r
\right)
\Gamma^{1/2}
\right]
\geqslant0,
\end{align}
which proves Eq.~\eqref{eq:app-variance-transfer}.
\end{proof}

Any lower bound $\Tr\Gamma_\rho(\bm X)\geqslant t_0$ therefore transfers immediately to the observables $\bm A=M\bm X$ as
\begin{align}
\Tr\Gamma_\rho(\bm A)
\geqslant\smin(M)^2\Tr\Gamma_\rho(\bm X)
\geqslant
\smin(M)^2 t_0.
\end{align}

To bound the hybrid uncertainty, we use the second covariance invariant $C_2$.
For self-adjoint observables $X$ and $Y$, the pairwise Robertson-Schr\"odinger uncertainty relation gives
\begin{align}
(\Delta_\rho X)^2(\Delta_\rho Y)^2
-
\operatorname{Cov}_\rho(X,Y)^2
\geqslant
\frac14
\left|
\langle[X,Y]\rangle_\rho
\right|^2.
\label{eq:RS-pairwise}
\end{align}
The left-hand side is the corresponding $2\times2$ principal covariance minor.
Summing over all pairs gives a lower bound on $C_2$, namely the second characteristic uncertainty relation~\cite[Eq.~(9)]{TrifonovDonev1998}.
The full family of characteristic relations includes Robertson's determinant inequality for several observables~\cite{Robertson1934}.
For the unbounded generators in Subsec.~\ref{sec:su11-weight-states}, Eq.~\eqref{eq:RS-pairwise} is interpreted in the quadratic form sense on the admissible state class defined there.

To transfer bounds on $C_2$ from $\bm X$ to $\bm A=M\bm X$, we use the $2\times2$ minors of $M$.
For $r\geqslant2$, these form the second compound matrix $\comp{M}\in\mathbb R^{\binom N2\times\binom r2}$, with entries
\begin{align}
[\comp{M}]_{(p,s),(a,b)}
=
m_{pa}m_{sb}-m_{pb}m_{sa},
\quad p<s,\quad a<b.
\label{eq:second-compound-definition}
\end{align}
For matrices $A$ and $B$ of compatible dimensions, the second compound matrices satisfy

\begin{align}
\comp{(AB)}
&=
\comp{A}\comp{B},
\label{eq:id1}\\[4pt]
\comp{(A^{\T})}
&=
(\comp{A})^{\T}.
\label{eq:id2}
\end{align}
The product identity follows from the Cauchy-Binet formula~\cite{HornJohnson2013}, and the transpose identity follows directly from the definition.
Both identities hold without any rank assumption.

For a matrix $M\in\mathbb R^{N\times r}$ of full column rank, with $r\geqslant2$, the singular values of $\comp{M}$ follow directly from those of $M$.
Let
\begin{align}
M=U\Sigma W^{\T}
\end{align}
be a singular value decomposition, where $U\in\mathrm O(N)$, $W\in\mathrm O(r)$, and the nonzero diagonal entries of $\Sigma$ are $0<\sigma_1(M)\leqslant\cdots\leqslant\sigma_r(M)$.
Using Eqs.~\eqref{eq:id1} and \eqref{eq:id2},
\begin{align}
\comp{M}
=
\comp{U}\comp{\Sigma}(\comp{W})^{\T}.
\end{align}
Since $\comp{U}$ and $\comp{W}$ are orthogonal, $\comp{M}$ and $\comp{\Sigma}$ have the same singular values.
The nonzero singular values of $\comp{\Sigma}$ are the pairwise products $\sigma_a(M)\sigma_b(M)$ with $1\leqslant a<b\leqslant r$.
Hence
\begin{align}
\label{eq:compound_min_singular}
\smin(\comp{M})
=
\sigma_1(M)\sigma_2(M).
\end{align}

\begin{mylem}{Second Invariant Transfer}{second-invariant-transfer}
Let $G\in\mathbb R^{r\times r}$ be positive semidefinite, and let $M\in\mathbb R^{N\times r}$ have rank $r\geqslant2$.
Then
\begin{align}
C_2^{(N)}(MGM^{\T})
\geqslant
\bigl[\sigma_1(M)\sigma_2(M)\bigr]^2
C_2^{(r)}(G).
\label{eq:second-invariant-transfer}
\end{align}
\end{mylem}

\begin{proof}
Since $G\geqslant0$, its second compound matrix $\comp{G}$ is positive semidefinite, with eigenvalues given by the pairwise products of the eigenvalues of $G$.
Using Eqs.~\eqref{eq:id1} and \eqref{eq:id2}, together with the cyclicity of the trace, we obtain
\begin{align}
C_2^{(N)}(MGM^{\T})
=
\Tr\left[\comp{(MGM^{\T})}\right]
=
\Tr\left[
(\comp{M})^{\T}
\comp{M}\comp{G}\right].
\end{align}
Equation~\eqref{eq:compound_min_singular} implies
\begin{align}
(\comp{M})^{\T}\comp{M}
\geqslant
\bigl[\sigma_1(M)\sigma_2(M)\bigr]^2
\1_{\binom r2}.
\end{align}
Since $\comp{G}\geqslant0$ and $\Tr[\comp{G}]=C_2^{(r)}(G)$, taking the trace against $\comp{G}$ yields Eq.~\eqref{eq:second-invariant-transfer}.
\end{proof}

Combining Lem.~\ref{lem:second-invariant-transfer} with the covariance uncertainty chain in Eq.~\eqref{eq:covariance_chain} immediately yields a transfer bound for both hybrid uncertainty functionals.

\begin{mycor}{Transfer of Hybrid Uncertainty}{Hybrid_U_transfer}
Let $\bm X=(X_1,\ldots,X_r)^{\T}$ be a list of self-adjoint observables and let $\bm A=M\bm X$, where $M\in\mathbb R^{N\times r}$ has rank $r\geqslant2$.
Then
\begin{align}
\label{eq:compound_transfer_conclusion}
\Mfun_\rho(\bm A)
\geqslant
\Mfun_\rho^{\mathrm{sp}}(\bm A)
\geqslant
\sigma_1(M)\sigma_2(M)
\sqrt{C_2^{(r)}\left(\Gamma_\rho(\bm X)\right)}.
\end{align}
\end{mycor}

\begin{proof}
Apply Lem.~\ref{lem:second-invariant-transfer} to $G=\Gamma_\rho(\bm X)$ and combine the result with Eq.~\eqref{eq:covariance_chain}.
\end{proof}

Consequently, any lower bound $C_2^{(r)}\left(\Gamma_\rho(\bm X)\right)\geqslant c_0$ immediately yields
\begin{align}
\Mfun_\rho(\bm A)
\geqslant
\Mfun_\rho^{\mathrm{sp}}(\bm A)
\geqslant
\sigma_1(M)\sigma_2(M)\sqrt{c_0}.
\end{align}
When $M$ is rank-deficient, the covariance identity $\Gamma_\rho(\bm A) = M\Gamma_\rho(\bm X)M^{\T}$ still holds, but the positive transfer bounds do not apply to the full reference covariance $\Gamma_\rho(\bm X)$ because $M^{\T}M$ has a nontrivial kernel.
Let $q\coloneqq\operatorname{rank}M<r$, and choose $Q\in\mathbb R^{r\times q}$ with orthonormal columns spanning $\operatorname{range}(M^{\T})$.
Defining the effective observable list $\bm Y\coloneqq Q^{\T}\bm X$, we have
\begin{align}
\bm A
=
(MQ)\bm Y,
\quad
\Gamma_\rho(\bm Y)
=
Q^{\T}\Gamma_\rho(\bm X)Q,
\end{align}
where $MQ$ has full column rank $q$.
For $q\geqslant2$, Cor.~\ref{cor:Hybrid_U_transfer} therefore gives
\begin{align}
\Mfun_\rho(\bm A)
\geqslant
\Mfun_\rho^{\mathrm{sp}}(\bm A)
\geqslant
\sigma_1(MQ)\sigma_2(MQ)
\sqrt{
C_2^{(q)}
\left(
Q^{\T}\Gamma_\rho(\bm X)Q
\right)
}.
\label{eq:rank-deficient-transfer}
\end{align}

For unbounded self-adjoint observables, the preceding covariance identities and transfer bounds require compatible operator domains and finite second moments.
We assume a common dense domain containing the eigenvectors with nonzero weights in a spectral decomposition of $\rho$.
On this domain, the chosen self-adjoint realizations of the components of $\bm A=M\bm X$ agree with the corresponding real linear combinations of the components of $\bm X$, and all quadratic forms entering $\Gamma_\rho(\bm X)$ and $\Gamma_\rho(\bm A)$ are finite.
These conditions are verified explicitly for the positive discrete series in Subsec.~\ref{sec:su11-weight-states}.

The transfer bounds above also apply to observable lists with $N>r$, provided that $M$ has full column rank.
In Appendices~\ref{app:su2-proofs} and~\ref{app:su11-proofs}, $\bm X$ consists of three linearly independent self-adjoint generators, so $r=3$.
Combining these transfer bounds with the covariance constraints imposed by each representation yields explicit uncertainty bounds.

%%%%%%%%%%%%%%%%%%%%%%%%%%%%%%%%%%%%%%%%%%%%%%%%%%%%%%%%%%%%%%%%%%%%%%%%%%%%%%%%%%%%%%%%%%%%%%%%%%%%%%%%%%%%%%%%%%

\section{\texorpdfstring{Spin-$j$ Observables}{Spin-j Observables}}
\label{app:su2-proofs}

We apply the covariance bounds of Appendix~\ref{Covariance_based_inequalities} to irreducible spin-$j$ representations.
Angular momentum is a standard setting for minimum uncertainty states, planar uncertainty relations, and multicomponent bounds~\cite{Nieto1967,HeEtAl2011,Dammeier2015,ShabbirBjork2016,MaEtAl2017}.
Subsec.~\ref{app:spin-matrices} fixes the representation and derives the coherent state covariance.
Subsec.~\ref{app:spin-covariance-proof} proves the mean spin refinement of the second covariance invariant and the exact spectral minimum for spanning lists of spin observables.
Subsec.~\ref{app:angular-proof} then treats an exactly solvable list of nonorthogonal qubit observables.

\subsection{\texorpdfstring{Spin-$j$ Representation and Coherent State Covariance}{Spin-j Representation and Coherent State Covariance}}
\label{app:spin-matrices}

We first specify the self-adjoint spin observables associated with the irreducible spin-$j$ representation and derive the coherent state covariance used to establish the optimal uncertainty bounds in Subsec.~\ref{app:spin-covariance-proof}.
Throughout, $j\in\{1/2,1,3/2,\ldots\}$ and $\hbar=1$.
In a unitary representation of the compact real form $\mathfrak{su}(2)$, the infinitesimal generators are anti-Hermitian; we work with the corresponding self-adjoint spin observables $\bm J=(J_x,J_y,J_z)^{\T}$ obtained by multiplication by $\ii$ in the complexified representation.
They satisfy
\begin{align}
[J_a,J_b]
=
\ii\sum_c\epsilon_{abc}J_c,
\label{eq:spin-commutation-relations}
\end{align}
together with the Casimir identity
\begin{align}
\sum_a J_a^2
=
j(j+1)\1_{2j+1}.
\label{eq:spin-casimir}
\end{align}
Here $\epsilon_{abc}$ is the Levi-Civita symbol with $\epsilon_{xyz}=1$.
The Hilbert space is finite-dimensional, so all operator moments are finite and no domain issues arise.

For a unit vector $\bm n\in\mathbb R^3$, define the spin component $J_{\bm n}\coloneqq\bm n\cdot\bm J$.
For a state $\rho$, let
\begin{align}
\bm m\coloneqq\langle\bm J\rangle_\rho
\end{align}
denote the mean spin vector, whose length satisfies
\begin{align}
r\coloneqq\|\bm m\|_2\leqslant j.
\end{align}
Define the polarization deficit by
\begin{align}
\delta\coloneqq j^2-r^2\geqslant0.
\end{align}
For $j=1/2$, the vector $2\bm m$ coincides with the Bloch vector of the qubit state.

Equation~\eqref{eq:spin-casimir} fixes the total spin variance in terms of the mean spin length:
\begin{align}
\label{eq:spin-total-variance}
\sum_{a=x,y,z}(\Delta_\rho J_a)^2
=
\Tr\Gamma_\rho(\bm J)
=
j(j+1)-r^2
=
j+\delta
\geqslant
j.
\end{align}
Maximal polarization, $\delta=0$, therefore minimizes the total spin variance.
This constraint does not determine how the variance is distributed among different spin directions.
For the spectral hybrid uncertainty, we also need a lower bound on the second covariance invariant $C_2^{(3)}\left(\Gamma_\rho(\bm J)\right)$.

To fix conventions, we use the standard $J_z$ eigenbasis
\begin{align}
\label{eq:spin-ordered-basis}
\{|j,m\rangle:m=j,j-1,\ldots,-j\},
\end{align}
with
\begin{align}
J_z|j,m\rangle
&=
m|j,m\rangle.
\end{align}
Defining $J_\pm\coloneqq J_x\pm\ii J_y$ and choosing the phases of the basis states such that the ladder coefficients are nonnegative, we have~\cite{Perelomov1986}
\begin{align}
J_+|j,m\rangle
&=
\sqrt{(j-m)(j+m+1)} |j,m+1\rangle,\\[4pt]
J_-|j,m\rangle
&=
\sqrt{(j+m)(j-m+1)} |j,m-1\rangle.
\label{eq:spin-ladder-action}
\end{align}
In particular, $J_+|j,j\rangle=0$ and $J_-|j,-j\rangle=0$.

For $j=1/2$, the spin observables reduce to the Pauli representation
\begin{align}
J_a=\frac12\sigma_a,
\quad
a=x,y,z.
\label{eq:spin-half-pauli}
\end{align}
This normalization fixes the qubit convention in Subsec.~\ref{app:angular-proof}.
For the highest weight state $|j,j\rangle$, Eq.~\eqref{eq:spin-ladder-action} gives
\begin{align}
J_x|j,j\rangle
=
\sqrt{\frac j2}|j,j-1\rangle,
\quad
J_y|j,j\rangle
=
\ii\sqrt{\frac j2}|j,j-1\rangle,
\quad
J_z|j,j\rangle
=
j|j,j\rangle.
\end{align}
Hence
\begin{align}
\langle\bm J\rangle
=
(0,0,j)^{\T},
\quad
(\Delta J_x)^2
=
(\Delta J_y)^2
=
\frac j2,
\quad
(\Delta J_z)^2
=
0.
\end{align}
The real cross covariances vanish, and therefore
\begin{align}
\Gamma_{|j,j\rangle}(\bm J)
=
\frac j2
\begin{pmatrix}
1&0&0\\
0&1&0\\
0&0&0
\end{pmatrix}.
\label{eq:highest-weight-covariance}
\end{align}

A spin coherent state $|j,\bm u\rangle$ is obtained by rotating $|j,j\rangle$ so that its polarization points along a unit vector $\bm u\in\mathbb R^3$~\cite{Radcliffe1971,Arecchi1972}.
Equivalently, it is the normalized eigenstate of $J_{\bm u}$ with eigenvalue $j$, up to an overall phase.
Rotational covariance then gives
\begin{align}
\langle\bm J\rangle_{|j,\bm u\rangle}
&=
j\bm u,
\\
\Gamma_{|j,\bm u\rangle}(\bm J)
&=
\frac j2
\left(
\1_3-\bm u\bm u^{\T}
\right).
\label{eq:spin-coherent-covariance}
\end{align}
Thus the variance is $j/2$ along every unit direction orthogonal to $\bm u$ and vanishes along $\bm u$.
The covariance eigenvalues are $j/2$, $j/2$, and $0$, so
\begin{align}
C_2^{(3)}
\left(
\Gamma_{|j,\bm u\rangle}(\bm J)
\right)
=
\frac{j^2}{4}.
\end{align}
Every spin coherent state therefore saturates the state-independent bound on the second covariance invariant.
For the Cartesian observables $\bm J=(J_x,J_y,J_z)^{\T}$, the hybrid uncertainty reaches $j/2$ only when $\bm u$ is aligned with a Cartesian axis, so that the covariance matrix is diagonal.
We use this covariance to prove attainability in Thm.~\ref{thm:spin-frame-minimum}.

%%%%%%%%%%%%%%%%%%%%%%%%%%%%%%%%%%%%%%%%%%%%%%%%%%%%%%%%%%%%%%%%%%%%%%%%%%%%%%%%%%%%%%%%%%%%%%%%%%%%%%%%%%%%%%%%%%

\subsection{Spin Covariance Refinement and Optimal Bounds}
\label{app:spin-covariance-proof}

We derive a lower bound on the second covariance invariant that depends on the mean spin length and refines the sharp state-independent bound of Ref.~\cite{ShabbirBjork2016}.
Together with the transfer bounds of Subsec.~\ref{Cov_transfer_UR} and the covariance of spin coherent states, this yields the exact minimum of the spectral hybrid uncertainty over all states for any observable list $\bm A=M\bm J$ with $\operatorname{rank}M=3$.

Write $m_a\coloneqq\langle J_a\rangle_\rho$ for the components of $\bm m=\langle\bm J\rangle_\rho$, and define the centered spin observables $\overline J_{a,\rho}\coloneqq J_a-m_a\1_{2j+1}$.
Alongside the real covariance matrix $\Gamma_\rho(\bm J)$, consider the Hermitian matrix of ordered centered moments
\begin{align}
G_{ab}
\coloneqq
\left\langle
\overline J_{a,\rho}\overline J_{b,\rho}
\right\rangle_\rho.
\end{align}
Using Eq.~\eqref{eq:spin-commutation-relations},
\begin{align}
G_{ab}
=
\Gamma_\rho(\bm J)_{ab}
+
\frac12
\langle[J_a,J_b]\rangle_\rho
=
\Gamma_\rho(\bm J)_{ab}
+
\frac{\ii}{2}
\sum_c\epsilon_{abc}m_c.
\end{align}
In the ordered Cartesian basis $(J_x,J_y,J_z)$, define
\begin{align}
E(\bm m)
\coloneqq
\begin{pmatrix}
0&m_z&-m_y\\
-m_z&0&m_x\\
m_y&-m_x&0
\end{pmatrix}.
\end{align}
The complex covariance matrix can then be written as
\begin{align}
G
=
\Gamma_\rho(\bm J)
+
\frac{\ii}{2}E(\bm m).
\label{eq:spin-complex-covariance}
\end{align}

\begin{mythm}{Spin Covariance Refinement}{Spin_covariance_refinement}
For every state $\rho$ in the irreducible spin-$j$ representation with $j\geqslant1$, the second covariance invariant satisfies
\begin{align}
\label{eq:spin-covariance-refinement}
C_2^{(3)}\left(\Gamma_\rho(\bm J)\right)
\geqslant
\frac{j^2}{4}
+\frac{3\delta}{4}
+\frac{j-1}{j^2}\delta^2,
\end{align}
where $\delta\coloneqq j^2-\|\langle\bm J\rangle_\rho\|_2^2\geqslant0$ is the deficit in the squared mean spin length relative to its maximal value $j^2$.
\end{mythm}

\begin{proof}
Write $\bm m\coloneqq\langle\bm J\rangle_\rho$, $r\coloneqq\|\bm m\|_2$, and $\Gamma\coloneqq\Gamma_\rho(\bm J)$.
If $r\neq0$, let $\bm n\coloneqq\bm m/r$.
Since the largest eigenvalue of $J_{\bm n}$ is $j$, we have $r=\langle J_{\bm n}\rangle_\rho\leqslant j$; the same bound is trivial for $r=0$.
Hence $\delta=j^2-r^2\geqslant0$.

Both $r$ and $C_2^{(3)}(\Gamma)$ are rotationally invariant.
We may therefore choose the $z$ axis along $\bm m$ and then rotate within the transverse plane so that $\operatorname{Cov}_\rho(J_x,J_y)=0$.
In these coordinates,
\begin{align}
\bm m=(0,0,r)^{\T},
\quad
\Gamma=
\begin{pmatrix}
x&0&u\\
0&y&v\\
u&v&z
\end{pmatrix},
\end{align}
where
\begin{align}
x\coloneqq(\Delta_\rho J_x)^2,
\quad
y\coloneqq(\Delta_\rho J_y)^2,
\quad
z\coloneqq(\Delta_\rho J_z)^2,
\quad
u\coloneqq\operatorname{Cov}_\rho(J_x,J_z),
\quad
v\coloneqq\operatorname{Cov}_\rho(J_y,J_z).
\end{align}
Equation~\eqref{eq:spin-total-variance} gives
\begin{align}
x+y+z
=
j+\delta,
\label{eq:spin-trace-frame}
\end{align}
while
\begin{align}
C_2^{(3)}(\Gamma)
=
xy+z(x+y)-u^2-v^2.
\label{eq:spin-C2-frame}
\end{align}
It remains to control the cross covariances $u$ and $v$.

In these coordinates, the complex covariance matrix in Eq.~\eqref{eq:spin-complex-covariance} becomes
\begin{align}
G
=
\begin{pmatrix}
x&\ii r/2&u\\
-\ii r/2&y&v\\
u&v&z
\end{pmatrix}
\geqslant0.
\end{align}
Positivity of its transverse principal minor gives
\begin{align}
a
\coloneqq
xy-\frac{r^2}{4}
\geqslant0.
\label{eq:spin-a-def}
\end{align}
Moreover,
\begin{align}
\det G
=
za-yu^2-xv^2
\geqslant0,
\end{align}
and hence
\begin{align}
\label{eq:spin-first-cross-constraint}
yu^2+xv^2
\leqslant
za.
\end{align}

To obtain the required bound on $u^2+v^2$, we also use the restriction imposed by the finite spin-$j$ representation.
Introduce an auxiliary spin-$1$ representation in the Cartesian basis, with generators
\begin{align}
(S_c)_{ab}
=
-\ii\epsilon_{cab},
\end{align}
and define on $\mathbb C^3\otimes\mathbb C^{2j+1}$
\begin{align}
\mathcal T
\coloneqq
\sum_c S_c\otimes J_c.
\end{align}
Let
\begin{align}
\bm L
=
\bm S\otimes\1_{2j+1}
+
\1_3\otimes\bm J
\end{align}
denote the total angular momentum.
Then
\begin{align}
\mathcal T
=
\frac12
\left[
\bm L^2
-
\left(
2+j(j+1)
\right)
\1_{3(2j+1)}
\right].
\end{align}
For $j\geqslant1$, angular momentum addition gives total spin sectors $j+1$, $j$, and $j-1$, on which $\mathcal T$ has eigenvalues $j$, $-1$, and $-j-1$, respectively.
Therefore
\begin{align}
\mathcal P
\coloneqq
\left(
\mathcal T+\1_{3(2j+1)}
\right)
\left(
\mathcal T-j\1_{3(2j+1)}
\right)
\geqslant0,
\label{eq:spin-P-positive}
\end{align}
since its eigenvalues on these sectors are $0$, $0$, and $j(2j+1)$.

To express this positivity condition in terms of spin moments, note that
\begin{align}
\mathcal T_{ab}
=
-\ii\sum_c\epsilon_{abc}J_c.
\end{align}
Using
\begin{align}
\sum_\ell
\epsilon_{a\ell c}\epsilon_{\ell b d}
=
\delta_{ad}\delta_{cb}
-
\delta_{ab}\delta_{cd},
\end{align}
together with Eq.~\eqref{eq:spin-casimir}, we obtain
\begin{align}
(\mathcal T^2)_{ab}
=
j(j+1)\delta_{ab}\1_{2j+1}
-
J_bJ_a.
\end{align}
Expanding $\mathcal P=\mathcal T^2+(1-j)\mathcal T-j\1_{3(2j+1)}$ and using Eq.~\eqref{eq:spin-commutation-relations} gives
\begin{align}
\mathcal P_{ab}
=
j^2\delta_{ab}\1_{2j+1}
-
\frac12\{J_a,J_b\}
+
\ii\left(j-\frac12\right)
\sum_c\epsilon_{abc}J_c.
\label{eq:spin-P-block}
\end{align}
Taking the expectation value of these blocks defines the Hermitian matrix
\begin{align}
H
\coloneqq
j^2\1_3
-
\Gamma
-
\bm m\bm m^{\T}
+
\ii\left(j-\frac12\right)E(\bm m).
\label{eq:spin-H}
\end{align}
For any $\bm w\in\mathbb C^3$, positivity of $\mathcal P$ implies
\begin{align}
\bm w^\dagger H\bm w
=
\Tr\left[
\rho
\sum_{a,b}
\overline w_a\mathcal P_{ab}w_b
\right]
\geqslant0,
\end{align}
so $H\geqslant0$.

This representation constraint also follows from the spin moment compatibility framework of Ref.~\cite[Prop.~3.2]{MoroderKeylLutkenhaus2008}.
In the symmetric $2j$ qubit realization, let $\sigma$ be the two qubit reduced state written in the Cartesian spin-$1$ basis.
Their reduction formulas give
\begin{align}
\sigma^{\T}=\frac{H}{j(2j-1)},\quad j\geqslant1.
\end{align}
Positivity of the reduced state therefore gives a necessary condition for compatibility of the spin moments.
The auxiliary spin construction establishes the same positivity condition without using the reduced-state representation.
We use it to prove the mean-dependent refinement in Eq.~\eqref{eq:spin-covariance-refinement}.

With this choice of axes,
\begin{align}
H
=
\begin{pmatrix}
j^2-x&\ii\left(j-\frac12\right)r&-u\\
-\ii\left(j-\frac12\right)r&j^2-y&-v\\
-u&-v&\delta-z
\end{pmatrix}.
\end{align}
Its diagonal entries imply
\begin{align}
0\leqslant z\leqslant\delta.
\end{align}
The transverse principal minor gives
\begin{align}
b
\coloneqq
(j^2-x)(j^2-y)
-
\left(j-\frac12\right)^2r^2
=
a-j\delta+j^2z
\geqslant0,
\label{eq:spin-b-def}
\end{align}
where Eq.~\eqref{eq:spin-trace-frame} and $r^2=j^2-\delta$ were used in the second equality.
The determinant condition $\det H\geqslant0$ yields
\begin{align}
(j^2-y)u^2+(j^2-x)v^2
\leqslant
(\delta-z)b.
\label{eq:spin-second-cross-constraint}
\end{align}
Adding Eqs.~\eqref{eq:spin-first-cross-constraint} and \eqref{eq:spin-second-cross-constraint}, we obtain
\begin{align}
j^2(u^2+v^2)
\leqslant
za+(\delta-z)b.
\label{eq:spin-cross-combined}
\end{align}

Substituting Eq.~\eqref{eq:spin-cross-combined} into Eq.~\eqref{eq:spin-C2-frame} gives
\begin{align}
\label{eq:spin-invariant-intermediate}
C_2^{(3)}(\Gamma)
\geqslant
xy+z(x+y)
-
\frac{za+(\delta-z)b}{j^2}
=
\frac{r^2}{4}
+
\frac{r^2}{j^2}(a+jz)
+
\frac{\delta^2}{j}.
\end{align}
Here we used
\begin{align}
xy
=
a+\frac{r^2}{4},\quad
x+y
=
j+\delta-z,	\quad
b
=
a-j\delta+j^2z.
\end{align}

It remains to bound $a+jz$.
From $b\geqslant0$,
\begin{align}
a+j^2z
\geqslant
j\delta.
\end{align}
Since $a\geqslant0$ and $j\geqslant1$,
\begin{align}
a+jz
=
\frac{a+j^2z}{j}
+
\left(
1-\frac1j
\right)a
\geqslant
\delta.
\end{align}
Using this estimate in Eq.~\eqref{eq:spin-invariant-intermediate} and recalling $r^2=j^2-\delta$, we obtain
\begin{align}
C_2^{(3)}(\Gamma)
\geqslant
\frac{r^2}{4}
+
\frac{r^2\delta}{j^2}
+
\frac{\delta^2}{j}
=
\frac{j^2}{4}
+
\frac{3\delta}{4}
+
\frac{j-1}{j^2}\delta^2.
\end{align}
This proves Eq.~\eqref{eq:spin-covariance-refinement}.
The argument does not require inversion of the covariance matrix or division by any variance or principal minor, and therefore also applies when $\Gamma$ is singular.
\end{proof}

Since $\delta\geqslant0$, Thm.~\ref{thm:Spin_covariance_refinement} immediately yields a state-independent lower bound for $j\geqslant1$.
Together with the qubit case, this gives the following optimal bounds for all irreducible spin representations.

\begin{mycor}{Tight Spin Bounds}{state_independent_spin_bounds}
For every state $\rho$ in an irreducible spin-$j$ representation, with $j\in\{1/2,1,3/2,\ldots\}$,
\begin{align}
\label{eq:spin-state-independent-invariant}
C_2^{(3)}\left(\Gamma_\rho(\bm J)\right)
\geqslant
\frac{j^2}{4},
\end{align}
and consequently
\begin{align}
\label{eq:spin-state-independent-hybrid}
\Mfun_\rho(\bm J)
\geqslant
\Mfun_\rho^{\mathrm{sp}}(\bm J)
\geqslant
\frac j2.
\end{align}
The invariant and spectral bounds are attained if and only if $\rho$ is a spin coherent state.
The Cartesian hybrid bound is attained if and only if $\rho$ is a spin coherent state polarized along a Cartesian axis.
\end{mycor}

\begin{proof}
For $j\geqslant1$, Eq.~\eqref{eq:spin-state-independent-invariant} follows directly from Thm.~\ref{thm:Spin_covariance_refinement} because $\delta\geqslant0$.
For $j=1/2$, Eq.~\eqref{eq:spin-half-pauli} gives
\begin{align}
\Gamma_\rho(\bm J)=\frac14\1_3-\bm m\bm m^{\T},
\end{align}
where $\bm m=\langle\bm J\rangle_\rho$ and $r=\|\bm m\|_2\leqslant1/2$.
Its eigenvalues are $1/4$, $1/4$, and $1/4-r^2$, so
\begin{align}
C_2^{(3)}\left(\Gamma_\rho(\bm J)\right)=\frac1{16}+\frac12\left(\frac14-r^2\right)\geqslant\frac1{16}.
\end{align}
In both cases, equality in Eq.~\eqref{eq:spin-state-independent-invariant} requires $r=j$.
Writing $\bm n=\bm m/r$, the condition $\langle J_{\bm n}\rangle_\rho=j$ and positivity of $j\1_{2j+1}-J_{\bm n}$ force $\rho$ to be the rank-one projector $|j,\bm n\rangle\langle j,\bm n|$ onto the eigenspace with eigenvalue $j$.
Conversely, Eq.~\eqref{eq:spin-coherent-covariance} shows that every spin coherent state has covariance eigenvalues $j/2,j/2,0$ and hence attains the invariant and spectral bounds.
Equation~\eqref{eq:spin-state-independent-hybrid} then follows from Eq.~\eqref{eq:covariance_chain}.
For a coherent state, the covariance matrix of the Cartesian spin components is diagonal if and only if $\bm n$ is aligned with a Cartesian axis, which gives the stated equality condition for $\Mfun_\rho(\bm J)$.
\end{proof}

The state-independent invariant bound in Eq.~\eqref{eq:spin-state-independent-invariant} agrees with Shabbir and Bj\"ork~\cite[Eq.~(5)]{ShabbirBjork2016}.
In their fixed $N=2j$ photon sector, $S_a=2J_a$ and $C_2^{(3)}(\Gamma(\bm S))\geqslant N^2$.
Covariance eigenvalues scale by four and $C_2^{(3)}$ by sixteen, giving $C_2^{(3)}(\Gamma(\bm J))\geqslant j^2/4$ for pure and mixed states.
For $j\geqslant1$, Thm.~\ref{thm:Spin_covariance_refinement} strengthens this state-independent result by retaining the mean spin length, yielding a strictly larger lower bound whenever $\delta>0$.

For pure spin-$1$ states, the curve $C_2^{(3)}=1-3r^2/4$ already follows from the canonical covariance in Ref.~\cite[Eq.~(19)]{Dammeier2015}; the pure state orbit was also analyzed in Ref.~\cite[Sec.~IV]{ShabbirBjork2016}.
The refinement proves that this curve is the sharp lower bound over all pure and mixed states at each fixed mean spin length.

\begin{mycor}{Exact Spin-$1$ Bound}{spin_one_fixed_mean}
Let $\bm J=(J_x,J_y,J_z)^{\T}$ be the angular momentum generators in the spin-$1$ irreducible representation.
Then, for every $r\in[0,1]$,
\begin{align}
\min_{\substack{\rho\\
\|\langle\bm J\rangle_\rho\|_2=r}}
C_2^{(3)}\left(\Gamma_\rho(\bm J)\right)
=
1-\frac34r^2.
\label{eq:spin-one-constrained-minimum}
\end{align}
The minimum is attained by $|\psi_\chi\rangle=\cos\chi |1,1\rangle+\sin\chi |1,-1\rangle$, with $\chi=\frac12\arccos r$.
\end{mycor}

\begin{proof}
For $j=1$, Thm.~\ref{thm:Spin_covariance_refinement} gives
\begin{align}
C_2^{(3)}\left(\Gamma_\rho(\bm J)\right)
\geqslant
1-\frac34r^2.
\end{align}
It remains to show that this bound is attainable for every $r\in[0,1]$.
Choose
\begin{align}
\chi=\frac12\arccos r,
\quad
0\leqslant\chi\leqslant\frac{\pi}{4},
\end{align}
and consider the pure state
\begin{align}
|\psi_\chi\rangle
=
\cos\chi |1,1\rangle
+
\sin\chi |1,-1\rangle.
\end{align}
Its mean spin is
\begin{align}
\langle\bm J\rangle_{\psi_\chi}
=
(0,0,\cos2\chi)^{\T}
=
(0,0,r)^{\T},
\end{align}
so the prescribed mean spin length is realized.
The covariance matrix is
\begin{align}
\Gamma_{\psi_\chi}(\bm J)
=
\operatorname{diag}\left(
\frac{1+\sin2\chi}{2},
\frac{1-\sin2\chi}{2},
\sin^2 2\chi
\right).
\end{align}
Writing $t\coloneqq\sin2\chi=\sqrt{1-r^2}$, we obtain
\begin{align}
C_2^{(3)}\left(\Gamma_{\psi_\chi}(\bm J)\right)
=
\frac{1-t^2}{4}
+t^2
=
1-\frac34r^2.
\end{align}
Thus $|\psi_\chi\rangle$ attains the lower bound for every $r\in[0,1]$, proving Eq.~\eqref{eq:spin-one-constrained-minimum}.
\end{proof}

\begin{figure}[t]
\centering
\includegraphics[width=0.48\textwidth]{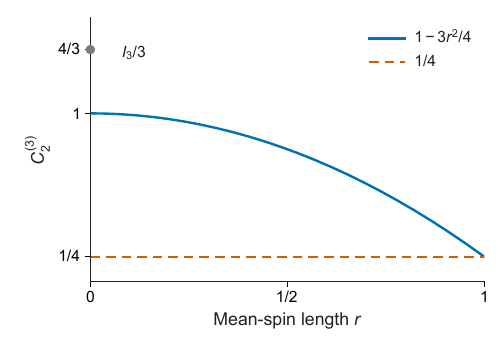}
\caption{\textbf{Exact spin-$1$ covariance bound at fixed mean spin.}
The blue solid curve shows the exact minimum of $C_2^{(3)}\left(\Gamma_\rho(\bm J)\right)$ at fixed mean spin length $r=\|\langle\bm J\rangle_\rho\|_2$, given by $1-3r^2/4$ in Eq.~\eqref{eq:spin-one-constrained-minimum}.
The orange dashed line is the state-independent bound $1/4$ from Eq.~\eqref{eq:spin-state-independent-invariant}.
The two bounds coincide at $r=1$.
The gray point marks the maximally mixed state $\rho_{\mathrm{mix}}=\1_3/3$, for which $(r,C_2^{(3)})=(0,4/3)$; it lies above the constrained minimum at $r=0$.}
\label{fig:spin-one-refinement}
\end{figure}

Equation~\eqref{eq:spin-one-constrained-minimum} gives the sharp lower bound of the second covariance invariant at fixed mean spin length; it does not imply that the covariance spectrum is determined by $r$.
For example, both $|1,0\rangle$ and the maximally mixed state $\rho_{\mathrm{mix}}=\1_3/3$ have $r=0$, but their second moments are different.
For $|1,0\rangle$,
\begin{align}
\Gamma_{|1,0\rangle}(\bm J)
=
\operatorname{diag}(1,1,0),
\quad
C_2^{(3)}\left(\Gamma_{|1,0\rangle}(\bm J)\right)
=
1,
\end{align}
which attains the constrained minimum at $r=0$.
By contrast,
\begin{align}
\Gamma_{\rho_{\mathrm{mix}}}(\bm J)
=
\frac23\1_3,
\quad
C_2^{(3)}\left(\Gamma_{\rho_{\mathrm{mix}}}(\bm J)\right)
=
\frac43.
\end{align}
Thus states with the same mean spin length can have different covariance spectra and different values of the invariant.
Figure~\ref{fig:spin-one-refinement} shows the sharp lower bound in Eq.~\eqref{eq:spin-one-constrained-minimum} together with the state-independent bound.

We now minimize over states for a fixed list of spin observables.
Corollary~\ref{cor:Hybrid_U_transfer} and Eq.~\eqref{eq:spin-state-independent-invariant} give a spectral lower bound for any spanning list of spin observables.
Equation~\eqref{eq:spin-coherent-covariance} shows that this bound is attainable by aligning the spin polarization with a right singular direction associated with the largest singular value of the coefficient matrix.

\begin{mythm}{Exact Spin Spectral Minimum}{spin-frame-minimum-appendix}
Let $\bm A=M\bm J$, where $M\in\mathbb R^{N\times3}$ has rank three, and order the singular values of $M$ as
\begin{align}
0<\sigma_1(M)\leqslant\sigma_2(M)\leqslant\sigma_3(M).
\end{align}
Then
\begin{align*}
\min_\rho\Mfun_\rho^{\mathrm{sp}}(\bm A)
=
\frac j2\sigma_1(M)\sigma_2(M),
\end{align*}
where the minimum is over all states in the irreducible spin-$j$ representation.
The minimum is attained by any spin coherent state polarized along a unit right singular vector associated with $\sigma_3(M)$.
\end{mythm}

\begin{proof}
Cor.~\ref{cor:Hybrid_U_transfer} and Eq.~\eqref{eq:spin-state-independent-invariant} give
\begin{align}
\Mfun_\rho^{\mathrm{sp}}(M\bm J)\geqslant\frac j2\sigma_1(M)\sigma_2(M).
\end{align}
To attain the bound, let $\bm w_3$ be a unit right singular vector associated with $\sigma_3(M)$ and choose the spin coherent state $\rho_*=|j,\bm w_3\rangle\langle j,\bm w_3|$.
By Eq.~\eqref{eq:spin-coherent-covariance} and the singular value decomposition of $M$, the only nonzero eigenvalues of $\Gamma_{\rho_*}(M\bm J)$ are $j\sigma_1(M)^2/2$ and $j\sigma_2(M)^2/2$.
Hence
\begin{align}
\Mfun_{\rho_*}^{\mathrm{sp}}(M\bm J)=\frac j2\sigma_1(M)\sigma_2(M),
\end{align}
which proves Eq.~\eqref{eq:main-spin-spectral-minimum}.
\end{proof}

For a weighted orthogonal triple, the three coefficient rows of $M$ are mutually orthogonal and their normalized directions are right singular vectors of $M$.
Choose a row of maximal norm and a spin coherent state polarized along its direction.
By Eq.~\eqref{eq:spin-coherent-covariance}, the corresponding observable then has zero variance, while the other two variances are $j\sigma_1(M)^2/2$ and $j\sigma_2(M)^2/2$.
The listed covariance matrix is therefore diagonal, so the first inequality in Eq.~\eqref{eq:covariance_chain} is also saturated.
Combining this observation with Thm.~\ref{thm:spin-frame-minimum} gives
\begin{align}
\min_\rho\Mfun_\rho(\bm A)
=
\min_\rho\Mfun_\rho^{\mathrm{sp}}(\bm A)
=
\frac j2\sigma_1(M)\sigma_2(M).
\end{align}
If $\sigma_3(M)$ is degenerate, every coherent state polarized within the corresponding right singular subspace attains the spectral minimum, but a generic polarization in that subspace need not diagonalize the listed covariance.
For the hybrid minimum, polarization along a coefficient row of maximal norm is sufficient.

%%%%%%%%%%%%%%%%%%%%%%%%%%%%%%%%%%%%%%%%%%%%%%%%%%%%%%%%%%%%%%%%%%%%%%%%%%%%%%%%%%%%%%%%%%%%%%%%%%%%%%%%%%%%%%%%%%

\subsection{Nonorthogonal Spin Observables}
\label{app:angular-proof}

A qubit example illustrates the distinction between the two minimizations.
For $0<\theta<\pi/2$, the listed and spectral minima are attained by different pure states.

Let $j=1/2$, and write a qubit state as $\rho=(\1_2+\bm s\cdot\bm\sigma)/2$, where $\bm s=(s_x,s_y,s_z)^{\T}$ is the Bloch vector with $\|\bm s\|_2\leqslant1$.
For $0<\theta\leqslant\pi/2$, consider
\begin{align}
A_1=J_x,
\quad
A_2=\cos\theta J_x+\sin\theta J_y,
\quad
A_3=J_z.
\label{eq:angular-qubit-frame}
\end{align}
The corresponding coefficient matrix is
\begin{align}
M_\theta
=
\begin{pmatrix}
1&0&0\\
\cos\theta&\sin\theta&0\\
0&0&1
\end{pmatrix},
\end{align}
with unit row directions
\begin{align}
\bm n_1=(1,0,0)^{\T},
\quad
\bm n_2=(\cos\theta,\sin\theta,0)^{\T},
\quad
\bm n_3=(0,0,1)^{\T}.
\end{align}
Using Eq.~\eqref{eq:spin-half-pauli}, for $i=1,2,3$ define
\begin{align}
d_i
\coloneqq
2\Delta_\rho A_i
=
\sqrt{1-(\bm n_i\cdot\bm s)^2}.
\end{align}
Then
\begin{align}
4\Mfun_\rho(A_1,A_2,A_3)
=
d_1d_2+d_1d_3+d_2d_3.
\label{eq:angular-hybrid-scaled}
\end{align}

\begin{mythm}{Nonorthogonal Qubit Minima}{nonorthogonal_qubit_minima}
For the qubit observables in Eq.~\eqref{eq:angular-qubit-frame}, with $0<\theta\leqslant\pi/2$,
\begin{align}
\min_\rho\Mfun_\rho(A_1,A_2,A_3)
=
\frac{\sin\theta}{4},
\end{align}
and
\begin{align}
\min_\rho\Mfun_\rho^{\mathrm{sp}}(A_1,A_2,A_3)
=
\frac{\sqrt{1-\cos\theta}}{4}.
\end{align}
For $0<\theta<\pi/2$, the listed minimum is attained precisely by the eigenstates of $A_1$ or $A_2$, whereas the spectral minimum is attained precisely by the two pure states with Bloch vectors $\bm s_\pm = \pm \left( \cos\frac{\theta}{2}, \sin\frac{\theta}{2}, 0 \right)^{\T}$.
At $\theta=\pi/2$, every pure qubit state attains the spectral minimum, while the listed minimum is attained precisely by eigenstates of one of the three Cartesian spin components.
\end{mythm}

\begin{proof}
We first determine the minimum of the listed hybrid uncertainty.
Write the Bloch vector as $\bm s=q\widehat{\bm s}$, where $0\leqslant q\leqslant1$ and $\|\widehat{\bm s}\|_2=1$.
For a fixed direction $\widehat{\bm s}$, the scaled standard deviations are
\begin{align}
d_i(q)
=
\sqrt{
1-q^2(\bm n_i\cdot\widehat{\bm s})^2
},
\quad
i=1,2,3.
\end{align}
Each $d_i(q)$ is nonincreasing in $q$.
Since the three measurement directions span $\mathbb R^3$, at least one scalar product $\bm n_i\cdot\widehat{\bm s}$ is nonzero, so at least one $d_i(q)$ strictly decreases as $q$ increases.
For $q<1$, all three $d_i(q)$ are strictly positive.
Consequently, the sum of products in Eq.~\eqref{eq:angular-hybrid-scaled} strictly decreases on $0\leqslant q<1$.
Increasing the Bloch radius therefore lowers the listed hybrid uncertainty of every mixed state, so all minimizers of this functional are pure.
We may restrict the remaining optimization to $\|\bm s\|_2=1$.

We now relate the standard deviations along the two nonorthogonal directions in the $xy$ plane.
Set
\begin{align}
a
\coloneqq
\bm n_1\cdot\bm s,
\quad
b
\coloneqq
\bm n_2\cdot\bm s.
\end{align}
For a pure state,
\begin{align}
d_1
=
\sqrt{1-a^2}
=
\|\bm n_1\times\bm s\|_2,
\end{align}
whereas
\begin{align}
d_2
=
\sqrt{1-b^2}
=
\|\bm n_2-b\bm s\|_2.
\end{align}
Using
\begin{align}
\bm n_1\times\bm n_2
=
\bm n_1\times(\bm n_2-b\bm s)
+
b(\bm n_1\times\bm s),
\end{align}
and the triangle inequality, we obtain
\begin{align}
\label{eq:spin-angular-deviation-sum}
\sin\theta
=
\|\bm n_1\times\bm n_2\|_2
\leqslant
d_2+|b|d_1
\leqslant
d_1+d_2.
\end{align}

To incorporate the third observable, define
\begin{align}
t
\coloneqq
\sqrt{s_x^2+s_y^2},
\quad
0\leqslant t\leqslant1.
\end{align}
Since $\bm s$ is a unit Bloch vector,
\begin{align}
d_3
=
\sqrt{1-s_z^2}
=
t.
\label{eq:angular-d3}
\end{align}
Moreover, $\bm n_1$ and $\bm n_2$ lie in the $xy$ plane, so $|\bm n_i\cdot\bm s|\leqslant t$ for $i=1,2$.
Consequently,
\begin{align}
d_1d_2
\geqslant
1-t^2.
\label{eq:angular-product-bound}
\end{align}

For $t>0$, introduce the unit vector
\begin{align}
\bm u
\coloneqq
\frac{(s_x,s_y,0)^{\T}}{t}.
\end{align}
For $i=1,2$,
\begin{align}
d_i^2
=
1-t^2(\bm n_i\cdot\bm u)^2
=
1-t^2
+
t^2
\left[
1-(\bm n_i\cdot\bm u)^2
\right],
\end{align}
and hence
\begin{align}
d_i
\geqslant
t
\sqrt{
1-(\bm n_i\cdot\bm u)^2
}.
\end{align}
Applying Eq.~\eqref{eq:spin-angular-deviation-sum} to the unit vector $\bm u$ gives
\begin{align}
d_1+d_2
\geqslant
t\sin\theta.
\label{eq:angular-sum-bound}
\end{align}
If $t=0$, this inequality is immediate because its right-hand side vanishes.

Combining Eqs.~\eqref{eq:angular-d3}, \eqref{eq:angular-product-bound}, and \eqref{eq:angular-sum-bound}, we find
\begin{align}
4\Mfun_\rho(A_1,A_2,A_3)
&=
d_1d_2+d_3(d_1+d_2)\\
&\geqslant
1-t^2+t^2\sin\theta\\
&=
\sin\theta
+
(1-\sin\theta)(1-t^2)\\
&\geqslant
\sin\theta.
\end{align}
An eigenstate of $A_1$ has
\begin{align}
(d_1,d_2,d_3)
=
(0,\sin\theta,1),
\end{align}
and therefore attains the lower bound.
An eigenstate of $A_2$ gives the same value with $d_1$ and $d_2$ interchanged.
Thus
\begin{align*}
\min_\rho\Mfun_\rho(A_1,A_2,A_3)
=
\frac{\sin\theta}{4}.
\end{align*}

We next determine the spectral minimum.
The eigenvalues of $M_\theta^{\T}M_\theta$ are
\begin{align}
1-\cos\theta,
\quad
1,
\quad
1+\cos\theta,
\end{align}
so the ordered singular values satisfy
\begin{align}
\sigma_1(M_\theta)
=
\sqrt{1-\cos\theta},
\quad
\sigma_2(M_\theta)
=
1.
\end{align}
Applying Thm.~\ref{thm:spin-frame-minimum} with $j=1/2$ therefore gives
\begin{align}
\min_\rho
\Mfun_\rho^{\mathrm{sp}}(A_1,A_2,A_3)
=
\frac14
\sigma_1(M_\theta)\sigma_2(M_\theta)
=
\frac{\sqrt{1-\cos\theta}}{4}.
\end{align}
This proves the two minimum values.

It remains to characterize the minimizing states.
For $0<\theta<\pi/2$, equality in the hybrid bound requires $t=1$, together with
\begin{align}
d_1d_2=0,
\quad
d_1+d_2=\sin\theta.
\end{align}
Hence either $d_1=0$ or $d_2=0$.
Since $d_i=0$ for a pure qubit state if and only if its Bloch vector is parallel or antiparallel to $\bm n_i$, the hybrid minimum is attained precisely by the eigenstates of $A_1$ or $A_2$.

We next characterize all states attaining the spectral minimum.
Writing $\bm s=q\widehat{\bm s}$ gives
\begin{align}
\Gamma_\rho(\bm J)
=\frac14\left(\1_3-q^2\widehat{\bm s}\widehat{\bm s}^{\T}\right),
\quad C_2^{(3)} \left(\Gamma_\rho(\bm J)\right)
=\frac{3-2q^2}{16}.
\end{align}
For $q<1$, Cor.~\ref{cor:Hybrid_U_transfer} therefore gives
\begin{align}
\Mfun_\rho^{\mathrm{sp}}(M_\theta\bm J)
\geqslant
\frac{\sigma_1(M_\theta)\sigma_2(M_\theta)}4\sqrt{3-2q^2}
>
\frac{\sigma_1(M_\theta)\sigma_2(M_\theta)}4,
\end{align}
so every spectral minimizer is pure.
For a pure qubit state with unit Bloch vector $\bm s$ and any invertible real $3\times3$ matrix $M$, the transformed covariance has rank two.
The second compound identity then yields
\begin{align}
\Mfun_\rho^{\mathrm{sp}}(M\bm J)
=\sqrt{C_2^{(3)}\left(M\Gamma_\rho(\bm J)M^{\T}\right)}
=\frac{|\det M|}{4}\left\|M^{-\T}\bm s\right\|_2
\geqslant\frac{\sigma_1(M)\sigma_2(M)}4.
\end{align}
Equality holds if and only if $\bm s$ belongs to the right singular subspace associated with $\sigma_3(M)$.
For $0<\theta<\pi/2$, the largest eigenvalue of $M_\theta^{\T}M_\theta$ is $1+\cos\theta$ and is nondegenerate.
Its unit eigenvectors are exactly the vectors $\bm s_\pm$ in the theorem statement, proving that these two pure states are precisely the spectral minimizers.

At $\theta=\pi/2$, the observable list becomes the orthogonal Cartesian triple $(J_x,J_y,J_z)^{\T}$.
Every pure qubit state then has covariance eigenvalues $1/4$, $1/4$, and $0$, and hence every pure state attains the spectral minimum.
For a pure state, the scaled standard deviations satisfy $0\leqslant d_i\leqslant1$ and $\sum_i d_i^2=2$.
Hence $\sum_i d_i\geqslant\sum_i d_i^2=2$, with equality exactly when $(d_1,d_2,d_3)$ is a permutation of $(0,1,1)$.
Using Eq.~\eqref{eq:angular-hybrid-scaled}, the listed minimum is therefore attained precisely by eigenstates of one of the three Cartesian spin components.
\end{proof}

For every nonorthogonal observable list in this family, $0<\theta<\pi/2$, the two exact minima are strictly separated.
Indeed,
\begin{align}
\sin^2\theta
=
(1-\cos\theta)(1+\cos\theta)
>
1-\cos\theta,
\end{align}
and therefore
\begin{align}
\min_\rho\Mfun_\rho^{\mathrm{sp}}(A_1,A_2,A_3)
<
\min_\rho\Mfun_\rho(A_1,A_2,A_3).
\end{align}
Thus orthogonal mixing can reduce the hybrid uncertainty below the minimum attainable in the prescribed nonorthogonal list.
At $\theta=\pi/2$, the observable list is orthogonal and the two minimum values coincide.

We next compare the exact hybrid minimum with the bound obtained from the three variances alone.
Let
\begin{align}
v_i
\coloneqq
(\Delta_\rho A_i)^2,
\quad
i=1,2,3.
\end{align}
For a qubit spin component, $0\leqslant v_i\leqslant1/4$.
Furthermore,
\begin{align}
\sum_{i=1}^3v_i
=
\frac{
3-\bm s^{\T}M_\theta^{\T}M_\theta\bm s
}{4}
\geqslant
\frac{2-\cos\theta}{4},
\end{align}
because $\|\bm s\|_2\leqslant1$ and $\lambda_{\max}(M_\theta^{\T}M_\theta)=1+\cos\theta$.

We minimize $\sum_{r<s}\sqrt{v_rv_s}$ subject only to these scalar constraints.
Since decreasing all three $v_i$ by a common factor decreases the objective, the minimum is attained when $\sum_i v_i=(2-\cos\theta)/4$.
The function
\begin{align}
(v_1,v_2,v_3)
\longmapsto
\sum_{r<s}\sqrt{v_rv_s}
\end{align}
is concave on $[0,\infty)^3$.
Hence its minimum over the resulting capped simplex is attained at an extreme point.
Because
\begin{align}
\frac14
\leqslant
\frac{2-\cos\theta}{4}
\leqslant
\frac12,
\end{align}
the relevant extreme points are the permutations of
\begin{align}
\left(
\frac14,
\frac{1-\cos\theta}{4},
0
\right).
\end{align}
Consequently, the scalar relaxation gives
\begin{align}
\Mfun_\rho(A_1,A_2,A_3)
\geqslant
\frac{\sqrt{1-\cos\theta}}{4}.
\end{align}
This value coincides with the exact spectral minimum, but for $0<\theta<\pi/2$ it is strictly smaller than the exact hybrid minimum.
Thus the optimal value of the scalar relaxation need not be attainable by the prescribed nonorthogonal observables.
The remaining gap comes from covariance constraints that are absent from the scalar relaxation.

For the observables in Eq.~\eqref{eq:su11-nonorthogonal-frame}, Thm.~\ref{thm:su11_quantitative_quantum_gap} establishes a strict gap above the positive optimized commutator bound, even though the minimizing mean is realized by a quantum state.
It also gives an explicit positive lower bound on this gap.

%%%%%%%%%%%%%%%%%%%%%%%%%%%%%%%%%%%%%%%%%%%%%%%%%%%%%%%%%%%%%%%%%%%%%%%%%%%%%%%%%%%%%%%%%%%%%%%%%%%%%%%%%%%%%%%%%%
\section{Positive Discrete Series Observables}
\label{app:su11-proofs}

The positive discrete series of $\mathfrak{su}(1,1)$ has unbounded generators.
Throughout this appendix, we restrict to states with finite second moments of the three generators and interpret commutator expectations as quadratic forms.
The representation restricts the first moments and covariance matrices that a quantum state can realize.
We determine when the optimized commutator bound equals the infimum of hybrid uncertainty over admissible states.

Subsec.~\ref{sec:su11-weight-states} fixes the representation conventions, establishes the domain and moment properties, and evaluates the moments of weight states.
Subsec.~\ref{sec:su11-covariance-bounds} derives the future timelike constraint on the mean vector and state-independent covariance bounds.
Subsec.~\ref{sec:su11-conic-bounds} formulates the first-moment optimization as a second-order cone program and proves the resulting hybrid uncertainty bound.
Subsec.~\ref{sec:su11-lorentz-frames} obtains exact minima for weighted observables transformed by Lorentz matrices.
Subsec.~\ref{sec:su11-conic-criterion} gives a necessary and sufficient condition for the optimized commutator bound to equal the quantum infimum and proves strict separation when this condition fails.
Finally, Subsec.~\ref{sec:su11-nonorthogonal-frame} gives a nonorthogonal example with a quantitative gap, a pure minimizer, and a strict separation from coherent state mixtures.

%%%%%%%%%%%%%%%%%%%%%%%%%%%%%%%%%%%%%%%%%%%%%%%%%%%%%%%%%%%%%%%%%%%%%%%%%%%%%%%%%%%%%%%%%%%%%%%%%%%%%%%%%%%%%%%%%%

\subsection{Positive Discrete Series and Moments of Weight States}\label{sec:su11-weight-states}

We first fix the representation conventions and derive the first and second moments needed in the uncertainty bounds.
Let $\mathcal D_k^+$ denote the positive discrete series representation of $\mathfrak{su}(1,1)$ with Bargmann index $k>0$, integrated to a strongly continuous unitary representation $\pi_k$ of the universal covering group $\widetilde{\mathrm{SU}(1,1)}$~\cite{BarutFronsdal1965,Perelomov1986}.
For arbitrary $k>0$, all group implementations in this appendix are understood on this covering group; when $\pi_k$ descends to $\mathrm{SU}(1,1)$, the usual group terminology applies.

\begin{mydef}{Positive Discrete Series Generators}{positive_discrete_series_generators}
For the positive discrete series $\mathcal D_k^+$ with $k>0$, let $\bm K\coloneqq(K_x,K_y,K_z)^{\T}$ denote the operator vector of self-adjoint generators.
On their common invariant dense domain,
\begin{align}
[K_x,K_y]&=-\ii K_z,
\quad
[K_y,K_z]=\ii K_x,
\quad
[K_z,K_x]=\ii K_y.
\end{align}
\end{mydef}

We use the standard lowest weight basis $\{\ket{k,m}:m=0,1,2,\ldots\}$, in which
\begin{align}
K_z\ket{k,m}
=
(k+m)\ket{k,m}.
\label{eq:su11-Kz-action}
\end{align}
Hence
\begin{align}
\operatorname{spec}(K_z)
=
\{k,k+1,k+2,\ldots\}.
\label{eq:su11-Kz-spectrum}
\end{align}
The finite linear span
\begin{align}
\mathcal D_0
\coloneqq
\operatorname{span}
\left\{
\ket{k,m}:m=0,1,2,\ldots
\right\}
\end{align}
is a common invariant dense domain for the generator polynomials in this appendix.

Define the ladder operators by $K_\pm\coloneqq K_x\pm\ii K_y$.
Choosing the phases of the weight states such that the ladder coefficients are positive,
\begin{align}
K_+\ket{k,m}
&=
c_m\ket{k,m+1},
\\[4pt]
K_-\ket{k,m}
&=
c_{m-1}\ket{k,m-1},
\label{eq:su11-ladder-action}
\end{align}
where
\begin{align}
c_m
=
\sqrt{(m+1)(2k+m)},
\quad
c_{-1}\coloneqq0.
\end{align}
In particular, $K_-\ket{k,0}=0$.

The ladder products are diagonal in the weight basis:
\begin{align}
K_+K_-\ket{k,m}
&=
m(2k+m-1)\ket{k,m},
\\[4pt]
K_-K_+\ket{k,m}
&=
(m+1)(2k+m)\ket{k,m}.
\label{eq:su11-ladder-products}
\end{align}
Their difference reproduces
\begin{align}
[K_-,K_+]
=
2K_z,
\end{align}
while the weight shift gives
\begin{align}
[K_z,K_\pm]
=
\pm K_\pm.
\end{align}
Together with $K_x=(K_++K_-)/2$ and $K_y=(K_+-K_-)/(2\ii)$, these relations are equivalent to the commutation relations in Def.~\ref{def:positive_discrete_series_generators}.

The same ladder products give
\begin{align}
K_x^2+K_y^2
=
\frac12
\left(
K_+K_-+K_-K_+
\right)
\end{align}
on $\mathcal D_0$.
Consequently,
\begin{align}
K_z^2-K_x^2-K_y^2
=
k(k-1)\1
\label{eq:su11-Casimir}
\end{align}
on this common invariant domain.
Indeed, acting on $\ket{k,m}$ gives
\begin{align}
(k+m)^2
-
\frac12
\left[
m(2k+m-1)
+
(m+1)(2k+m)
\right]
=
k(k-1),
\end{align}
independently of $m$.

\begin{mycor}{Moments of Weight States}{su11_weight_state_moments}
For $k>0$ and $m=0,1,2,\ldots$, let 
$\rho_{k,m}
\coloneqq
\ket{k,m}\bra{k,m}$, then
\begin{align}
\langle\bm K\rangle_{\rho_{k,m}}
=
\left(
\langle K_x\rangle_{\rho_{k,m}},
\langle K_y\rangle_{\rho_{k,m}},
\langle K_z\rangle_{\rho_{k,m}}
\right)^{\T}
=
(0,0,k+m)^{\T},
\label{eq:su11-weight-mean}
\end{align}
and
\begin{align}
\Gamma_{\rho_{k,m}}(\bm K)
=
\operatorname{diag}(\frac{m^2+2km+k}{2},\frac{m^2+2km+k}{2},0).
\label{eq:su11-weight-covariance}
\end{align}
\end{mycor}

In particular, the lowest weight state $\rho_{k,0}$ has covariance
\begin{align}
\Gamma_{\rho_{k,0}}(\bm K)
=
\operatorname{diag}
\left(
\frac k2,
\frac k2,
0
\right).
\label{eq:su11-lowest-weight-covariance}
\end{align}

\begin{proof}
Since $K_x$ and $K_y$ connect $\ket{k,m}$ only to the neighboring weight states,
\begin{align}
\langle K_x\rangle_{\rho_{k,m}}
=
\langle K_y\rangle_{\rho_{k,m}}
=
0.
\end{align}
Together with Eq.~\eqref{eq:su11-Kz-action}, this gives Eq.~\eqref{eq:su11-weight-mean}.
Using Eq.~\eqref{eq:su11-ladder-products},
\begin{align}
\langle K_x^2\rangle_{\rho_{k,m}}
=
\langle K_y^2\rangle_{\rho_{k,m}}
=
\frac14
\left(
c_m^2+c_{m-1}^2
\right)
=
\frac{m^2+2km+k}{2}.
\end{align}
Since $\rho_{k,m}$ is an eigenstate of $K_z$,
\begin{align}
(\Delta_{\rho_{k,m}}K_z)^2
=
0.
\end{align}
The symmetrized expectation $\langle K_xK_y+K_yK_x\rangle_{\rho_{k,m}}$ vanishes, and the centered covariances involving $K_z$ vanish because $K_x$ and $K_y$ change the weight.
Hence the real covariance matrix is diagonal and Eq.~\eqref{eq:su11-weight-covariance} follows.
Setting $m=0$ gives Eq.~\eqref{eq:su11-lowest-weight-covariance}.
\end{proof}

To extend the covariance analysis from weight states to general mixed states, we must specify how second moments of the unbounded generators are evaluated.
Let
\begin{align}
\rho
=
\sum_{\ell}p_\ell
\ket{\psi_\ell}\bra{\psi_\ell},
\quad
p_\ell>0,
\quad
\sum_\ell p_\ell=1
\end{align}
be the spectral decomposition of a density operator.
For a self-adjoint operator $A$, let $\operatorname{Dom}(A)$ denote its operator domain.
We define its second moment in the quadratic form sense by
\begin{align}
\langle A^2\rangle_\rho
\coloneqq
\sum_\ell
p_\ell
\|A\psi_\ell\|^2,
\label{eq:quadratic-second-moment}
\end{align}
whenever $\psi_\ell\in\operatorname{Dom}(A)$ for every $p_\ell>0$ and the sum is finite.
This definition requires $\psi_\ell\in\operatorname{Dom}(A)$, but not $\psi_\ell\in\operatorname{Dom}(A^2)$.

\begin{mydef}{Admissible States in the Positive Discrete Series }{admissible_discrete_series_states}
The admissible state space is
\begin{align}
\mathcal S_2(\mathcal D_k^+)
\coloneqq
\left\{
\rho   \middle|
\langle K_a^2\rangle_\rho<\infty
\ \text{for}\
a=x,y,z
\right\},
\end{align}
where $\rho$ is a density operator on the Hilbert space carrying $\mathcal D_k^+$ and the second moments are understood as in Eq.~\eqref{eq:quadratic-second-moment}.
\end{mydef}

For every $\rho\in\mathcal S_2(\mathcal D_k^+)$, the first moments of $K_x$, $K_y$, and $K_z$ are finite by the Cauchy-Schwarz inequality.
Hence the mean vector $\langle\bm K\rangle_\rho$ and the real covariance matrix $\Gamma_\rho(\bm K)$ are well defined.

For $\bm u=(u_x,u_y,u_z)^{\T}\in\mathbb R^3$, let $K_{\bm u}$ denote the self-adjoint generator associated with $\bm u$ in the representation $\pi_k$.
On $\mathcal D_0$, its action is given by
\begin{align}
K_{\bm u}\varphi
&=
u_xK_x\varphi+u_yK_y\varphi+u_zK_z\varphi,
\quad \varphi\in\mathcal D_0.
\label{eq:su11-directional-generator}
\end{align}
The ladder coefficients give the estimate
\begin{align}
\|K_{\bm u}\varphi\|
&\leqslant C_{\bm u,k}
\bigl(\|K_z\varphi\|+\|\varphi\|\bigr),
\quad \varphi\in\mathcal D_0,
\label{eq:su11-directional-graph-bound}
\end{align}
where $C_{\bm u,k}$ is a finite constant depending only on $\bm u$ and $k$.
Indeed, for $m\geqslant0$, $c_m^2\leqslant2(1+k^{-1})(m+k)^2$ and $c_{m-1}^2\leqslant(m+k)^2$.
These inequalities bound the norms of the ladder operators in terms of $\|K_z\varphi\|$ and imply Eq.~\eqref{eq:su11-directional-graph-bound}.
Let $\Pi_J$ be the projection onto $\operatorname{span}\{\ket{k,0},\ldots,\ket{k,J}\}$.
For $\psi\in\operatorname{Dom}(K_z)$, the vectors $\Pi_J\psi$ converge to $\psi$ in the graph norm of $K_z$.
By Eq.~\eqref{eq:su11-directional-graph-bound}, the sequence $K_{\bm u}\Pi_J\psi$ is Cauchy.
Since $K_{\bm u}$ is closed, it follows that $\psi\in\operatorname{Dom}(K_{\bm u})$ and $K_{\bm u}\Pi_J\psi\to K_{\bm u}\psi$.
Applying the same argument to the coordinate directions gives $K_a\Pi_J\psi\to K_a\psi$ for $a=x,y,z$.
Applying Eq.~\eqref{eq:su11-directional-generator} to $\Pi_J\psi$ and taking $J\to\infty$ therefore yields
\begin{align}
K_{\bm u}\psi
&=\sum_{a=x,y,z}u_aK_a\psi,
\quad \psi\in\operatorname{Dom}(K_z).
\label{eq:su11-directional-domain-identity}
\end{align}
Every eigenvector of an admissible state with positive eigenvalue has a finite quadratic second moment of $K_z$ and hence belongs to $\operatorname{Dom}(K_z)$.
Equation~\eqref{eq:su11-directional-domain-identity} therefore applies to each such vector.
No membership in $\operatorname{Dom}(K_a^2)$ is required.
We take the Hilbert space inner product to be linear in its second argument.

For the unbounded generators considered here, the covariance transformation law requires control of operator domains and second moments.
The following lemma establishes these conditions for every real linear combination of the generators.
\begin{mylem}{Covariance Transfer}{su11_covariance_transfer}
For every $\rho\in\mathcal S_2(\mathcal D_k^+)$, each operator $K_{\bm u}$ has finite second moment.
Consequently, for any real matrix $M\in\mathbb R^{N\times3}$,
\begin{align}
\Gamma_\rho(M\bm K)
=
M\Gamma_\rho(\bm K)M^{\T}.
\label{eq:su11-covariance-transfer}
\end{align}
\end{mylem}

\begin{proof}
Let $\rho=\sum_\ell p_\ell|\psi_\ell\rangle\langle\psi_\ell|$ be its spectral decomposition and let $\bm u=(u_x,u_y,u_z)^{\T}\in\mathbb R^3$.
For every component with $p_\ell>0$,
\begin{align}
\|K_{\bm u}\psi_\ell\|^2
&=
\left\|
\sum_{a=x,y,z}u_aK_a\psi_\ell
\right\|^2
\leqslant
\|\bm u\|_2^2
\sum_{a=x,y,z}\|K_a\psi_\ell\|^2.
\end{align}
Summing with weights $p_\ell$ gives
\begin{align}
\langle K_{\bm u}^2\rangle_\rho
\leqslant
\|\bm u\|_2^2
\sum_{a=x,y,z}\langle K_a^2\rangle_\rho
<\infty.
\end{align}
Thus every operator $K_{\bm u}$ with real $\bm u$ has finite second moment.

Now write the $p$th component of $M\bm K$ as $A_p=\sum_a M_{pa}K_a$.
The mixed quadratic forms are finite by the Cauchy-Schwarz inequality, so bilinearity of the real covariance gives
\begin{align}
\operatorname{Cov}_\rho(A_p,A_q)
=
\sum_{a,b}M_{pa}M_{qb}\operatorname{Cov}_\rho(K_a,K_b).
\end{align}
This is precisely Eq.~\eqref{eq:su11-covariance-transfer}.
\end{proof}

The Casimir identity also extends from $\mathcal D_0$ to the admissible state class in the quadratic form sense.
For every eigenvector $\psi_\ell$ with $p_\ell>0$, finite quadratic second moments give $\psi_\ell\in\operatorname{Dom}(K_z)$.
The preceding graph norm argument therefore gives $\Pi_J\psi_\ell\to\psi_\ell$ and $K_a\Pi_J\psi_\ell\to K_a\psi_\ell$ for $a=x,y,z$.
Applying Eq.~\eqref{eq:su11-Casimir} to $\Pi_J\psi_\ell$ and passing to the limit therefore gives the Casimir relation as a quadratic form identity on every admissible eigenvector.
Summing over the spectral decomposition of $\rho$ yields
\begin{align}
\langle K_z^2\rangle_\rho
-
\langle K_x^2\rangle_\rho
-
\langle K_y^2\rangle_\rho
=
k(k-1),
\label{eq:su11-Casimir-form}
\end{align}
for every $\rho\in\mathcal S_2(\mathcal D_k^+)$.

For an admissible state $\rho$ and operators $A$ and $B$ given by real linear combinations of $\bm K$, define the commutator expectation in the quadratic form sense by
\begin{align}
\langle[A,B]\rangle_\rho
&\coloneqq
\sum_\ell p_\ell
\left[
\langle A\psi_\ell,B\psi_\ell\rangle
-
\langle B\psi_\ell,A\psi_\ell\rangle
\right].
\label{eq:su11-commutator-form}
\end{align}
With the convention that the Hilbert space inner product is linear in its second argument, this expression agrees with $\sum_\ell p_\ell\langle\psi_\ell,[A,B]\psi_\ell\rangle$ whenever $\psi_\ell\in\operatorname{Dom}(AB)\cap\operatorname{Dom}(BA)$ for every $p_\ell>0$.

The truncation argument used to establish Eq.~\eqref{eq:su11-directional-domain-identity} gives $A\Pi_J\psi_\ell\to A\psi_\ell$ and $B\Pi_J\psi_\ell\to B\psi_\ell$ for every $p_\ell>0$.
The first moments of the generators also converge under this truncation.
Applying the commutation relations in Def.~\ref{def:positive_discrete_series_generators} to $\Pi_J\psi_\ell$ and taking $J\to\infty$ therefore gives the corresponding quadratic form identities on $\psi_\ell$.
Summing these identities with weights $p_\ell$ shows that the Lie commutation relations remain valid for the expectations defined in Eq.~\eqref{eq:su11-commutator-form}.
The series in Eq.~\eqref{eq:su11-commutator-form} is absolutely convergent by the Cauchy-Schwarz inequality and the finite second moments of $A$ and $B$.
Positivity of the centered complex Gram matrix then gives the Robertson-Schr\"odinger relation in Eq.~\eqref{eq:RS-pairwise}, with the commutator expectation interpreted as in Eq.~\eqref{eq:su11-commutator-form}.

%%%%%%%%%%%%%%%%%%%%%%%%%%%%%%%%%%%%%%%%%%%%%%%%%%%%%%%%%%%%%%%%%%%%%%%%%%%%%%%%%%%%%%%%%%%%%%%%%%%%%%%%%%%%%%%%%%%%%%

\subsection{Mean Constraint and State-Independent Bounds}
\label{sec:su11-covariance-bounds}

The positive discrete series imposes a nontrivial constraint on the first moments of the generators.
For $\rho\in\mathcal S_2(\mathcal D_k^+)$, define
\begin{align}
\bm\mu
\coloneqq
\langle\bm K\rangle_\rho
=
(\mu_x,\mu_y,\mu_z)^{\T}.
\end{align}
The Bargmann index $k$ fixes the boundary of the allowed first-moment region.
For pure states, the equivalent Lorentzian constraint was obtained in Ref.~\cite{deGuise2018}; Lem.~\ref{lem:su11_mean_constraint} establishes the same bound for every state in the admissible class $\mathcal S_2(\mathcal D_k^+)$.

\begin{mylem}{Future Timelike Mean Constraint}{su11_mean_constraint}
For every $\rho\in\mathcal S_2(\mathcal D_k^+)$, the mean vector $\bm\mu\coloneqq\langle\bm K\rangle_\rho$ satisfies
\begin{align}
\mu_z
&\geqslant
\sqrt{k^2+\mu_x^2+\mu_y^2}.
\label{eq:su11-mean-constraint}
\end{align}
\end{mylem}

\begin{proof}
For $\tau\geqslant0$ and $\phi\in[0,2\pi)$, define
\begin{align}
K_{\tau,\phi}
&\coloneqq
\cosh\tau K_z
-
\sinh\tau
\left(
\cos\phi K_x+\sin\phi K_y
\right).
\end{align}
The displayed sum is initially defined on the space $\mathcal H_k^\infty$ of smooth vectors for $\pi_k$, and $K_{\tau,\phi}$ denotes its self-adjoint closure.
We first show that $K_{\tau,\phi}$ is unitarily equivalent to $K_z$.

The space $\mathcal H_k^\infty$ is dense, invariant under $\pi_k$, and a common core for the generators and their real linear combinations~\cite[Sec.~5.1]{NeebRusso2025}.
Let $d\pi_k$ denote the derived representation of the Lie algebra, defined on $\mathcal H_k^\infty$ by
\begin{align}
d\pi_k(X)\psi
=
\left.\frac{d}{dt}\pi_k(\exp(tX))\psi\right|_{t=0}.
\end{align}
Choose a real Lie algebra basis $X_a$ such that $d\pi_k(X_a)=\ii K_a$ on $\mathcal H_k^\infty$.
On this space, adjoint covariance gives
\begin{align}
\pi_k(g) d\pi_k(X) \pi_k(g)^\dagger
&=
d\pi_k(\operatorname{Ad}_gX),
\quad X\in\mathfrak{su}(1,1).
\end{align}
With the commutation conventions of Def.~\ref{def:positive_discrete_series_generators}, the adjoint action satisfies
\begin{align}
\operatorname{Ad}_{\exp(\tau X_y)}X_z
&=\cosh\tau X_z-\sinh\tau X_x,\\
\operatorname{Ad}_{\exp(-\phi X_z)}X_x
&=\cos\phi X_x+\sin\phi X_y.
\end{align}
Set
\begin{align}
U_\tau&\coloneqq\pi_k(\exp(\tau X_y))=\exp(\ii\tau K_y)\\
R_\phi&\coloneqq\pi_k(\exp(-\phi X_z))=\exp(-\ii\phi K_z).
\end{align}
The corresponding operator identities on $\mathcal H_k^\infty$ are
\begin{align}
U_\tau K_zU_\tau^\dagger
&=
\cosh\tau K_z-\sinh\tau K_x,\\
R_\phi K_xR_\phi^\dagger
&=
\cos\phi K_x+\sin\phi K_y.
\end{align}
Since $R_\phi$ commutes with $K_z$, these identities give $K_{\tau,\phi}=V_{\tau,\phi}K_zV_{\tau,\phi}^\dagger$ on $\mathcal H_k^\infty$, where $V_{\tau,\phi}\coloneqq R_\phi U_\tau$.
The unitary $V_{\tau,\phi}$ preserves this common core.
Taking self-adjoint closures therefore yields
\begin{align}
K_{\tau,\phi}
&=
V_{\tau,\phi}K_zV_{\tau,\phi}^\dagger.
\label{eq:su11-boost-unitary-equivalence}
\end{align}
This argument uses the invariant smooth core and does not require $\mathcal D_0$ to be invariant under the exponentials.

Equations~\eqref{eq:su11-Kz-spectrum} and~\eqref{eq:su11-boost-unitary-equivalence} imply $K_{\tau,\phi}\geqslant k\1$.
Taking the expectation in $\rho$ and using Eq.~\eqref{eq:su11-directional-domain-identity} gives
\begin{align}
\mu_z\cosh\tau
-
\left(
\mu_x\cos\phi+\mu_y\sin\phi
\right)\sinh\tau
&\geqslant k.
\label{eq:su11-boost-mean-bound}
\end{align}

Let $\mu_\perp\coloneqq\sqrt{\mu_x^2+\mu_y^2}$.
For $\mu_\perp>0$, choose $\phi$ such that $\cos\phi=\mu_x/\mu_\perp$ and $\sin\phi=\mu_y/\mu_\perp$.
For $\mu_\perp=0$, any $\phi$ gives the same expression.
Equation~\eqref{eq:su11-boost-mean-bound} then becomes
\begin{align}
\mu_z\cosh\tau-\mu_\perp\sinh\tau
&\geqslant k,
\quad \tau\geqslant0.
\end{align}
Since $k>0$, this inequality implies $\mu_z>\mu_\perp$.
Indeed, if $\mu_z<\mu_\perp$, its left-hand side tends to $-\infty$ as $\tau\to\infty$, whereas if $\mu_z=\mu_\perp$, it tends to zero.
The minimum over $\tau\geqslant0$ is attained at $\tanh\tau_*=\mu_\perp/\mu_z$ and equals $\sqrt{\mu_z^2-\mu_\perp^2}$.
Consequently,
\begin{align}
\mu_z^2-\mu_x^2-\mu_y^2
&\geqslant k^2.
\end{align}
Since $\mu_z>0$, this is equivalent to Eq.~\eqref{eq:su11-mean-constraint}.
\end{proof}

The admissible mean vectors satisfy
\begin{align}
\mu_z>0,
\quad
\mu_z^2-\mu_x^2-\mu_y^2
\geqslant
k^2,
\end{align}
and therefore lie in the future timelike region bounded by the hyperboloid $\mu_z^2-\mu_x^2-\mu_y^2=k^2$.
In particular, the mean constraint implies $\mu_z\geqslant k$, which sets the state-independent scale in Cor.~\ref{cor:su11_state_independent_bounds}.

Combining $\mu_z\geqslant k$ with the Robertson--Schr\"odinger inequality for $K_x$ and $K_y$ gives a state-independent covariance bound.

\begin{mycor}{Discrete Series Uncertainty Bounds}{su11_state_independent_bounds}
For every $\rho\in\mathcal S_2(\mathcal D_k^+)$ in Def.~\ref{def:admissible_discrete_series_states}, we have
\begin{align}
C_2^{(3)}
\left(
\Gamma_\rho(\bm K)
\right)
\geqslant
\frac{\mu_z^2}{4}
\geqslant
\frac{k^2}{4}.
\label{eq:su11-C2-bound}
\end{align}
Consequently,
\begin{align}
\Mfun_\rho(\bm K)
\geqslant
\Mfun_\rho^{\mathrm{sp}}(\bm K)
\geqslant
\frac{k}{2}.
\label{eq:su11-hybrid-bound}
\end{align}
The state-independent lower bounds in Eqs.~\eqref{eq:su11-C2-bound} and \eqref{eq:su11-hybrid-bound} are attained precisely by the lowest weight state $\rho_{k,0}$.
\end{mycor}

\begin{proof}
Let $\Gamma\coloneqq\Gamma_\rho(\bm K)$.
Applying Eq.~\eqref{eq:RS-pairwise} to $K_x$ and $K_y$ and using $[K_x,K_y]=-\ii K_z$ gives
\begin{align}
\Gamma_{xx}\Gamma_{yy}-\Gamma_{xy}^2\geqslant\frac{\mu_z^2}{4}.
\label{eq:su11-xy-principal-minor}
\end{align}
Since $\Gamma\geqslant0$, every other second-order principal minor is nonnegative, and Eq.~\eqref{eq:second-covariance-principal-minors} therefore gives $C_2^{(3)}(\Gamma)\geqslant\mu_z^2/4$.
Equation~\eqref{eq:su11-mean-constraint} implies $\mu_z\geqslant k$, proving Eq.~\eqref{eq:su11-C2-bound}, while Eq.~\eqref{eq:covariance_chain} immediately gives Eq.~\eqref{eq:su11-hybrid-bound}.
Equation~\eqref{eq:su11-lowest-weight-covariance} shows that $\rho_{k,0}$ attains all these lower bounds.
Conversely, attaining any of the state-independent lower bounds forces $\mu_z=k$.
Because $K_z-k\1\geqslant0$ and $\ker(K_z-k\1)=\operatorname{span}\{|k,0\rangle\}$ by Eq.~\eqref{eq:su11-Kz-action}, the condition $\Tr[\rho(K_z-k\1)]=0$ implies $\rho=\rho_{k,0}$.
\end{proof}

We next consider the weighted observable list
\begin{align}
\bm A_{a,b,c}
\coloneqq
\left(
aK_x,
bK_y,
cK_z
\right)^{\T},
\quad
a,b,c>0.
\label{eq:su11-weighted-standard-frame}
\end{align}

\begin{mycor}{Weighted Observable Minima}{su11_weighted_standard_frame}
For the observable list $\bm A_{a,b,c}$ defined in Eq.~\eqref{eq:su11-weighted-standard-frame}, we have
\begin{align}
\min_{\rho\in\mathcal S_2(\mathcal D_k^+)}
\Mfun_\rho(\bm A_{a,b,c})
=
\frac{abk}{2},
\label{eq:su11-weighted-hybrid-minimum}
\end{align}
and
\begin{align}
\min_{\rho\in\mathcal S_2(\mathcal D_k^+)}
\Mfun_\rho^{\mathrm{sp}}(\bm A_{a,b,c})
=
\frac{abk}{2}.
\label{eq:su11-weighted-spectral-minimum}
\end{align}
Both minima are attained by the lowest weight state $\rho_{k,0}$.
\end{mycor}

\begin{proof}
Let $D\coloneqq\operatorname{diag}(a,b,c)$ and $\Gamma\coloneqq\Gamma_\rho(\bm K)$.
By Lem.~\ref{lem:su11_covariance_transfer}, $\Gamma_\rho(\bm A_{a,b,c})=D\Gamma D^{\T}$.
Using Eq.~\eqref{eq:second-covariance-principal-minors}, positivity of the principal minors of $\Gamma$, and Eq.~\eqref{eq:su11-xy-principal-minor}, we obtain
\begin{align}
C_2^{(3)}\left(D\Gamma D^{\T}\right)\geqslant a^2b^2\left(\Gamma_{xx}\Gamma_{yy}-\Gamma_{xy}^2\right)\geqslant\frac{a^2b^2k^2}{4}.
\end{align}
Equation~\eqref{eq:covariance_chain} therefore gives
\begin{align}
\Mfun_\rho(\bm A_{a,b,c})\geqslant\Mfun_\rho^{\mathrm{sp}}(\bm A_{a,b,c})\geqslant\frac{abk}{2}.
\end{align}
For $\rho_{k,0}$, Lem.~\ref{lem:su11_covariance_transfer} and Eq.~\eqref{eq:su11-lowest-weight-covariance} give
\begin{align}
\Gamma_{\rho_{k,0}}(\bm A_{a,b,c})=\operatorname{diag}\left(\frac{a^2k}{2},\frac{b^2k}{2},0\right),
\end{align}
for which both hybrid functionals equal $abk/2$.
Thus Eqs.~\eqref{eq:su11-weighted-hybrid-minimum} and \eqref{eq:su11-weighted-spectral-minimum} are exact and are attained by $\rho_{k,0}$.
\end{proof}

The absence of $c$ from Eqs.~\eqref{eq:su11-weighted-hybrid-minimum} and \eqref{eq:su11-weighted-spectral-minimum} has a simple origin: $K_z$ has zero variance in the lowest weight state, whereas the transverse variances remain fixed at $k/2$.
Thus rescaling the longitudinal generator does not change the minimum.

\subsection{Optimized Commutator Bounds}
\label{sec:su11-conic-bounds}

For every $\rho\in\mathcal S_2(\mathcal D_k^+)$, Lem.~\ref{lem:su11_covariance_transfer} ensures that $\Gamma_\rho(\bm K)$ is finite and that covariance transfer holds for every real coefficient matrix.
Applying the transfer bound in Eq.~\eqref{eq:compound_transfer_conclusion} and then Corollary~\ref{cor:su11_state_independent_bounds}, we obtain, for every $M\in\mathbb R^{N\times3}$ with rank three,
\begin{align}
\Mfun_\rho(M\bm K)
\geqslant
\Mfun_\rho^{\mathrm{sp}}(M\bm K)
\geqslant
\frac{k}{2}
\sigma_1(M)\sigma_2(M).
\label{eq:su11-singular-frame-bound}
\end{align}
Here the singular values are ordered as $\sigma_1(M)\leqslant\sigma_2(M)\leqslant\sigma_3(M)$.
The bound in Eq.~\eqref{eq:su11-singular-frame-bound} depends only on the Euclidean singular values of the coefficient matrix.
For the listed hybrid uncertainty, we also use a bound that retains the orientation of the pairwise commutators relative to the first moment constraint region.

Let
\begin{align}
M\in\mathbb R^{N\times3},
\quad
N\geqslant3,
\quad
\operatorname{rank}(M)=3,
\end{align}
and assume that no row of $M$ vanishes.
Write the $r$th row as $\bm m_r^{\T}$, with $\bm m_r\in\mathbb R^3$, so that
\begin{align}
A_r
=
\bm m_r\cdot\bm K,
\quad
\bm A
=
M\bm K.
\label{eq:su11-general-frame}
\end{align}
For each pair $r<s$, define
\begin{align}
\bm c_{rs}
\coloneqq
\eta_3
\left(
\bm m_r\times\bm m_s
\right),
\quad
\eta_3
\coloneqq
\operatorname{diag}(1,1,-1).
\label{eq:su11-commutator-vector}
\end{align}
The $\mathfrak{su}(1,1)$ commutation relations then give
\begin{align}
[A_r,A_s]
=
\ii \bm c_{rs}\cdot\bm K.
\label{eq:su11-frame-commutator}
\end{align}
The minus sign in the third entry of $\eta_3$ reflects $[K_x,K_y]=-\ii K_z$.

For the normalized mean
\begin{align}
\bm u
\coloneqq
\frac{\bm\mu}{k},
\end{align}
Lemma~\ref{lem:su11_mean_constraint} implies
\begin{align}
\bm u\in\mathcal C_1,
\quad
\mathcal C_1
\coloneqq
\left\{
\bm u=(u_x,u_y,u_z)^{\T}\in\mathbb R^3
\middle|
u_z\geqslant
\sqrt{1+u_x^2+u_y^2}
\right\}.
\label{eq:su11-normalized-mean-region}
\end{align}
Let $B_N(M)$ be the $\binom{N}{2}\times3$ matrix whose row associated with the pair $(r,s)$ is $\bm c_{rs}^{\T}$.

\begin{mydef}{Optimized Commutator Coefficient}{su11_conic_frame_coefficient}
The optimized commutator coefficient of $M$ is
\begin{align}
\gamma_N(M)
\coloneqq
\min_{\bm u\in\mathcal C_1}
\sum_{r<s}
\left|
\bm c_{rs}\cdot\bm u
\right|.
\label{eq:su11-conic-frame-coefficient}
\end{align}
\end{mydef}

By the definition of $B_N(M)$, Eq.~\eqref{eq:su11-conic-frame-coefficient} can equivalently be written as
\begin{align}
\gamma_N(M)
=
\min_{\bm u\in\mathcal C_1}
\left\|
B_N(M)\bm u
\right\|_1,
\end{align}
where $\|\bm v\|_1\coloneqq\sum_j|v_j|$.

The matrix $B_N(M)$ is directly related to the second compound matrix defined in Eq.~\eqref{eq:second-compound-definition}.
With the same ordering of row pairs,
\begin{align}
B_N(M)
=
\comp M Q,
\quad
Q
\coloneqq
\begin{pmatrix}
0&0&-1\\
0&-1&0\\
1&0&0
\end{pmatrix}.
\end{align}
Since $Q^{\T}Q=\1_3$, the matrices $B_N(M)$ and $\comp M$ have the same singular values.
Therefore, by Eq.~\eqref{eq:compound_min_singular},
\begin{align}
\sigma_{\min}\left(B_N(M)\right)
=
\sigma_1(M)\sigma_2(M)
>
0.
\label{eq:su11-B-smallest-singular}
\end{align}

\begin{mylem}{Positive Commutator Minimum}{su11_conic_attainment}
For every $M\in\mathbb R^{N\times3}$ with rank three, the minimum in Eq.~\eqref{eq:su11-conic-frame-coefficient} is attained, and
\begin{align}
\gamma_N(M)
&\geqslant
\sigma_1(M)\sigma_2(M)
>
0.
\label{eq:su11-conic-coefficient-bound}
\end{align}
\end{mylem}

\begin{proof}
Write $F_M(\bm u)\coloneqq\|B_N(M)\bm u\|_1$.
By Eq.~\eqref{eq:su11-B-smallest-singular}, every $\bm u\in\mathcal C_1$ satisfies
\begin{align}
F_M(\bm u)
&\geqslant
\|B_N(M)\bm u\|_2
\geqslant
\sigma_1(M)\sigma_2(M)\|\bm u\|_2.
\label{eq:su11-conic-coercive-bound}
\end{align}
Equation~\eqref{eq:su11-normalized-mean-region} gives $u_z\geqslant1$ and hence $\|\bm u\|_2\geqslant1$.
Thus $F_M(\bm u)\geqslant\sigma_1(M)\sigma_2(M)>0$ throughout $\mathcal C_1$.

Equation~\eqref{eq:su11-conic-coercive-bound} also implies that $F_M(\bm u)\to\infty$ as $\|\bm u\|_2\to\infty$.
The feasible set $\mathcal C_1$ is closed and nonempty, since it contains $(0,0,1)^{\T}$.
Every minimizing sequence is therefore bounded and has a convergent subsequence with limit in $\mathcal C_1$.
Continuity of $F_M$ shows that this limit attains the minimum.
The preceding lower bound then gives Eq.~\eqref{eq:su11-conic-coefficient-bound}.
\end{proof}

Positive homogeneity shows that every minimizer lies on the boundary of $\mathcal C_1$.
Let $q(\bm u)\coloneqq u_z^2-u_x^2-u_y^2$.
For any $\bm u\in\mathcal C_1$ with $q(\bm u)>1$, the vector $\bm u'\coloneqq\bm u/\sqrt{q(\bm u)}$ belongs to $\mathcal C_1$ and satisfies $q(\bm u')=1$.
Moreover,
\begin{align}
F_M(\bm u')
&=
\frac{F_M(\bm u)}{\sqrt{q(\bm u)}}
<
F_M(\bm u),
\end{align}
where strictness follows from $F_M(\bm u)>0$.
Every minimizer therefore lies on the future unit hyperboloid $q(\bm u)=1$, $u_z>0$.

Each such normalized mean is realized by a Lorentz transform of the lowest weight state.
Choose a proper orthochronous Lorentz transformation $L$ satisfying $L(0,0,1)^{\T}=\bm u$, with a unitary implementation $U_L$ supplied by $\pi_k$ as in Eq.~\eqref{eq:su11-lorentz-unitary}.
By Lem.~\ref{lem:su11_lorentz_invariance}, the state $\rho_{\bm u}\coloneqq U_L^\dagger\rho_{k,0}U_L$ is admissible.
Its mean satisfies
\begin{align}
\langle\bm K\rangle_{\rho_{\bm u}}
=
L\langle\bm K\rangle_{\rho_{k,0}}
=
k\bm u.
\end{align}
Thus every minimizing mean is realized by an admissible coherent state.
It remains to determine whether the covariance constraints imposed by the representation allow $\Mfun_\rho(M\bm K)$ to attain or approach the optimized commutator lower bound $k\gamma_N(M)/2$.

The minimization in Def.~\ref{def:su11_conic_frame_coefficient} admits a standard second-order cone formulation.
Introduce an auxiliary variable $t_{rs}\geqslant0$ for each pair $r<s$.
Then
\begin{align}
\gamma_N(M)
=
\min_{\bm u,\{t_{rs}\}}
\sum_{r<s}t_{rs},
\label{eq:su11-conic-program}
\end{align}
subject to
\begin{align}
-t_{rs}
\leqslant
\bm c_{rs}\cdot\bm u
\leqslant
t_{rs},
\quad
r<s,
\end{align}
and
\begin{align}
\left\|
(1,u_x,u_y)^{\T}
\right\|_2
\leqslant
u_z.
\label{eq:su11-conic-cone-constraint}
\end{align}
By Eq.~\eqref{eq:su11-normalized-mean-region}, the constraint in Eq.~\eqref{eq:su11-conic-cone-constraint} is equivalent to $\bm u\in\mathcal C_1$.
Hence Eq.~\eqref{eq:su11-conic-program} is a convex optimization problem with a linear objective, linear inequalities, and one second-order cone constraint.

Combining the optimized commutator coefficient with the pairwise uncertainty relations gives the following bound.

\begin{mythm}{Optimized Commutator Bound}{su11_conic_hybrid_bound}
Let $M$ have rank three and no zero row, and set $\bm A=M\bm K$.
For every $\rho\in\mathcal S_2(\mathcal D_k^+)$ in Def.~\ref{def:admissible_discrete_series_states}, we have
\begin{align}
\Mfun_\rho(M\bm K)
\geqslant
\frac{k}{2}\gamma_N(M)
\geqslant
\frac{k}{2}\sigma_1(M)\sigma_2(M).
\label{eq:su11-conic-hybrid-bound}
\end{align}
Equality in the first inequality holds if and only if the normalized mean $\bm\mu/k$ attains the minimum in Def.~\ref{def:su11_conic_frame_coefficient} and, for every $r<s$, the Robertson-Schr\"odinger relation is saturated with $\operatorname{Cov}_\rho(A_r,A_s)=0$.
\end{mythm}

\begin{proof}
Write $A_r=\bm m_r\cdot\bm K$ as defined in Eq.~\eqref{eq:su11-general-frame}.
For each pair $r<s$, the Robertson-Schr\"odinger relation in Eq.~\eqref{eq:RS-pairwise}, together with Eq.~\eqref{eq:su11-frame-commutator}, gives
\begin{align}
\Delta_\rho A_r
\Delta_\rho A_s
\geqslant
\frac12
\left|
\langle[A_r,A_s]\rangle_\rho
\right|
=
\frac12
\left|
\bm c_{rs}\cdot\bm\mu
\right|.
\label{eq:su11-pairwise-commutator-bound}
\end{align}
Summing Eq.~\eqref{eq:su11-pairwise-commutator-bound} over all unordered pairs yields
\begin{align}
\Mfun_\rho(M\bm K)
\geqslant
\frac12
\sum_{r<s}
\left|
\bm c_{rs}\cdot\bm\mu
\right|
=
\frac{k}{2}
\sum_{r<s}
\left|
\bm c_{rs}\cdot
\frac{\bm\mu}{k}
\right|.
\end{align}
By Lem.~\ref{lem:su11_mean_constraint}, together with Eq.~\eqref{eq:su11-normalized-mean-region}, the normalized mean $\bm\mu/k$ belongs to $\mathcal C_1$.
Definition~\ref{def:su11_conic_frame_coefficient} therefore gives
\begin{align}
\Mfun_\rho(M\bm K)
\geqslant
\frac{k}{2}\gamma_N(M).
\end{align}
Finally, Lem.~\ref{lem:su11_conic_attainment}, in particular Eq.~\eqref{eq:su11-conic-coefficient-bound}, gives
\begin{align}
\gamma_N(M)
\geqslant
\sigma_1(M)\sigma_2(M),
\end{align}
which proves Eq.~\eqref{eq:su11-conic-hybrid-bound}.

Equality in the first inequality requires equality in every pairwise bound and in the minimization over the normalized mean.
Thus the normalized mean $\bm\mu/k$ must attain the minimum in Def.~\ref{def:su11_conic_frame_coefficient}, and
\begin{align}
\Delta_\rho A_r
\Delta_\rho A_s
=
\frac12
\left|
\bm c_{rs}\cdot\bm\mu
\right|,
\quad
r<s.
\end{align}
By Eq.~\eqref{eq:RS-pairwise}, this is equivalent to saturation of the Robertson-Schr\"odinger relation together with $\operatorname{Cov}_\rho(A_r,A_s)=0$ for every pair $r<s$.
\end{proof}

Thus a mean vector that minimizes the commutator objective does not by itself guarantee saturation of the quantum uncertainty bound.
Theorem~\ref{thm:su11_conic_hybrid_bound} applies to the listed hybrid uncertainty.
It does not imply the same optimized commutator bound for $\Mfun_\rho^{\mathrm{sp}}(M\bm K)$, which involves a further minimization over orthogonal mixing of the observables.

For the weighted observable list
\begin{align}
D
\coloneqq
\operatorname{diag}(a,b,c),
\quad
a,b,c>0,
\end{align}
the three commutator vectors defined in Eq.~\eqref{eq:su11-commutator-vector} are
\begin{align}
\bm c_{12}
&=
(0,0,-ab)^{\T},
\quad
\bm c_{13}
=
(0,-ac,0)^{\T},
\quad
\bm c_{23}
=
(bc,0,0)^{\T}.
\end{align}
Hence Def.~\ref{def:su11_conic_frame_coefficient} gives
\begin{align}
\gamma_3(D)
=
\min_{\bm u\in\mathcal C_1}
\left(
bc|u_x|
+
ac|u_y|
+
ab u_z
\right),
\end{align}
where $u_z\geqslant1$ follows from Eq.~\eqref{eq:su11-normalized-mean-region}.
The minimum is therefore attained at
\begin{align}
\bm u_0
=
(0,0,1)^{\T},
\end{align}
with
\begin{align}
\gamma_3(D)
=
ab.
\label{eq:su11-diagonal-conic-coefficient}
\end{align}

Thm.~\ref{thm:su11_conic_hybrid_bound} then yields
\begin{align}
\Mfun_\rho(D\bm K)
\geqslant
\frac{abk}{2}.
\end{align}
By Cor.~\ref{cor:su11_weighted_standard_frame}, this value is the exact minimum of the listed hybrid uncertainty.
Moreover, the normalized mean of the lowest weight state is
\begin{align}
\frac{
\langle\bm K\rangle_{\rho_{k,0}}
}{k}
=
(0,0,1)^{\T},
\end{align}
by Eq.~\eqref{eq:su11-weight-mean}, and Cor.~\ref{cor:su11_weighted_standard_frame} shows that $\rho_{k,0}$ attains the minimum.
Thus the optimized commutator bound is exact for the weighted observable list.

\subsection{Exact Minima under Lorentz Transformations}
\label{sec:su11-lorentz-frames}

We next consider real transformations of the generator coordinates that preserve the normalized mean region in Eq.~\eqref{eq:su11-normalized-mean-region}.
Let $L\in\mathbb R^{3\times3}$ satisfy
\begin{align}
L^{\T}\eta_3L
=
\eta_3,
\quad
\det L
=
1,
\quad
L\mathcal C_1
=
\mathcal C_1,
\label{eq:su11-proper-lorentz}
\end{align}
where $\eta_3$ is defined in Eq.~\eqref{eq:su11-commutator-vector} and $\mathcal C_1$ in Eq.~\eqref{eq:su11-normalized-mean-region}.

We use unitary implementations supplied by $\pi_k$ on the covering group introduced in Subsec.~\ref{sec:su11-weight-states}, and restrict to transformations satisfying
\begin{align}
U_L\bm K U_L^\dagger
=
L\bm K.
\label{eq:su11-lorentz-unitary}
\end{align}
We first show that these transformations leave the optimized commutator coefficient invariant and preserve the admissible state class.
Together with Cor.~\ref{cor:su11_weighted_standard_frame}, this will transfer the exact minima of the weighted observables to their Lorentz transforms.

\begin{mylem}{Lorentz Invariant Coefficient}{su11_lorentz_invariance}
Let $M\in\mathbb R^{N\times3}$ have rank three, and let $L$ satisfy Eq.~\eqref{eq:su11-proper-lorentz}.
Then
\begin{align}
\gamma_N(ML)
=
\gamma_N(M).
\label{eq:su11-conic-lorentz-invariance}
\end{align}
If, in addition, $L$ admits the unitary implementation in Eq.~\eqref{eq:su11-lorentz-unitary}, then unitary conjugation by $U_L$ preserves $\mathcal S_2(\mathcal D_k^+)$ bijectively.
\end{mylem}

\begin{proof}
Let $\widetilde{\bm c}_{rs}$ denote the commutator vector obtained after replacing $M$ by $ML$.
The $r$th row of $ML$ corresponds to the column vector $L^{\T}\bm m_r$.
Using Eq.~\eqref{eq:su11-commutator-vector}, $\det L=1$, and $L^{\T}\eta_3L=\eta_3$, we obtain
\begin{align}
\widetilde{\bm c}_{rs}
=
\eta_3
\left[
(L^{\T}\bm m_r)
\times
(L^{\T}\bm m_s)
\right]
=
L^{\T}\eta_3
(\bm m_r\times\bm m_s)
=
L^{\T}\bm c_{rs}.
\end{align}
In the second equality, we used
\begin{align}
(L^{\T}\bm x)\times(L^{\T}\bm y)=L^{-1}(\bm x\times\bm y).
\end{align}
Therefore, for every $\bm u\in\mathcal C_1$,
\begin{align}
\sum_{r<s}
\left|
\widetilde{\bm c}_{rs}\cdot\bm u
\right|
=
\sum_{r<s}
\left|
\bm c_{rs}\cdot L\bm u
\right|.
\end{align}
Since $L\mathcal C_1=\mathcal C_1$ by Eq.~\eqref{eq:su11-proper-lorentz}, Def.~\ref{def:su11_conic_frame_coefficient} gives Eq.~\eqref{eq:su11-conic-lorentz-invariance}.

To prove preservation of the admissible state class, let $\rho\in\mathcal S_2(\mathcal D_k^+)$ and set $\rho_L\coloneqq U_L\rho U_L^\dagger$.
Equation~\eqref{eq:su11-lorentz-unitary} implies
\begin{align}
U_L^\dagger\bm K U_L
=
L^{-1}\bm K.
\end{align}
Hence each second moment of a generator in $\rho_L$ is the second moment, in $\rho$, of a real linear combination of the generators.
By Lem.~\ref{lem:su11_covariance_transfer}, these moments are finite, so $\rho_L\in\mathcal S_2(\mathcal D_k^+)$.
The same argument applied to $U_L^\dagger$ proves the converse and hence bijectivity.
\end{proof}

Let $ D \coloneqq \operatorname{diag}(a,b,c), a,b,c>0, $ and consider the coefficient matrix
\begin{align}
M
\coloneqq
DL,
\label{eq:su11-weighted-lorentz-frame}
\end{align}
where $L$ satisfies Eq.~\eqref{eq:su11-proper-lorentz}.
By Eq.~\eqref{eq:su11-diagonal-conic-coefficient} and Lem.~\ref{lem:su11_lorentz_invariance}, we have
\begin{align}
\gamma_3(M)
=
\gamma_3(D)
=
ab.
\end{align}

\begin{mycor}{Lorentz Transformed Minima}{su11_lorentz_frame_minima}
Let $M=DL$ be the coefficient matrix in Eq.~\eqref{eq:su11-weighted-lorentz-frame}, where $L$ satisfies Eq.~\eqref{eq:su11-proper-lorentz} and admits the unitary implementation in Eq.~\eqref{eq:su11-lorentz-unitary}.
Then
\begin{align}
\min_{\rho\in\mathcal S_2(\mathcal D_k^+)}
\Mfun_\rho(M\bm K)
=
\min_{\rho\in\mathcal S_2(\mathcal D_k^+)}
\Mfun_\rho^{\mathrm{sp}}(M\bm K)
=
\frac{abk}{2}.
\label{eq:su11-lorentz-frame-minima}
\end{align}
Both minima are attained by
\begin{align}
\rho_L
\coloneqq
U_L\rho_{k,0}U_L^\dagger.
\label{eq:su11-lorentz-minimizer}
\end{align}
\end{mycor}

\begin{proof}
By Eq.~\eqref{eq:su11-lorentz-unitary} and Lem.~\ref{lem:su11_lorentz_invariance}, the map $\rho_0\mapsto U_L\rho_0U_L^\dagger$ is a bijection of $\mathcal S_2(\mathcal D_k^+)$.
Under this map, the moments of $DL\bm K$ in the transformed state equal those of $D\bm K$ in the original state.
The two minimizations therefore reduce exactly to Cor.~\ref{cor:su11_weighted_standard_frame}, which gives Eq.~\eqref{eq:su11-lorentz-frame-minima}; choosing $\rho_0=\rho_{k,0}$ gives the minimizer in Eq.~\eqref{eq:su11-lorentz-minimizer}.
\end{proof}

A useful one-parameter family is the boost in the $xz$ plane,
\begin{align}
L_\tau
\coloneqq
\begin{pmatrix}
\cosh\tau & 0 & -\sinh\tau\\
0 & 1 & 0\\
-\sinh\tau & 0 & \cosh\tau
\end{pmatrix},
\quad
\tau\geqslant0.
\end{align}
As established in the proof of Lem.~\ref{lem:su11_mean_constraint}, the unitary
\begin{align}
U_\tau
=
\exp(\ii\tau K_y)
\end{align}
implements this transformation,
\begin{align}
U_\tau\bm K U_\tau^\dagger
=
L_\tau\bm K.
\label{eq:su11-boost-implementation}
\end{align}
Direct calculation shows that $L_\tau$ satisfies Eq.~\eqref{eq:su11-proper-lorentz}.

The singular values of $L_\tau$ are
\begin{align}
e^{-\tau},
\quad
1,
\quad
e^\tau.
\label{eq:su11-boost-singular-values}
\end{align}
Hence the singular value estimate in Eq.~\eqref{eq:su11-singular-frame-bound} gives
\begin{align}
\Mfun_\rho(L_\tau\bm K)
\geqslant
\Mfun_\rho^{\mathrm{sp}}(L_\tau\bm K)
\geqslant
\frac{k}{2}e^{-\tau}.
\end{align}

By contrast, Eq.~\eqref{eq:su11-diagonal-conic-coefficient} gives $\gamma_3(\1_3)=1$, and Lem.~\ref{lem:su11_lorentz_invariance} therefore yields
\begin{align}
\gamma_3(L_\tau)
=
1.
\end{align}
Thm.~\ref{thm:su11_conic_hybrid_bound} then gives
\begin{align}
\Mfun_\rho(L_\tau\bm K)
\geqslant
\frac{k}{2}.
\end{align}
Moreover, Cor.~\ref{cor:su11_lorentz_frame_minima}, with $a=b=c=1$, shows that this value is the exact listed hybrid minimum:
\begin{align}
\min_{\rho\in\mathcal S_2(\mathcal D_k^+)}
\Mfun_\rho(L_\tau\bm K)
=
\frac{k}{2}.
\end{align}
Thus the optimized commutator bound remains exact for all $\tau\geqslant0$, whereas the singular value bound decreases as $e^{-\tau}$.
The ratio of the exact minimum to the singular value bound is $e^\tau$.

\subsection{An Exactness Criterion for Spanning Observable Lists}
\label{sec:su11-conic-criterion}

We determine when the optimized commutator bound equals the quantum infimum.
Throughout this subsection, we fix $k>0$ and work in the irreducible representation $\mathcal D_k^+$.
Let $M\in\mathbb R^{N\times3}$ have rank three and no zero row, and write $\bm A=M\bm K$.
Denote the $r$th row of $M$ by $\bm m_r^{\T}$, so that $A_r=\bm m_r\cdot\bm K$.
We first identify the minimizing mean and the state that realizes it.

Write $F_M(\bm u)\coloneqq\|B_N(M)\bm u\|_1$ and $q(\bm u)\coloneqq u_z^2-u_x^2-u_y^2$.
By Lem.~\ref{lem:su11_conic_attainment} and the positive homogeneity argument in Subsec.~\ref{sec:su11-conic-bounds}, the minimum of $F_M$ over $\mathcal C_1$ is attained, and every minimizer lies on the future unit hyperboloid.
The minimizer is unique.
Indeed, suppose that $\bm u$ and $\bm v$ are distinct minimizers.
Strict convexity of $(u_x,u_y)\mapsto\sqrt{1+u_x^2+u_y^2}$ implies that their midpoint $\bm w=(\bm u+\bm v)/2$ satisfies $q(\bm w)>1$.
Convexity gives $F_M(\bm w)\leqslant\gamma_N(M)$, whereas positive homogeneity yields
\begin{align}
F_M \left(\frac{\bm w}{\sqrt{q(\bm w)}}\right)
&=
\frac{F_M(\bm w)}{\sqrt{q(\bm w)}}
<
\gamma_N(M),
\end{align}
contradicting optimality.
We denote the unique minimizer by $\bm u_M$; it satisfies
\begin{align}
u_{M,z}^2-u_{M,x}^2-u_{M,y}^2
&=1,
\quad u_{M,z}>0.
\end{align}

A boundary mean also determines the quantum state uniquely.
For any future unit vector $\bm u$, define the self-adjoint operator
\begin{align}
H_{\bm u}
&\coloneqq
u_zK_z-u_xK_x-u_yK_y.
\end{align}
As in the proof of Lem.~\ref{lem:su11_mean_constraint}, this operator is unitarily equivalent to $K_z$ and has the simple spectrum $k,k+1,k+2,\ldots$.
Let $\rho_{\bm u}$ denote its rank-one spectral projection at eigenvalue $k$.
Any admissible state $\rho$ with $\langle\bm K\rangle_\rho=k\bm u$ satisfies
\begin{align}
\Tr(\rho H_{\bm u})
&=
k(u_z^2-u_x^2-u_y^2)
=
k.
\end{align}
Positivity of $H_{\bm u}-k\1$ then forces $\rho$ to be supported on its one-dimensional kernel, so $\rho=\rho_{\bm u}$.
The Lorentz construction in Subsec.~\ref{sec:su11-conic-bounds} provides an admissible coherent state with this mean.
Thus $\rho_{\bm u}$ is precisely that Perelomov coherent state~\cite{Perelomov1986}.

Set $\rho_M\coloneqq\rho_{\bm u_M}$ and define
\begin{align}
T_M
&\coloneqq
M(\eta_3+\bm u_M\bm u_M^{\T})M^{\T}.
\label{eq:su11-conic-covariance-test}
\end{align}
This matrix determines whether the optimized commutator bound is exact.

\begin{mythm}{Exactness Criterion}{su11_conic_criterion_appendix}
Let $M\in\mathbb R^{N\times3}$ have rank three and no zero row, and let $T_M$ be defined by Eq.~\eqref{eq:su11-conic-covariance-test}.
Then
\begin{align*}
\inf_{\rho\in\mathcal S_2(\mathcal D_k^+)}
\Mfun_\rho(M\bm K)
=
\frac{k}{2}\gamma_N(M)
\quad\Longleftrightarrow\quad
T_M\text{ is diagonal}.
\end{align*}
\end{mythm}

\begin{proof}
\emph{Exact attainment.}
In the lowest weight state, the complex Gram matrix of the centered generators is
\begin{align}
G_0
&=
\frac{k}{2}
\begin{pmatrix}
1&-\ii&0\\
\ii&1&0\\
0&0&0
\end{pmatrix}.
\end{align}
It has rank one.
Lorentz transformations of the state and real linear transformations of the observables act on this matrix by real matrix congruence.
The resulting Gram matrix therefore has rank at most one, and all its $2\times2$ principal minors vanish.
Consequently, every pair of observables in a coherent state saturates the Robertson-Schr\"odinger relation in Eq.~\eqref{eq:RS-pairwise}, consistently with Refs.~\cite{TrifonovDonev1998,Trifonov2000}.

Transferring the lowest weight covariance gives
\begin{align}
\Gamma_{\rho_{\bm u}}(M\bm K)
&=
\frac{k}{2}
M(\eta_3+\bm u\bm u^{\T})M^{\T}.
\label{eq:su11-boundary-frame-covariance}
\end{align}
In particular, $\Gamma_{\rho_M}(M\bm K)=kT_M/2$.
Since the full Robertson-Schr\"odinger relation is saturated for every pair, the corresponding commutator bound is saturated exactly when the pair covariance vanishes.
Thus $\rho_M$ attains the optimized commutator bound if and only if $T_M$ is diagonal.

Conversely, any admissible state attaining the optimized commutator bound must have normalized mean $\bm u_M$ by Thm.~\ref{thm:su11_conic_hybrid_bound} and uniqueness of the minimizing mean.
Uniqueness of the state with this boundary mean then identifies it with $\rho_M$.
The equality conditions therefore require $T_M$ to be diagonal.

\emph{Exclusion of approximating sequences.}
Suppose that admissible states $\rho_n$ satisfy $\Mfun_{\rho_n}(M\bm K)\to k\gamma_N(M)/2$.
Write $\bm u_n\coloneqq\langle\bm K\rangle_{\rho_n}/k$.
The pairwise commutator bound gives
\begin{align}
\gamma_N(M)
\leqslant
F_M(\bm u_n)
\leqslant
\frac{2}{k}\Mfun_{\rho_n}(M\bm K)
\longrightarrow
\gamma_N(M).
\end{align}
The lower bound in Eq.~\eqref{eq:su11-conic-coercive-bound} makes $\bm u_n$ bounded.
Every convergent subsequence has a limit in $\mathcal C_1$, and continuity of $F_M$ makes that limit a minimizing mean.
Uniqueness of the minimizer therefore implies $\bm u_n\to\bm u_M$.

Set $H\coloneqq H_{\bm u_M}$.
The spectral gap of $H$ above its lowest eigenvalue is one, so $H-k\1\geqslant\1-\rho_M$.
Convergence of the means implies $\Tr(\rho_n H)\to k$, and hence
\begin{align}
\delta_n
\coloneqq
1-\Tr(\rho_n\rho_M)
\leqslant
\Tr(\rho_n H)-k
\longrightarrow0.
\end{align}
The gentle operator lemma~\cite[Lem.~9.4.2]{Wilde2017}, together with $\rho_M\rho_n\rho_M=(1-\delta_n)\rho_M$, gives
\begin{align}
\|\rho_n-\rho_M\|_1
&\leqslant
2\sqrt{\delta_n}+\delta_n
\longrightarrow0.
\end{align}

For each $A_r$, trace norm convergence applied to bounded spectral truncations of $A_r^2$ gives
\begin{align}
\Tr(\rho_M A_r^2)
&\leqslant
\liminf_{n\to\infty}\Tr(\rho_n A_r^2).
\end{align}
Thus the second moment is lower semicontinuous along the sequence.
Moreover, $\langle A_r\rangle_{\rho_n}=k\bm m_r\cdot\bm u_n$ converges to $\langle A_r\rangle_{\rho_M}$.
Subtracting $\langle A_r\rangle_{\rho_n}^2$, which tends to $\langle A_r\rangle_{\rho_M}^2$, from the second moment therefore gives
\begin{align}
(\Delta_{\rho_M}A_r)^2
\leqslant
\liminf_{n\to\infty}(\Delta_{\rho_n}A_r)^2.
\end{align}
Because all variances are nonnegative and the sum contains finitely many pairs, these bounds imply
\begin{align}
\Mfun_{\rho_M}(M\bm K)
&\leqslant
\liminf_{n\to\infty}\Mfun_{\rho_n}(M\bm K)
=
\frac{k}{2}\gamma_N(M).
\label{eq:su11-boundary-hybrid-lsc}
\end{align}
If $T_M$ is not diagonal, the exact-attainment argument gives $\Mfun_{\rho_M}(M\bm K)>k\gamma_N(M)/2$, contradicting Eq.~\eqref{eq:su11-boundary-hybrid-lsc}.
Thus no admissible sequence can approach the optimized commutator bound in this case.
Together with the optimized commutator bound and admissibility of $\rho_M$, this proves the equivalence.
\end{proof}

\begin{mycor}{Attainment and Strict Gap}{su11_conic_attainment_gap}
Under the assumptions of Thm.~\ref{thm:su11_conic_criterion}, if $T_M$ is diagonal, the quantum minimum equals $k\gamma_N(M)/2$ and is attained uniquely by $\rho_M$.
If $T_M$ is not diagonal, then
\begin{align}
\inf_{\rho\in\mathcal S_2(\mathcal D_k^+)}
\Mfun_\rho(M\bm K)
&>
\frac{k}{2}\gamma_N(M).
\end{align}
\end{mycor}

\begin{proof}
Attainment and uniqueness in the diagonal case follow from the exact-attainment argument in the theorem.
In the nondiagonal case, the optimized commutator bound and Eq.~\eqref{eq:main-conic-criterion} give the strict inequality.
\end{proof}

The strict gap holds for each fixed $k$ and $M$.
Its magnitude and attainment of the quantum infimum in the nondiagonal case require further analysis.
The criterion recovers the exact minima for weighted Lorentz transformations.
For $M=DL$ in Eq.~\eqref{eq:su11-weighted-lorentz-frame}, invariance of the commutator objective gives
\begin{align}
\bm u_M
&=
L^{-1}(0,0,1)^{\T}.
\end{align}
Lorentz covariance then yields
\begin{align}
T_M
&=
\operatorname{diag}(a^2,b^2,0).
\end{align}
Theorem~\ref{thm:su11_conic_criterion} therefore recovers the exact quantum minimum $abk/2$ in Cor.~\ref{cor:su11_lorentz_frame_minima}.

For the nonorthogonal list $(K_x,K_y,K_x+K_z)^{\T}$, the same calculation gives
\begin{align}
T_M
&=
\frac13
\begin{pmatrix}
4&0&2\\
0&3&0\\
2&0&1
\end{pmatrix}.
\end{align}
Since $(T_M)_{13}=2/3$, the quantum infimum is strictly larger than the optimized commutator bound.

The criterion also gives a sufficient condition expressed directly in terms of the rows of $M$.

\begin{mycor}{Sufficient Gap Condition}{su11_spacelike_frame_gap}
Let $M\in\mathbb R^{N\times3}$ have rank three and no zero row.
If every row $\bm m_r^{\T}$ satisfies $\bm m_r^{\T}\eta_3\bm m_r\geqslant0$, then the quantum infimum is strictly larger than $k\gamma_N(M)/2$ for every fixed $k>0$.
\end{mycor}

\begin{proof}
For any future unit vector $\bm u$, Eq.~\eqref{eq:su11-boundary-frame-covariance} gives
\begin{align}
\frac{2}{k}\Delta_{\rho_{\bm u}}^2(\bm m_r\cdot\bm K)
&=
\bm m_r^{\T}\eta_3\bm m_r
+
(\bm m_r\cdot\bm u)^2.
\end{align}
This expression is strictly positive whenever $\bm m_r^{\T}\eta_3\bm m_r>0$.

It remains to consider a nonzero null row.
Write $\bm m_{r,\perp}\coloneqq(m_{r,x},m_{r,y})^{\T}$ and $\bm u_\perp\coloneqq(u_x,u_y)^{\T}$.
The null condition gives $|m_{r,z}|=\|\bm m_{r,\perp}\|_2>0$.
If $\bm m_r\cdot\bm u=0$, the Cauchy-Schwarz inequality and $u_z>\|\bm u_\perp\|_2$ would imply
\begin{align}
|m_{r,z}|u_z
&\leqslant
\|\bm m_{r,\perp}\|_2 \|\bm u_\perp\|_2
<
|m_{r,z}|u_z,
\end{align}
a contradiction.
Thus the variance is also strictly positive for every nonzero null row.

Applying this conclusion at $\bm u=\bm u_M$ shows that every diagonal entry of $T_M$ is positive.
On the other hand, $\eta_3+\bm u_M\bm u_M^{\T}$ has rank two.
Since $M$ has full column rank, it follows that $\operatorname{rank}T_M=2$.
A diagonal matrix with all diagonal entries positive would have rank $N\geqslant3$.
Hence $T_M$ is not diagonal, and Thm.~\ref{thm:su11_conic_criterion} gives the strict gap.
\end{proof}

The row condition is sufficient; the necessary and sufficient criterion remains the diagonal condition on $T_M$.
As an application, consider
\begin{align}
\bm A_\alpha
&\coloneqq
(K_x,K_y,K_x+\alpha K_z)^{\T}
=
M_\alpha\bm K,
\quad 0<\alpha\leqslant1,
\label{eq:su11-alpha-gap-family}\\
M_\alpha
&\coloneqq
\begin{pmatrix}
1&0&0\\
0&1&0\\
1&0&\alpha
\end{pmatrix}.
\end{align}
Since $\det M_\alpha=\alpha>0$, this matrix has rank three.
The three values of $\bm m_r^{\T}\eta_3\bm m_r$ are $1$, $1$, and $1-\alpha^2$.
Corollary~\ref{cor:su11_spacelike_frame_gap} therefore gives a strict gap for every fixed $k>0$ and $0<\alpha\leqslant1$.

The optimized commutator coefficient can also be evaluated explicitly.
For $\bm u\in\mathcal C_1$, the inequality $u_z>|u_x|$ implies $u_z+\alpha u_x>0$, so
\begin{align}
F_{M_\alpha}(\bm u)
&=
2u_z+\alpha u_x+\alpha|u_y|.
\end{align}
At fixed $u_x$, minimization over $u_y$ and $u_z$ gives $u_y=0$ and $u_z=\sqrt{1+u_x^2}$.
Consequently,
\begin{align}
\gamma_3(M_\alpha)
&=
\min_{u_x\in\mathbb R}
\left(
2\sqrt{1+u_x^2}+\alpha u_x
\right).
\end{align}
Minimizing this strictly convex function gives
\begin{align}
\gamma_3(M_\alpha)
&=
\sqrt{4-\alpha^2},
\label{eq:su11-alpha-conic-coefficient}\\
\bm u_{M_\alpha}
&=
\frac{(-\alpha,0,2)^{\T}}{\sqrt{4-\alpha^2}}.
\label{eq:su11-alpha-conic-optimizer}
\end{align}

For $\alpha=1$, Thm.~\ref{thm:su11_quantitative_quantum_gap} gives an explicit positive lower bound on the gap and proves existence of the quantum minimum.
Corollary~\ref{cor:su11_coherent_state_limitation} further shows that this minimum lies strictly below the minimum over admissible coherent state mixtures.

\subsection{Nonorthogonal Observables and a Strict Gap}
\label{sec:su11-nonorthogonal-frame}

We give an explicit bound on the gap for a nonorthogonal list.
In this subsection, we write $\bm A$ for the list $\bm A_*$ used in the main text.
Let
\begin{align}
\bm A
\coloneqq
\left(
K_x,
K_y,
K_x+K_z
\right)^{\T}
=
M\bm K,
\quad
M
\coloneqq
\begin{pmatrix}
1&0&0\\
0&1&0\\
1&0&1
\end{pmatrix}.
\label{eq:su11-nonorthogonal-frame}
\end{align}
Using Eq.~\eqref{eq:su11-commutator-vector}, the commutator vectors associated with the pairs $(1,2)$, $(1,3)$, and $(2,3)$ are
\begin{align}
\bm c_{12}
&=
(0,0,-1)^{\T},
\quad
\bm c_{13}
=
(0,-1,0)^{\T},
\quad
\bm c_{23}
=
(1,0,1)^{\T}.
\end{align}
This is the case $\alpha=1$ of the family in Eq.~\eqref{eq:su11-alpha-gap-family}.
Equations~\eqref{eq:su11-alpha-conic-coefficient} and~\eqref{eq:su11-alpha-conic-optimizer} give
\begin{align}
\gamma_3(M)=\sqrt3,
\quad
\bm u_*\coloneqq\bm u_M=\frac{(-1,0,2)^{\T}}{\sqrt3}.
\end{align}
The corresponding optimized commutator bound is
\begin{align}
\Mfun_\rho(\bm A)\geqslant\frac{\sqrt3 k}{2}.
\end{align}

The minimizing vector $\bm u_*$ is physically attainable at the level of first moments.
Specifically, let
\begin{align}
\tau_*\coloneqq\operatorname{arsinh}(1/\sqrt3),
\quad
\rho_{\mathrm m}\coloneqq U_{\tau_*}^{\dagger}\rho_{k,0}U_{\tau_*}\,.
\end{align}
With the convention $U_\tau\bm K U_\tau^\dagger=L_\tau\bm K$ in Eq.~\eqref{eq:su11-boost-implementation}, the expectation values transform as
\begin{align}
\frac{\langle\bm K\rangle_{\rho_{\mathrm m}}}{k}
&=L_{\tau_*}(0,0,1)^{\T}=\bm u_*.
\end{align}
The lowest weight covariance and covariance transfer further give
\begin{align}
\Gamma_{\rho_{\mathrm m}}(\bm K)
=\frac{k}{6}
\begin{pmatrix}4&0&-2\\0&3&0\\-2&0&1\end{pmatrix},
\quad
\Gamma_{\rho_{\mathrm m}}(\bm A)
=\frac{k}{6}
\begin{pmatrix}4&0&2\\0&3&0\\2&0&1\end{pmatrix}.
\end{align}
All three Robertson--Schr\"odinger inequalities in Eq.~\eqref{eq:RS-pairwise} are simultaneously saturated in this state.
The variance products for the pairs $(1,2)$, $(1,3)$, and $(2,3)$ are, respectively, $k^2/3$, $k^2/9$, and $k^2/12$.
For the three pairs, equality is attained in the Robertson--Schr\"odinger inequality; the pair $(1,3)$ has a nonzero covariance contribution but a vanishing commutator contribution.
However,
\begin{align}
\Delta_{\rho_{\mathrm m}}A_1\Delta_{\rho_{\mathrm m}}A_3
=\operatorname{Cov}_{\rho_{\mathrm m}}(A_1,A_3)=\frac{k}{3},
\quad
\frac12\left|\langle[A_1,A_3]\rangle_{\rho_{\mathrm m}}\right|=0.
\end{align}
Thus simultaneous saturation of the
Robertson-Schr\"odinger inequalities
does not imply saturation of the commutator bound.
In fact,
\begin{align}
\Mfun_{\rho_{\mathrm m}}(\bm A)
&=k\left(\frac{\sqrt3}{2}+\frac13\right),
\end{align}
which exceeds the optimized commutator bound even though the mean is the minimizing mean.

For convenience, define
\begin{align}
\epsilon_k
\coloneqq
\frac{1}{
4\left(
1+\sqrt{(3k+6)/2}
\right)
},
\quad
k>0.
\label{eq:su11-gap-coefficient}
\end{align}
The following theorem bounds the gap above the optimized commutator bound and proves that the quantum minimum is attained.

\begin{mythm}{Quantitative Quantum Gap}{su11_quantitative_quantum_gap_appendix}
For the observable list $\bm A$ defined in Eq.~\eqref{eq:su11-nonorthogonal-frame}, every $\rho\in\mathcal S_2(\mathcal D_k^+)$ satisfies
\begin{align}
\Mfun_\rho(\bm A)
\geqslant
k
\left(
\frac{\sqrt3}{2}
+
\epsilon_k
\right).
\label{eq:su11-quantitative-gap}
\end{align}
Moreover, the minimum of $\Mfun_\rho(\bm A)$ over $\mathcal S_2(\mathcal D_k^+)$ is attained by a pure state.
\end{mythm}

\begin{proof}
\emph{First moment constraints.}
Set
\begin{align}
p
\coloneqq
\Delta_\rho K_x,
\quad
q
\coloneqq
\Delta_\rho K_y,
\quad
r
\coloneqq
\Delta_\rho(K_x+K_z),\quad
P
\coloneqq
pq,
\quad
Q
\coloneqq
qr,
\quad
R
\coloneqq
pr,
\quad
B
\coloneqq
P+Q.
\end{align}
For the observable in Eq.~\eqref{eq:su11-nonorthogonal-frame},
\begin{align}
\Mfun_\rho(\bm A)
=
B+R.
\label{eq:su11-gap-hybrid-decomposition}
\end{align}
Write
\begin{align}
\bm\mu
&\coloneqq
\langle\bm K\rangle_\rho,
\quad
n
\coloneqq
\mu_x+\mu_z,
\quad
\nu
\coloneqq
\mu_z^2-\mu_x^2-\mu_y^2.
\end{align}
By Lem.~\ref{lem:su11_mean_constraint}, $\nu\geqslant k^2$ and $\mu_z>0$.
Hence
\begin{align}
\mu_z
\geqslant
k,
\quad
\mu_z>|\mu_x|,
\quad
n>0.
\end{align}

Using Eq.~\eqref{eq:RS-pairwise} and the commutation relations of Def.~\ref{def:positive_discrete_series_generators}, we obtain
\begin{align}
P
\geqslant
\frac{\mu_z}{2},
\quad
Q
\geqslant
\frac{n}{2}.
\label{eq:su11-gap-Robertson-products}
\end{align}
Since
\begin{align}
\nu
=
2\mu_z n-n^2-\mu_y^2,
\label{eq:su11-gap-nu-representation}
\end{align}
we also have
\begin{align}
\frac{(\mu_z+n)^2}{3}-\nu
=
\frac{
(\mu_z-2n)^2+3\mu_y^2
}{3}
\geqslant
0.
\label{eq:su11-gap-mean-identity}
\end{align}
Equations~\eqref{eq:su11-gap-Robertson-products} and \eqref{eq:su11-gap-mean-identity} therefore give
\begin{align}
B
&\geqslant
\frac{\mu_z+n}{2}
\geqslant
\frac{\sqrt{3\nu}}{2}
\geqslant
\frac{\sqrt3 k}{2}.
\label{eq:su11-gap-conic-part}
\end{align}
Moreover, Eq.~\eqref{eq:su11-gap-nu-representation} implies $2\mu_z n\geqslant\nu\geqslant k^2$.
Thus
\begin{align}
PQ
\geqslant
\frac{\mu_z n}{4}
\geqslant
\frac{k^2}{8}.
\label{eq:su11-gap-product-bound}
\end{align}
In particular, $p$, $q$, $r$, and $R$ are strictly positive.

\emph{Second moment compatibility.}
Let
\begin{align}
c
&\coloneqq
\operatorname{Cov}_\rho
\left(
K_x,K_x+K_z
\right),
\quad
\chi
\coloneqq
\nu-k(k-1).
\end{align}
Writing the second moments in Eq.~\eqref{eq:su11-Casimir-form} as variances plus squared means gives
\begin{align}
q^2+2c-r^2
=
\chi.
\label{eq:su11-gap-Casimir}
\end{align}
Positivity of the covariance matrix of $K_x$ and $K_x+K_z$ gives
\begin{align}
|c|
\leqslant
pr
=
R.
\label{eq:su11-gap-covariance-bound}
\end{align}
Since
\begin{align}
r^2
=
\frac{QR}{P},
\quad
Q
=
B-P,
\end{align}
and $P\geqslant k/2$ by Eq.~\eqref{eq:su11-gap-Robertson-products}, Eqs.~\eqref{eq:su11-gap-Casimir} and \eqref{eq:su11-gap-covariance-bound} imply
\begin{align}
q^2
\leqslant
\chi
+
\left(
1+\frac{2B}{k}
\right)R.
\label{eq:su11-gap-q-upper-first}
\end{align}
On the other hand, Eq.~\eqref{eq:su11-gap-mean-identity} and $\mu_z+n\leqslant2B$ give
\begin{align}
\chi
=
\nu-k(k-1)
\leqslant
\frac{4B^2}{3}
-
k(k-1).
\label{eq:su11-gap-chi-upper}
\end{align}

Define
\begin{align}
g
\coloneqq
\frac{\sqrt3 k}{2},
\quad
t
\coloneqq
\Mfun_\rho(\bm A)-g.
\label{eq:su11-gap-excess}
\end{align}
By Eqs.~\eqref{eq:su11-gap-hybrid-decomposition} and \eqref{eq:su11-gap-conic-part},
\begin{align}
0<R
\leqslant
t,
\quad
B
\leqslant
g+t.
\label{eq:su11-gap-BR-bounds}
\end{align}
Equation~\eqref{eq:su11-gap-product-bound}, together with $q^2=PQ/R$, yields
\begin{align}
q^2
\geqslant
\frac{k^2}{8t}.
\label{eq:su11-gap-q-lower}
\end{align}
Combining Eqs.~\eqref{eq:su11-gap-q-upper-first}, \eqref{eq:su11-gap-chi-upper}, and \eqref{eq:su11-gap-BR-bounds}, we obtain
\begin{align}
q^2
\leqslant
k
+
\left(
\frac{4\sqrt3}{3}k
+\sqrt3+1
\right)t
+
\left(
\frac43+\frac{2}{k}
\right)t^2.
\end{align}
If $0<t\leqslant k/3$, then
\begin{align}
q^2
\leqslant
k
+
\left[
\left(
\frac{4\sqrt3}{3}
+
\frac49
\right)k
+
\sqrt3
+
\frac53
\right]t
\leqslant
k+(3k+4)t.
\end{align}
Together with Eq.~\eqref{eq:su11-gap-q-lower}, this gives
\begin{align}
(3k+4)t^2
+
kt
-
\frac{k^2}{8}
\geqslant
0.
\label{eq:su11-gap-quadratic}
\end{align}
The positive root of the quadratic in Eq.~\eqref{eq:su11-gap-quadratic} is
\begin{align}
t
=
\frac{k}{
4\left(
1+\sqrt{(3k+6)/2}
\right)
}
=
k\epsilon_k.
\end{align}
Hence $t\geqslant k\epsilon_k$ whenever $0<t\leqslant k/3$.
If $t>k/3$, the same conclusion follows from $\epsilon_k<1/3$.
Using Eq.~\eqref{eq:su11-gap-excess} proves Eq.~\eqref{eq:su11-quantitative-gap}.

\emph{Existence and purity of a minimizing state.}
For some finite $\Lambda$, a minimizing sequence lies in the sublevel set
\begin{align}
\Mfun_\rho(\bm A)
\leqslant
\Lambda.
\label{eq:su11-gap-sublevel}
\end{align}
On this sublevel set,
\begin{align}
P,Q,R
\leqslant
\Lambda.
\end{align}
Equation~\eqref{eq:su11-gap-product-bound} then gives
\begin{align}
q^2
=
\frac{PQ}{R}
\geqslant
\frac{k^2}{8\Lambda}.
\end{align}
Independently, Eq.~\eqref{eq:su11-gap-product-bound} and $P\leqslant\Lambda$ give
\begin{align}
Q
\geqslant
\frac{k^2}{8P}
\geqslant
\frac{k^2}{8\Lambda}.
\end{align}
Using $p^2=PR/Q$, $r^2=QR/P$, and $P\geqslant k/2$, we obtain
\begin{align}
p^2
&\leqslant
\frac{8\Lambda^3}{k^2},
\quad
r^2
\leqslant
\frac{2\Lambda^2}{k}.
\end{align}
Moreover, Eqs.~\eqref{eq:su11-gap-q-upper-first} and \eqref{eq:su11-gap-chi-upper} give a uniform upper bound on $q^2$.

Equation~\eqref{eq:su11-gap-Robertson-products} also implies
\begin{align}
\mu_z
\leqslant
2\Lambda,
\quad
n
\leqslant
2\Lambda.
\end{align}
Together with Eq.~\eqref{eq:su11-gap-nu-representation}, these estimates uniformly bound all components of $\bm\mu$.
Consequently, the second moments of $K_x$, $K_y$, and $K_x+K_z$ are uniformly bounded on the sublevel set in Eq.~\eqref{eq:su11-gap-sublevel}.
Since $K_z=(K_x+K_z)-K_x$, we also have
\begin{align}
\langle K_z^2\rangle_\rho
\leqslant
\mu_z^2
+
2p^2
+
2r^2
\leqslant
4\Lambda^2
+
\frac{16\Lambda^3}{k^2}
+
\frac{4\Lambda^2}{k}.
\label{eq:su11-gap-Kz-second-moment}
\end{align}

Let $\Pi_J$ denote the projection onto $\operatorname{span}\{\ket{k,0},\ldots,\ket{k,J}\}$.
Since the eigenvalues of $K_z$ are $k,k+1,\ldots$, Eq.~\eqref{eq:su11-gap-Kz-second-moment} gives
\begin{align}
\operatorname{Tr}
\left[
(\1-\Pi_J)\rho
\right]
\leqslant
\frac{
\langle K_z^2\rangle_\rho
}{
(k+J+1)^2
}.
\end{align}
The right-hand side tends to zero uniformly on the sublevel set.
Applying the gentle operator lemma~\cite[Lem.~9.4.2]{Wilde2017} to $\Pi_J$ gives, for any density operator $\rho$,
\begin{align}
\left\|\rho-\Pi_J\rho\Pi_J\right\|_1
\leqslant
2\sqrt{\Tr \left[(\1-\Pi_J)\rho\right]}.
\end{align}
Hence the sublevel set is uniformly approximated in trace norm by finite-dimensional compressions and is therefore relatively compact.

Let $\rho_*$ be the trace norm limit of a minimizing subsequence.
Lower semicontinuity of the positive quadratic forms $\langle K_a^2\rangle_\rho$ gives $\rho_*\in\mathcal S_2(\mathcal D_k^+)$.
To control the first moments, for each self-adjoint component $A_r$ of $\bm A$ define the bounded spectral truncation
\begin{align}
A_r^{(T)}
\coloneqq
A_r
\1_{\{|A_r|\leqslant T\}}.
\end{align}
The uniform second moment bound gives
\begin{align}
\left|
\langle A_r\rangle_\rho
-
\langle A_r^{(T)}\rangle_\rho
\right|
\leqslant
\frac{
\langle A_r^2\rangle_\rho
}{T},
\end{align}
uniformly along the minimizing sequence.
Since $A_r^{(T)}$ is bounded, trace norm convergence implies convergence of its expectation values.
Letting $T\to\infty$ therefore gives convergence of the full first moments.
Along this sequence, convergence of the first moments and lower semicontinuity of the second moments imply lower semicontinuity of each variance and of $\Mfun_\rho(\bm A)$.
Thus $\rho_*$ attains the minimum.

Finally, write the spectral decomposition
\begin{align}
\rho_*
=
\sum_\ell
p_\ell
\ket{\psi_\ell}\bra{\psi_\ell}.
\end{align}
Finite second moments of the generators imply that every pure-state component with $p_\ell>0$ is admissible.
For each $A_r$, variance concavity follows directly from the total variance formula~\cite{TothPetz2013} and gives
\begin{align}
\Delta_{\rho_*}^2 A_r
\geqslant
\sum_\ell
p_\ell
\Delta_{\psi_\ell}^2 A_r.
\end{align}
Hence, by the Cauchy-Schwarz inequality,
\begin{align}
\Delta_{\rho_*}A_r
\Delta_{\rho_*}A_s
\geqslant
\sum_\ell
p_\ell
\Delta_{\psi_\ell}A_r
\Delta_{\psi_\ell}A_s.
\end{align}
Summing over $r<s$ gives
\begin{align}
\Mfun_{\rho_*}(\bm A)
\geqslant
\sum_\ell
p_\ell
\Mfun_{\psi_\ell}(\bm A).
\label{eq:su11-gap-mixture-bound}
\end{align}
Since $\rho_*$ is minimizing, the hybrid uncertainty of every pure component with positive weight is at least the same minimum.
Equation~\eqref{eq:su11-gap-mixture-bound} therefore forces every component with $p_\ell>0$ to attain that minimum.
In particular, a pure minimizing state exists.
\end{proof}

Theorem~\ref{thm:su11_quantitative_quantum_gap} rules out sequences of admissible states whose hybrid uncertainty approaches the optimized commutator bound $\sqrt3 k/2$.
We next compare the unrestricted quantum minimum with the minimum achievable by Perelomov coherent states and their probability mixtures.

We use the standard positive discrete series Perelomov coherent states~\cite{Perelomov1986}: they are unitary images of the lowest weight state $\rho_{k,0}$ under $\pi_k$, equivalently under the transformations introduced in Eq.~\eqref{eq:su11-lorentz-unitary}.
This definition uses the covering group for arbitrary $k>0$.
An admissible coherent state mixture is a probability mixture of such states that belongs to the state class $\mathcal S_2(\mathcal D_k^+)$ defined in Def.~\ref{def:admissible_discrete_series_states}.
We denote this class by $\mathcal C_{\mathrm{coh}}$.

\begin{mycor}{Coherent Mixture Minimum}{su11_coherent_state_limitation}
For the observable list $\bm A$ defined in Eq.~\eqref{eq:su11-nonorthogonal-frame}, the minimum over admissible coherent state mixtures is
\begin{align}
\min_{\rho\in\mathcal C_{\mathrm{coh}}}
\Mfun_\rho(\bm A)
=
\frac{19k}{16}.
\label{eq:su11-coherent-mixture-minimum}
\end{align}

The unrestricted quantum minimum is strictly smaller:
\begin{align}
\min_{\rho\in\mathcal S_2(\mathcal D_k^+)}
\Mfun_\rho(\bm A)
<
\frac{19k}{16}.
\label{eq:su11-coherent-strict-separation}
\end{align}
\end{mycor}

\begin{proof}
\emph{Coherent state minimum.}
Let $\rho_{\mathrm{coh}}$ be a Perelomov coherent state and write its normalized mean as
\begin{align}
\bm u
\coloneqq
\frac{
\langle\bm K\rangle_{\rho_{\mathrm{coh}}}
}{k}
=
(u_x,u_y,u_z)^{\T}.
\end{align}
Since coherent states are Lorentz transforms of the lowest weight state, Eq.~\eqref{eq:su11-normalized-mean-region} is saturated:
\begin{align}
u_z
=
\sqrt{1+u_x^2+u_y^2}.
\end{align}
Equation~\eqref{eq:su11-boundary-frame-covariance}, with $M=\1_3$, gives the covariance of $\rho_{\mathrm{coh}}$.
For the three operators in Eq.~\eqref{eq:su11-nonorthogonal-frame}, this gives
\begin{align}
(\Delta_{\rho_{\mathrm{coh}}}A_1)^2
&=
\frac{k}{2}(1+u_x^2),
\\
(\Delta_{\rho_{\mathrm{coh}}}A_2)^2
&=
\frac{k}{2}(1+u_y^2),
\\
(\Delta_{\rho_{\mathrm{coh}}}A_3)^2
&=
\frac{k}{2}(u_x+u_z)^2.
\end{align}

At fixed $u_x$, increasing $|u_y|$ increases both $(\Delta_{\rho_{\mathrm{coh}}}A_2)^2$ and $u_z$, and hence also $(\Delta_{\rho_{\mathrm{coh}}}A_3)^2$, since $u_x+u_z>0$.
The first variance is unchanged, so every minimizing coherent state has $u_y=0$.
Set
\begin{align}
v
\coloneqq
u_x+u_z
>
0.
\end{align}
Since $u_z^2-u_x^2=1$, we have
\begin{align}
u_z
=
\frac12
\left(
v+\frac1v
\right).
\end{align}
Substituting these variances into the definition of $\Mfun$ gives
\begin{align}
\Mfun_{\rho_{\mathrm{coh}}}(\bm A)
=
\frac{k}{4}
\left(
v^2+3v+1+\frac1v
\right).
\label{eq:su11-coherent-objective}
\end{align}
For $v>0$,
\begin{align}
\frac{\mathrm d}{\mathrm dv}
\left(
v^2+3v+1+\frac1v
\right)
=
2v+3-\frac1{v^2}.
\end{align}
The stationarity condition is equivalent to
\begin{align}
(v+1)^2(2v-1)
=
0,
\end{align}
so the unique positive stationary point is $v=1/2$.
Since the expression in Eq.~\eqref{eq:su11-coherent-objective} diverges as $v\to0^+$ and as $v\to\infty$, this point gives the global minimum,
\begin{align}
\Mfun_{\rho_{\mathrm{coh}}}(\bm A)
\geqslant
\frac{19k}{16}.
\label{eq:su11-pure-coherent-minimum}
\end{align}
Equality occurs for
\begin{align}
\bm u
=
\left(
-\frac34,
0,
\frac54
\right)^{\T}.
\label{eq:su11-coherent-minimizing-mean}
\end{align}

Now let $\rho$ be an admissible probability mixture of coherent states, written as
\begin{align}
\rho
=
\int
\rho_\lambda \mathrm dP(\lambda).
\end{align}
For each component $A_r$ of $\bm A$, the total variance formula gives
\begin{align}
\Delta_\rho^2 A_r
\geqslant
\int
\Delta_{\rho_\lambda}^2 A_r
\mathrm dP(\lambda).
\end{align}
Hence the Cauchy-Schwarz inequality gives, for every pair $r<s$,
\begin{align}
\Delta_\rho A_r
\Delta_\rho A_s
\geqslant
\int
\Delta_{\rho_\lambda}A_r
\Delta_{\rho_\lambda}A_s
\mathrm dP(\lambda).
\end{align}
After summing over $r<s$ and using Eq.~\eqref{eq:su11-pure-coherent-minimum}, we obtain
\begin{align}
\Mfun_\rho(\bm A)
\geqslant
\frac{19k}{16}.
\end{align}
The pure coherent state with the mean in Eq.~\eqref{eq:su11-coherent-minimizing-mean} attains this value, proving Eq.~\eqref{eq:su11-coherent-mixture-minimum}.

\emph{A pure-state perturbation that lowers the uncertainty.}
Let
\begin{align}
V
\coloneqq
\exp[-\ii(\ln2)K_y],
\end{align}
and, for real $\epsilon$, define
\begin{align}
\ket{\psi_\epsilon}
\coloneqq
V
\frac{
\ket{k,0}
+
\epsilon\ket{k,2}
}{
\sqrt{1+\epsilon^2}
}.
\label{eq:su11-lowering-perturbation}
\end{align}
The finite weight superposition before applying $V$ is admissible, and Lem.~\ref{lem:su11_lorentz_invariance} shows that unitary conjugation by $V$ preserves $\mathcal S_2(\mathcal D_k^+)$.
Thus every $\ket{\psi_\epsilon}$ is admissible.

At $\epsilon=0$, $\ket{\psi_0}$ is the coherent state with normalized mean given in Eq.~\eqref{eq:su11-coherent-minimizing-mean}.
Using the boost relation established in Lem.~\ref{lem:su11_mean_constraint}, we have
\begin{align}
V^\dagger A_1V
&=
\frac{5K_x-3K_z}{4},
\quad
V^\dagger A_2V
=
K_y,
\quad
V^\dagger A_3V
=
\frac{K_x+K_z}{2}.
\label{eq:su11-perturbed-observables}
\end{align}

Let
\begin{align}
\ket{\phi_\epsilon}
\coloneqq
\frac{
\ket{k,0}
+
\epsilon\ket{k,2}
}{
\sqrt{1+\epsilon^2}
}.
\end{align}
Using Eq.~\eqref{eq:su11-ladder-action}, we find
\begin{align}
\Delta_{\phi_\epsilon}^2K_x
&=
\frac{k}{2}
+
\epsilon\sqrt{k(2k+1)}
+
O(\epsilon^2),
\label{eq:su11-perturbed-x-variance}
\\
\Delta_{\phi_\epsilon}^2K_y
&=
\frac{k}{2}
-
\epsilon\sqrt{k(2k+1)}
+
O(\epsilon^2),
\label{eq:su11-perturbed-transverse-variances}
\end{align}
while
\begin{align}
\Delta_{\phi_\epsilon}^2K_z
=
O(\epsilon^2),
\quad
\operatorname{Cov}_{\phi_\epsilon}(K_x,K_z)
=
0.
\label{eq:su11-perturbed-longitudinal-moments}
\end{align}
Combining Eq.~\eqref{eq:su11-perturbed-observables} with Eqs.~\eqref{eq:su11-perturbed-x-variance},~\eqref{eq:su11-perturbed-transverse-variances}, and~\eqref{eq:su11-perturbed-longitudinal-moments} gives
\begin{align}
\begin{pmatrix}
\Delta_{\psi_\epsilon}^2A_1\\
\Delta_{\psi_\epsilon}^2A_2\\
\Delta_{\psi_\epsilon}^2A_3
\end{pmatrix}
&=
\begin{pmatrix}
25k/32\\
k/2\\
k/8
\end{pmatrix}
+
\epsilon\sqrt{k(2k+1)}
\begin{pmatrix}
25/16\\
-1\\
1/4
\end{pmatrix}
+
O(\epsilon^2).
\end{align}
At the unperturbed point $(25k/32,k/2,k/8)$, the gradient of $\Mfun$ with respect to the three listed variances is
\begin{align}
\left(
\frac35,
\frac78,
\frac94
\right).
\end{align}
Consequently,
\begin{align}
\Mfun_{\psi_\epsilon}(\bm A)
=
\frac{19k}{16}
+
\frac58
\epsilon
\sqrt{k(2k+1)}
+
O(\epsilon^2).
\end{align}
For every fixed $k>0$, a sufficiently small negative $\epsilon$ therefore gives
\begin{align}
\Mfun_{\psi_\epsilon}(\bm A)
<
\frac{19k}{16},
\end{align}
which proves Eq.~\eqref{eq:su11-coherent-strict-separation}.

\end{proof}

Combining Thm.~\ref{thm:su11_quantitative_quantum_gap} with Cor.~\ref{cor:su11_coherent_state_limitation} gives
\begin{align}
k
\left(
\frac{\sqrt3}{2}
+
\epsilon_k
\right)
\leqslant
\min_{\rho\in\mathcal S_2(\mathcal D_k^+)}
\Mfun_\rho(\bm A)
<
\frac{19k}{16}.
\end{align}
Existence of a minimizing state follows from Thm.~\ref{thm:su11_quantitative_quantum_gap}, but the exact minimum remains undetermined.
Equation~\eqref{eq:su11-quantitative-gap} gives a certified positive lower bound on the gap above the optimized commutator bound.

For the observables in Eq.~\eqref{eq:su11-nonorthogonal-frame}, Thm.~\ref{thm:su11_quantitative_quantum_gap} and Cor.~\ref{cor:su11_coherent_state_limitation} place the attained quantum minimum strictly between the optimized commutator bound and the minimum over coherent state mixtures.
The sampling bound in Eq.~\eqref{eq:su11-quantum-gap-sampling} does not require its exact value.

\section{Applications of Hybrid Uncertainty Relations}
\label{app:hybrid-applications}

We compare the estimation errors of two protocols that differ in whether the target is known before measurement.
The target is a random signed combination of the prescribed observables, and both protocols use the same number of independently prepared copies.
In the direct measurement protocol, the target is disclosed before measurement.
In the prescribed separate measurement protocol, the listed observables are measured first and the target expectation is reconstructed after disclosure.
After optimizing the copy allocation, the difference between the separate and direct mean-squared-error coefficients equals twice the listed hybrid uncertainty.
Throughout this appendix, we optimize the acquisition design using the true component variances and, when orthogonal mixing is allowed, the true covariance matrix.
The copy budget includes only the independently prepared systems used to estimate the target.
Learning this design information from an unknown state is outside the model.
Subsec.~\ref{sec:target-averaged-errors} defines the target-averaged task.
Subsec.~\ref{sec:optimal-allocation} derives the optimal allocation and converts uncertainty bounds into excess sampling costs.
Subsec.~\ref{sec:orthogonal-mixing} shows that optimizing orthogonal combinations gives a minimum excess mean squared error coefficient of $2\Mfun_\rho^{\mathrm{sp}}(\bm A)$ relative to direct measurement.
\subsection{Target-Averaged Sampling Errors}
\label{sec:target-averaged-errors}

Let $\rho$ be a fixed state and let $\bm A=(A_1,\ldots,A_N)^{\T}$ be a finite list of self-adjoint operators with finite second moments in $\rho$.
For each $r$, define
\begin{align}
a_r
&\coloneqq
\langle A_r\rangle_\rho,
\quad
d_r^2
\coloneqq
(\Delta_\rho A_r)^2.
\label{eq:sampling-component-moments}
\end{align}
Thus $d_r\geqslant0$ is the component standard deviation, and
\begin{align}
\Mfun_\rho(\bm A)
=
\sum_{r<s}d_rd_s.
\end{align}
For unbounded operators, the second moments in Eq.~\eqref{eq:sampling-component-moments} are understood in the quadratic form sense introduced in Eq.~\eqref{eq:quadratic-second-moment}.

\begin{mydef}{Sign Target Estimation Task}{sign_target_estimation}
For a sign vector
$\bm s
=
(s_1,\ldots,s_N)^{\T}
\in
\{-1,+1\}^{N}$,
define the target operator and its expectation by
\begin{align}
H_{\bm s}
&\coloneqq
\sum_{r=1}^{N}s_rA_r,
\quad
h_{\bm s}
\coloneqq
\langle H_{\bm s}\rangle_\rho
=
\sum_{r=1}^{N}s_ra_r.
\end{align}
The components $s_r$ are independent and uniformly distributed over $\{-1,+1\}$.
For each realized label $\bm s$, the task is to estimate $h_{\bm s}$ from $n$ independently prepared copies of $\rho$.
\end{mydef}

The target label is drawn independently of the preparation and, once drawn, is held fixed throughout a single estimation task.
For an estimator $\widehat h_{\bm s}$, define the target-averaged mean-squared error by
\begin{align}
\operatorname{MSE}(\widehat h;n,\rho)
\coloneqq
2^{-N}
\sum_{\bm s\in\{-1,+1\}^{N}}
\mathbb E_{\mathrm{data}\mid\bm s}
\left[
\left(
\widehat h_{\bm s}-h_{\bm s}
\right)^2
\right].
\label{eq:target-averaged-MSE}
\end{align}
The conditional expectation is over the measurement outcomes for a fixed target label, while the outer average is over the uniformly distributed sign targets, each evaluated with the same copy budget $n$.

For an unbounded operator list, assume a common dense domain $\mathcal D\subseteq\bigcap_r\operatorname{Dom}(A_r)$.
For a spectral decomposition
\begin{align}
\rho
=
\sum_{\ell}p_\ell
\ket{\psi_\ell} \bra{\psi_\ell},
\end{align}
we require $\ket{\psi_\ell}\in\mathcal D$ whenever $p_\ell>0$.
For each sign vector $\bm s$, the target $H_{\bm s}$ is taken to be a specified self-adjoint realization that agrees with $\sum_r s_rA_r$ on $\mathcal D$.
These conditions are automatic for bounded operators.
For the positive discrete series considered here, Eq.~\eqref{eq:su11-directional-generator} and Lem.~\ref{lem:su11_covariance_transfer} give the required self-adjoint realizations of these linear combinations and finite second moments.

The two protocols differ in when the target label is disclosed and in whether the target observable is measured directly or reconstructed from component measurements.
In the direct protocol, the target label $\bm s$ is revealed before measurement.
The observable $H_{\bm s}$ is measured spectrally on each of the $n$ copies, producing independent and identically distributed outcomes $Y_1,\ldots,Y_n$.
The estimator is
\begin{align}
\widehat h_{\bm s}^{ \mathrm{dir}}
\coloneqq
\frac1n
\sum_{i=1}^{n}Y_i.
\label{eq:direct-estimator}
\end{align}

In the prescribed separate protocol, the measurements are completed before the target label is revealed.
For $n\geqslant N$, the copies are divided into disjoint batches of positive integer sizes
\begin{align}
\bm n
\coloneqq
(n_1,\ldots,n_N)^{\T},
\quad
\sum_{r=1}^{N}n_r
=
n.
\label{eq:separate-batch-allocation}
\end{align}
For each $r$, the observable $A_r$ is measured on the $n_r$ copies in batch $r$, producing independent and identically distributed outcomes $X_{r1},\ldots,X_{rn_r}$.
Outcomes belonging to different batches are independent because the batches contain disjoint sets of independently prepared copies.
The component sample mean is
\begin{align}
\widehat a_r
\coloneqq
\frac1{n_r}
\sum_{\ell=1}^{n_r}
X_{r\ell}.
\label{eq:component-estimator}
\end{align}
After $\bm s$ is revealed, the target expectation is reconstructed as
\begin{align}
\widehat h_{\bm s}^{ \mathrm{sep}}
\coloneqq
\sum_{r=1}^{N}
s_r\widehat a_r.
\end{align}
The term ``prescribed'' emphasizes that the measured settings $\{A_r\}$, the allocation $\bm n$, and the reconstruction rule are fixed before $\bm s$ is known; only the realized signs are inserted after the target is disclosed.
Both protocols measure the copies individually and use only the resulting classical outcomes.
The two acquisition rules are illustrated in Fig.~\ref{fig:sampling-cost-protocol}(a,b).

Let $\operatorname{MSE}_{\mathrm{dir}}(n,\rho)$ and $\operatorname{MSE}_{\mathrm{sep}}(n,\rho;\bm n)$ denote the target-averaged mean squared error in Eq.~\eqref{eq:target-averaged-MSE} for the estimators $\widehat h_{\bm s}^{ \mathrm{dir}}$ and $\widehat h_{\bm s}^{ \mathrm{sep}}$, respectively.

\begin{mylem}{Direct and Separate Errors}{direct_separate_errors}
For the task in Def.~\ref{def:sign_target_estimation}, $n\geqslant N$, and a fixed allocation $\bm n$ satisfying Eq.~\eqref{eq:separate-batch-allocation}, we have
\begin{align}
\operatorname{MSE}_{\mathrm{dir}}(n,\rho)
&=
\frac1n
\sum_{r=1}^{N}d_r^2,
\label{eq:direct-MSE}
\\
\operatorname{MSE}_{\mathrm{sep}}(n,\rho;\bm n)
&=
\sum_{r=1}^{N}
\frac{d_r^2}{n_r}.
\label{eq:separate-MSE}
\end{align}
\end{mylem}

\begin{proof}
For a fixed target label $\bm s$, the spectral measurement of $H_{\bm s}$ satisfies
\begin{align}
\mathbb E_{\mathrm{data}\mid\bm s}[Y_i]
&=
h_{\bm s},
\\[4pt]
\mathbb E_{\mathrm{data}\mid\bm s}
\left[
(Y_i-h_{\bm s})^2
\right]
&=
(\Delta_\rho H_{\bm s})^2.
\end{align}
Since the $n$ copies are independent at fixed $\bm s$, Eq.~\eqref{eq:direct-estimator} gives
\begin{align}
\mathbb E_{\mathrm{data}\mid\bm s}
\left[
\left(
\widehat h_{\bm s}^{ \mathrm{dir}}
-
h_{\bm s}
\right)^2
\right]
=
\frac{
(\Delta_\rho H_{\bm s})^2
}{n}.
\label{eq:direct-conditional-MSE}
\end{align}
Expanding the target variance,
\begin{align}
(\Delta_\rho H_{\bm s})^2
=
\sum_{r=1}^{N}
d_r^2
+
2
\sum_{r<t}
s_rs_t
\operatorname{Cov}_\rho(A_r,A_t).
\label{eq:sign-target-variance}
\end{align}
The independent uniform signs satisfy
\begin{align}
\mathbb E_{\bm s}[s_rs_t]
=
\delta_{rt},
\end{align}
where $\delta_{rt}$ is the Kronecker delta.
Averaging Eq.~\eqref{eq:direct-conditional-MSE} over $\bm s$ therefore removes all covariance terms in Eq.~\eqref{eq:sign-target-variance}, proving Eq.~\eqref{eq:direct-MSE}.

For the separate protocol, the component estimator in Eq.~\eqref{eq:component-estimator} is unbiased and satisfies
\begin{align}
\mathbb E_{\mathrm{data}}
\left[
(\widehat a_r-a_r)^2
\right]
=
\frac{d_r^2}{n_r}.
\end{align}
The component estimates are obtained from disjoint sets of independently prepared copies and are therefore mutually independent.
Thus, for every fixed $\bm s$,
\begin{align}
\mathbb E_{\mathrm{data}\mid\bm s}
\left[
\left(
\widehat h_{\bm s}^{ \mathrm{sep}}
-
h_{\bm s}
\right)^2
\right]
=
\sum_{r=1}^{N}
\frac{d_r^2}{n_r}.
\end{align}
The right-hand side is independent of $\bm s$, so the target average leaves it unchanged and gives Eq.~\eqref{eq:separate-MSE}.
\end{proof}

The target average in Eq.~\eqref{eq:target-averaged-MSE} is essential because it cancels the cross covariance terms in the direct error.
For a fixed sign vector $\bm s$, the direct variance retains the covariance terms in Eq.~\eqref{eq:sign-target-variance}, whereas the separate error in Eq.~\eqref{eq:separate-MSE} is independent of $\bm s$.
Consequently,
\begin{align}
\Mfun_\rho(\bm A)
=
\frac12
\left[
\left(
\sum_{r=1}^{N}d_r
\right)^2
-
(\Delta_\rho H_{\bm s})^2
\right]
+
\sum_{r<t}
s_rs_t
\operatorname{Cov}_\rho(A_r,A_t).
\end{align}
The covariance contribution generally remains for a fixed target, but its average over the independent uniform signs is zero.
We next optimize the batch allocation in the prescribed separate protocol.

\subsection{Optimal Allocation and Hybrid Sampling Cost}
\label{sec:optimal-allocation}

Lem.~\ref{lem:direct_separate_errors} shows that the error of the prescribed separate protocol depends on the allocation of copies among the listed observables.
Allocating copies in proportion to the component standard deviations is the standard variance-minimizing rule~\cite{RubinBabbushMcClean2018,CrawfordEtAl2021}.
The asymptotic allocation analysis gives the exact relation to hybrid uncertainty in Eq.~\eqref{eq:main-hybrid-cost}.

For $n\geqslant N$, define the set of admissible integer allocations by
\begin{align}
\mathcal N_n
\coloneqq
\left\{
\bm n=(n_1,\ldots,n_N)^{\T} \middle|
n_r\in\mathbb N,
n_r\geqslant1,
\sum_{r=1}^{N}n_r=n
\right\}.
\end{align}
We also write
\begin{align}
S
\coloneqq
\sum_{r=1}^{N}d_r.
\end{align}
By Eq.~\eqref{eq:direct-MSE}, the coefficient of $1/n$ in the direct error is
\begin{align}
R_{\mathrm{dir}}
\coloneqq
n \operatorname{MSE}_{\mathrm{dir}}(n,\rho)
=
\sum_{r=1}^{N}d_r^2.
\label{eq:direct-error-coefficient}
\end{align}
For the prescribed separate protocol, define the optimized asymptotic coefficient by
\begin{align}
R_{\mathrm{sep}}
\coloneqq
\lim_{n\to\infty}
n
\min_{\bm n\in\mathcal N_n}
\operatorname{MSE}_{\mathrm{sep}}(n,\rho;\bm n),
\label{eq:separate-error-coefficient}
\end{align}
whenever the limit exists.

\begin{mythm}{Optimal Sampling Allocation}{optimal_allocation_hybrid_excess}
For the sign target estimation task in Def.~\ref{def:sign_target_estimation}, the limit in Eq.~\eqref{eq:separate-error-coefficient} exists and satisfies
\begin{align}
R_{\mathrm{sep}}
=
S^2.
\label{eq:optimal-separate-error-coefficient}
\end{align}
Consequently,
\begin{align}
\Mfun_\rho(\bm A)
=
\frac{1}{2}\left(
R_{\mathrm{sep}}-R_{\mathrm{dir}}\right)
.
\end{align}
If $S>0$, then for $r=1,\ldots,N$, every asymptotically optimal sequence of allocations satisfies
\begin{align}
\frac{n_r}{n}
\longrightarrow
\frac{d_r}{S}.
\label{eq:optimal-allocation-fractions}
\end{align}
\end{mythm}

\begin{proof}
If $S=0$, then $d_r=0$ for every $r$.
Equations~\eqref{eq:direct-MSE} and \eqref{eq:separate-MSE} give $R_{\mathrm{dir}}=R_{\mathrm{sep}}=0$, and Eq.~\eqref{eq:main-hybrid-cost} follows immediately.
We therefore assume $S>0$.

For any allocation $\bm n\in\mathcal N_n$, the Cauchy-Schwarz inequality gives
\begin{align}
S^2
=
\left(
\sum_{r=1}^{N}
\frac{d_r}{\sqrt{n_r}}
\sqrt{n_r}
\right)^2
\leqslant
\left(
\sum_{r=1}^{N}
\frac{d_r^2}{n_r}
\right)
\left(
\sum_{r=1}^{N}n_r
\right)
=
n
\operatorname{MSE}_{\mathrm{sep}}(n,\rho;\bm n),
\label{eq:allocation-CS-bound}
\end{align}
where the last equality uses Eq.~\eqref{eq:separate-MSE}.
Hence
\begin{align}
\liminf_{n\to\infty}
n
\min_{\bm n\in\mathcal N_n}
\operatorname{MSE}_{\mathrm{sep}}(n,\rho;\bm n)
\geqslant
S^2.
\label{eq:separate-liminf}
\end{align}

To attain this coefficient asymptotically, let
\begin{align}
N_0
\coloneqq
\#\{r  |  d_r=0\}.
\end{align}
Assign one copy to each setting with zero variance.
For every $r$ with $d_r>0$, choose positive integer counts satisfying
\begin{align}
n_r
=
(n-N_0)\frac{d_r}{S}
+
O(1),
\end{align}
with the bounded rounding corrections chosen so that $\sum_r n_r=n$.
Since $N$ is fixed,
\begin{align}
n_r
=
n\frac{d_r}{S}
+
O(1),
\quad
d_r>0.
\end{align}
Therefore,
\begin{align}
\lim_{n\to\infty}
n
\operatorname{MSE}_{\mathrm{sep}}(n,\rho;\bm n)
=
\sum_{d_r>0}
\frac{d_r^2}{d_r/S}
=
S^2.
\end{align}
Together with Eq.~\eqref{eq:separate-liminf}, this proves both the existence of the limit in Eq.~\eqref{eq:separate-error-coefficient} and Eq.~\eqref{eq:optimal-separate-error-coefficient}.

Using Eqs.~\eqref{eq:direct-error-coefficient} and \eqref{eq:optimal-separate-error-coefficient}, we obtain
\begin{align}
R_{\mathrm{sep}}-R_{\mathrm{dir}}
=
\left(
\sum_{r=1}^{N}d_r
\right)^2
-
\sum_{r=1}^{N}d_r^2
=
2
\sum_{r<s}
d_rd_s
=
2\Mfun_\rho(\bm A),
\end{align}
which proves Eq.~\eqref{eq:main-hybrid-cost}.

It remains to characterize asymptotically optimal allocations.
Set
\begin{align}
p_r
\coloneqq
\frac{n_r}{n},
\quad
p_r>0,
\quad
\sum_{r=1}^{N}p_r=1.
\end{align}
A direct calculation gives
\begin{align}
n
\operatorname{MSE}_{\mathrm{sep}}(n,\rho;\bm n)
-
S^2
=
\sum_{r=1}^{N}
\frac{
(d_r-Sp_r)^2
}{p_r}.
\end{align}
For any asymptotically optimal sequence, the left-hand side tends to zero.
Every term on the right-hand side is nonnegative, so
\begin{align}
d_r-Sp_r
\longrightarrow
0,
\end{align}
for every $r$.
Since $S>0$, this proves Eq.~\eqref{eq:optimal-allocation-fractions}.
\end{proof}

Equation~\eqref{eq:main-hybrid-cost} states that, with known component variances and optimal copy allocation, the listed hybrid uncertainty is half the excess mean squared error coefficient of the prescribed separate protocol relative to direct measurement.
The identity holds for the specified measurement settings, independent batches, and sample-mean estimator; it does not bound the cost of delayed disclosure for arbitrary measurement protocols.
The optimal separate protocol allocates copies in proportion to the component standard deviations, rather than their variances.

For a prescribed root-mean-square error tolerance $\varepsilon>0$, define the minimum copy counts
\begin{align}
n_{\mathrm{sep}}^*(\varepsilon)
&\coloneqq
\min
\left\{
n\geqslant N  \middle|\,
\min_{\bm n\in\mathcal N_n}
\operatorname{MSE}_{\mathrm{sep}}(n,\rho;\bm n)
\leqslant
\varepsilon^2
\right\},
\\[4pt]
n_{\mathrm{dir}}^*(\varepsilon)
&\coloneqq
\min
\left\{
n\geqslant1 \middle|\,
\operatorname{MSE}_{\mathrm{dir}}(n,\rho)
\leqslant
\varepsilon^2
\right\}.
\end{align}
The next result refines Eq.~\eqref{eq:main-hybrid-cost} by giving the minimum numbers of copies needed for a prescribed error tolerance, including integer batch sizes.
It counts the estimation copies only and excludes any calibration copies used to learn the variances.

\begin{mycor}{Finite Copy Requirements}{finite_copy_requirements}
For the sign target estimation task in Def.~\ref{def:sign_target_estimation},
\begin{align}
\frac{R_{\mathrm{sep}}}{\varepsilon^2}
\leqslant
n_{\mathrm{sep}}^*(\varepsilon)
\leqslant
\frac{R_{\mathrm{sep}}}{\varepsilon^2}
+
N.
\label{eq:finite-copy-separate-bounds}
\end{align}
Consequently,
\begin{align}
\frac{\varepsilon^2}{2}
\left[
n_{\mathrm{sep}}^*(\varepsilon)
-
n_{\mathrm{dir}}^*(\varepsilon)
-
N
\right]
\leqslant
\Mfun_\rho(\bm A)
\leqslant
\frac{\varepsilon^2}{2}
\left[
n_{\mathrm{sep}}^*(\varepsilon)
-
n_{\mathrm{dir}}^*(\varepsilon)
+
1
\right].
\label{eq:finite-copy-hybrid-bounds}
\end{align}
\end{mycor}

\begin{proof}
For every $\bm n\in\mathcal N_n$, Eq.~\eqref{eq:allocation-CS-bound} gives
\begin{align}
\operatorname{MSE}_{\mathrm{sep}}(n,\rho;\bm n)
\geqslant
\frac{R_{\mathrm{sep}}}{n}.
\end{align}
Hence any allocation achieving mean-squared error at most $\varepsilon^2$ must satisfy
\begin{align}
n
\geqslant
\frac{R_{\mathrm{sep}}}{\varepsilon^2},
\end{align}
which proves the lower bound in Eq.~\eqref{eq:finite-copy-separate-bounds}.

For the upper bound, first assume $S>0$.
Choose
\begin{align}
n_r
=
\begin{cases}
\left\lceil
\dfrac{d_r S}{\varepsilon^2}
\right\rceil,
&
d_r>0,
\\[1.2ex]
1,
&
d_r=0
\end{cases},
\end{align}
where $\lceil x\rceil$ denotes the smallest integer greater than or equal to $x$.
For every setting with positive variance,
\begin{align}
\frac{d_r^2}{n_r}
\leqslant
\varepsilon^2
\frac{d_r}{S}.
\end{align}
Summing over $r$ and using Eq.~\eqref{eq:separate-MSE} gives
\begin{align}
\operatorname{MSE}_{\mathrm{sep}}
\leqslant
\varepsilon^2.
\end{align}
Moreover,
\begin{align}
\sum_{r=1}^{N}n_r
\leqslant
\frac{S}{\varepsilon^2}
\sum_{d_r>0}d_r
+
N
=
\frac{S^2}{\varepsilon^2}
+
N
=
\frac{R_{\mathrm{sep}}}{\varepsilon^2}
+
N.
\end{align}
If $S=0$, one copy per setting gives zero separate error, so $n_{\mathrm{sep}}^*(\varepsilon)=N$ and Eq.~\eqref{eq:finite-copy-separate-bounds} again holds.

For the direct protocol, Eq.~\eqref{eq:direct-MSE} implies
\begin{align}
n_{\mathrm{dir}}^*(\varepsilon)
=
\begin{cases}
\left\lceil
\dfrac{R_{\mathrm{dir}}}{\varepsilon^2}
\right\rceil,
&
R_{\mathrm{dir}}>0,
\\[1.2ex]
1,
&
R_{\mathrm{dir}}=0.
\end{cases}.
\end{align}
Therefore
\begin{align}
0
\leqslant
n_{\mathrm{dir}}^*(\varepsilon)
-
\frac{R_{\mathrm{dir}}}{\varepsilon^2}
\leqslant
1.
\label{eq:finite-copy-direct-rounding}
\end{align}
Equation~\eqref{eq:finite-copy-separate-bounds} similarly implies
\begin{align}
n_{\mathrm{sep}}^*(\varepsilon)-N
\leqslant
\frac{R_{\mathrm{sep}}}{\varepsilon^2}
\leqslant
n_{\mathrm{sep}}^*(\varepsilon).
\label{eq:finite-copy-separate-rounding}
\end{align}
Combining Eqs.~\eqref{eq:finite-copy-direct-rounding} and \eqref{eq:finite-copy-separate-rounding} with the exact identity Eq.~\eqref{eq:main-hybrid-cost} gives Eq.~\eqref{eq:finite-copy-hybrid-bounds}.
\end{proof}

Within the stated design model, the finite copy difference determines $2\Mfun_\rho(\bm A)/\varepsilon^2$ up to an additive $O(1)$ correction from integer rounding and minimum batch sizes.
This bound does not include calibration.
A calibration cost that is $o(n)$ may still exceed $O(1)$, so the finite copy correction does not extend to an unknown-state adaptive protocol without further analysis.

Equation~\eqref{eq:main-hybrid-cost} converts every lower bound on hybrid uncertainty into a lower bound on the excess sampling cost of the prescribed separate protocol.
To make the dependence on the state and measured list explicit, we write
\begin{align}
R_{\mathrm{dir}}(\rho;\bm A)
&\coloneqq
\sum_{r=1}^{N}d_r^2,
\quad
R_{\mathrm{sep}}(\rho;\bm A)
\coloneqq
\left(
\sum_{r=1}^{N}d_r
\right)^2.
\end{align}
Thus Eq.~\eqref{eq:main-hybrid-cost} becomes
\begin{align*}
R_{\mathrm{sep}}(\rho;\bm A)
-
R_{\mathrm{dir}}(\rho;\bm A)
=
2\Mfun_\rho(\bm A).
\end{align*}

Equation~\eqref{eq:main-hybrid-cost} converts the uncertainty bounds in Appendices~\ref{app:su2-proofs} and~\ref{app:su11-proofs} directly into bounds on the excess sampling coefficient.
Let $M\in\mathbb R^{N\times3}$ have rank three, with singular values $0<\sigma_1(M)\leqslant\sigma_2(M)\leqslant\sigma_3(M)$.

\begin{mycor}{Spin Sampling Bound}{spin_separate_sampling_excess}
Let $M\in\mathbb R^{N\times3}$ have rank three.
For every state $\rho$ in the irreducible spin-$j$ representation with $j>0$,
\begin{align}
R_{\mathrm{sep}}(\rho;M\bm J)-R_{\mathrm{dir}}(\rho;M\bm J)\geqslant j \sigma_1(M)\sigma_2(M).
\label{eq:spin-sampling-excess}
\end{align}
\end{mycor}

\begin{proof}
Equation~\eqref{eq:main-hybrid-cost} gives
\begin{align*}
R_{\mathrm{sep}}(\rho;M\bm J)-R_{\mathrm{dir}}(\rho;M\bm J)=2\Mfun_\rho(M\bm J).
\end{align*}
By the covariance hierarchy in Eq.~\eqref{eq:covariance_chain} and Thm.~\ref{thm:spin-frame-minimum}, every state satisfies
\begin{align}
\Mfun_\rho(M\bm J)\geqslant\Mfun_\rho^{\mathrm{sp}}(M\bm J)\geqslant\frac j2\sigma_1(M)\sigma_2(M),
\end{align}
which proves Eq.~\eqref{eq:spin-sampling-excess}.
\end{proof}

\begin{mycor}{Discrete Series Sampling Bound}{su11_separate_sampling_excess}
Let $M\in\mathbb R^{N\times3}$ have rank three and no zero row.
For every admissible state $\rho\in\mathcal S_2(\mathcal D_k^+)$ with Bargmann index $k>0$,
\begin{align}
R_{\mathrm{sep}}(\rho;M\bm K)-R_{\mathrm{dir}}(\rho;M\bm K)\geqslant k \gamma_N(M)\geqslant k \sigma_1(M)\sigma_2(M).
\label{eq:su11-sampling-excess}
\end{align}
\end{mycor}

\begin{proof}
Equation~\eqref{eq:main-hybrid-cost} gives
\begin{align*}
R_{\mathrm{sep}}(\rho;M\bm K)-R_{\mathrm{dir}}(\rho;M\bm K)=2\Mfun_\rho(M\bm K).
\end{align*}
Thm.~\ref{thm:su11_conic_hybrid_bound} gives
\begin{align}
\Mfun_\rho(M\bm K)\geqslant\frac k2\gamma_N(M)\geqslant\frac k2\sigma_1(M)\sigma_2(M),
\end{align}
which proves Eq.~\eqref{eq:su11-sampling-excess}.
\end{proof}

The nonorthogonal discrete series triple
\begin{align}
\bm A_*\coloneqq(K_x,K_y,K_x+K_z)^{\T}
\end{align}
admits a strictly stronger excess bound than the commutator bound in Cor.~\ref{cor:su11_separate_sampling_excess}.
By Thm.~\ref{thm:su11_quantitative_quantum_gap} and Eq.~\eqref{eq:main-hybrid-cost},
\begin{align}
R_{\mathrm{sep}}(\rho;\bm A_*)-R_{\mathrm{dir}}(\rho;\bm A_*)\geqslant k\left(\sqrt3+2\epsilon_k\right),
\label{eq:su11-quantum-gap-sampling}
\end{align}
where $\epsilon_k>0$ is defined in Eq.~\eqref{eq:su11-gap-coefficient}.
The positive term $2k\epsilon_k$ certifies an additional sampling penalty beyond the first-moment commutator bound; the actual penalty may be larger.

For the same observable list, Cor.~\ref{cor:su11_coherent_state_limitation} compares coherent mixtures and the constructed pure states.
Every admissible coherent state mixture satisfies
\begin{align}
R_{\mathrm{sep}}(\rho;\bm A_*)-R_{\mathrm{dir}}(\rho;\bm A_*)\geqslant\frac{19k}{8},
\end{align}
whereas the perturbed states constructed in Eq.~\eqref{eq:su11-lowering-perturbation} attain a strictly smaller excess.
This comparison concerns the excess $R_{\mathrm{sep}}-R_{\mathrm{dir}}$, not the separate-measurement coefficient $R_{\mathrm{sep}}(\rho;\bm A_*)$ itself, because $R_{\mathrm{dir}}(\rho;\bm A_*)$ depends on the state.
Thus it does not imply an ordering of $R_{\mathrm{sep}}$ across different states.

The optimal allocation uses the true component variances as design information, while the sample means are evaluated from fresh measurement outcomes.
The cost of obtaining that information is excluded.
We next keep the target ensemble fixed and allow the measured list to change by orthogonal mixing.

\subsection{Orthogonal Mixing and Spectral Hybrid Costs}
\label{sec:orthogonal-mixing}

Subsec.~\ref{sec:optimal-allocation} optimizes the batch allocation for a fixed measured list $\bm A$.
A second design freedom is to replace that list, before the target label $\bm s$ is disclosed, by a real orthogonal mixture
\begin{align}
\bm B\coloneqq O\bm A,\quad O\in\mathrm O(N),
\end{align}
while the state $\rho$, the sign target task in Def.~\ref{def:sign_target_estimation}, and the total copy budget remain unchanged.
The matrix $O$ may depend on $\Gamma_\rho(\bm A)$, but it is fixed before $\bm s$ is disclosed and therefore cannot depend on the realized target label.
The true covariance matrix may be supplied for the design, while the estimator is evaluated on fresh measurement outcomes; learning the covariance matrix is not included in the copy budget.

For bounded observables, every component of $\bm B$ is self-adjoint.
For the positive discrete series, these components are real linear combinations of the generators as in Eq.~\eqref{eq:su11-directional-generator}, and Lem.~\ref{lem:su11_covariance_transfer} guarantees finite second moments for every admissible state.
For a general unbounded list, we impose the same common domain, self-adjoint realization, and finite second moment assumptions as in Subsec.~\ref{sec:target-averaged-errors}.

Set $\bm t\coloneqq O\bm s$.
Since $\bm A=O^{\T}\bm B$, the target operator itself is unchanged:
\begin{align}
H_{\bm s}
=
\bm s^{\T}\bm A
=
\bm t^{\T}\bm B,
\quad
h_{\bm s}
=
\bm s^{\T}\langle\bm A\rangle_\rho
=
\bm t^{\T}\langle\bm B\rangle_\rho.
\label{eq:orthogonal-target-invariance}
\end{align}

The components of $\bm t$ are generally neither signs nor independent random variables.
However, since the original signs are independent and uniform,
\begin{align}
\mathbb E_{\bm s}[\bm t\bm t^{\T}]
=
O \mathbb E_{\bm s}[\bm s\bm s^{\T}] O^{\T}
=
\1_N.
\label{eq:orthogonal-coefficient-second-moment}
\end{align}
Orthogonal mixing changes the measured observables but preserves both the target operators and the second moments of the target coefficients used in the averaged mean-squared error.

For a fixed allocation $\bm n=(n_1,\ldots,n_N)^{\T}\in\mathcal N_n$, measure each $B_p$ on its own batch of $n_p$ copies and denote the corresponding sample mean by $\widehat b_p$.
After the target label is disclosed, estimate $h_{\bm s}$ by
\begin{align}
\widehat h_{\bm s}^{ O}\coloneqq\sum_{p=1}^{N}t_p\widehat b_p.
\end{align}
Define the target-averaged mean-squared error of this protocol by
\begin{align}
\operatorname{MSE}_{\mathrm{sep}}^{O}(n,\rho;\bm n)\coloneqq 2^{-N}\sum_{\bm s}\mathbb E_{\mathrm{data}\mid\bm s}\left[\left(\widehat h_{\bm s}^{ O}-h_{\bm s}\right)^2\right].
\end{align}
Its optimized asymptotic coefficient is
\begin{align}
R_{\mathrm{sep}}^{O}(\rho;\bm A)\coloneqq\lim_{n\to\infty}n\min_{\bm n\in\mathcal N_n}\operatorname{MSE}_{\mathrm{sep}}^{O}(n,\rho;\bm n),
\label{eq:orthogonal-separate-coefficient-definition}
\end{align}
whenever the limit exists.
The argument $\bm A$ specifies the original target observables, while the measured list is $\bm B=O\bm A$.

Combining the standard allocation formula~\cite{RubinBabbushMcClean2018,CrawfordEtAl2021} with covariance diagonalization gives the following exact identity for the sign target task.

\begin{mythm}{Spectral Sampling Identity}{spectral_hybrid_sampling_excess}
For the sign target estimation task in Def.~\ref{def:sign_target_estimation}, the limit in Eq.~\eqref{eq:orthogonal-separate-coefficient-definition} exists for every $O\in\mathrm O(N)$, and
\begin{align*}
\Mfun_\rho^{\mathrm{sp}}(\bm A)=\frac12\min_{O\in\mathrm O(N)}\left[R_{\mathrm{sep}}^{O}(\rho;\bm A)-R_{\mathrm{dir}}(\rho;\bm A)\right].
\end{align*}
An orthogonal transformation $O$ attains the minimum if and only if $O\Gamma_\rho(\bm A)O^{\T}$ is diagonal.
\end{mythm}

\begin{proof}
For a fixed target label $\bm s$, the batch means $\widehat b_p$ are independent and unbiased, with
\begin{align}
\mathbb E_{\mathrm{data}}
\left[
\left(
\widehat b_p-\langle B_p\rangle_\rho
\right)^2
\right]
=
\frac{(\Delta_\rho B_p)^2}{n_p}.
\end{align}
Therefore,
\begin{align}
\mathbb E_{\mathrm{data}\mid\bm s}\left[\left(\widehat h_{\bm s}^{ O}-h_{\bm s}\right)^2\right]=\sum_{p=1}^{N}\frac{t_p^2(\Delta_\rho B_p)^2}{n_p}.
\end{align}
Averaging over the original sign ensemble and using Eq.~\eqref{eq:orthogonal-coefficient-second-moment} yields
\begin{align}
\operatorname{MSE}_{\mathrm{sep}}^{O}(n,\rho;\bm n)=\sum_{p=1}^{N}\frac{(\Delta_\rho B_p)^2}{n_p}.
\label{eq:orthogonal-target-averaged-MSE}
\end{align}
Equation~\eqref{eq:orthogonal-target-averaged-MSE} has the same allocation form as Eq.~\eqref{eq:separate-MSE}, so the allocation argument in the proof of Thm.~\ref{thm:optimal_allocation_hybrid_excess} applies with the component standard deviations $\Delta_\rho B_p$.
Hence the limit in Eq.~\eqref{eq:orthogonal-separate-coefficient-definition} exists and
\begin{align}
R_{\mathrm{sep}}^{O}(\rho;\bm A)=\left(\sum_{p=1}^{N}\Delta_\rho B_p\right)^2.
\label{eq:orthogonal-separate-coefficient}
\end{align}

The direct protocol is unchanged because Eq.~\eqref{eq:orthogonal-target-invariance} leaves every target operator $H_{\bm s}$ unchanged.
Moreover, $\Gamma_\rho(\bm B)=O\Gamma_\rho(\bm A)O^{\T}$, and therefore
\begin{align}
R_{\mathrm{dir}}(\rho;\bm A)=\Tr\Gamma_\rho(\bm A)=\Tr\Gamma_\rho(\bm B)=\sum_{p=1}^{N}(\Delta_\rho B_p)^2.
\label{eq:orthogonal-direct-invariance}
\end{align}
Combining Eqs.~\eqref{eq:orthogonal-separate-coefficient} and \eqref{eq:orthogonal-direct-invariance} gives
\begin{align}
R_{\mathrm{sep}}^{O}(\rho;\bm A)-R_{\mathrm{dir}}(\rho;\bm A)=2\sum_{p<q}\Delta_\rho B_p\Delta_\rho B_q=2\Mfun_\rho(\bm B)=2\Mfun_\rho(O\bm A).
\label{eq:orthogonal-hybrid-excess}
\end{align}
Minimizing Eq.~\eqref{eq:orthogonal-hybrid-excess} over $O\in\mathrm O(N)$ and applying Lem.~\ref{lem:Spectral_min_UR} proves Eq.~\eqref{eq:main-spectral-cost}.
The equality condition follows from the same lemma: the minimum is attained if and only if $O\Gamma_\rho(\bm A)O^{\T}$ is diagonal.
\end{proof}

Combining Eq.~\eqref{eq:main-hybrid-cost} with Thm.~\ref{thm:spectral_hybrid_sampling_excess} gives
\begin{align}
\Mfun_\rho(\bm A)-\Mfun_\rho^{\mathrm{sp}}(\bm A)=\frac12\left[R_{\mathrm{sep}}(\rho;\bm A)-\min_{O\in\mathrm O(N)}R_{\mathrm{sep}}^{O}(\rho;\bm A)\right].
\label{eq:orthogonal-mixing-cost-reduction}
\end{align}
Thus $\Mfun_\rho(\bm A)-\Mfun_\rho^{\mathrm{sp}}(\bm A)$ is exactly half the part of the excess sampling coefficient removed by orthogonal mixing at fixed state and fixed target task.
Equivalently, $\Mfun_\rho^{\mathrm{sp}}(\bm A)$ is half the minimum excess sampling coefficient relative to direct measurement, optimized over orthogonal combinations of the observables.
These identities describe the separate measurement protocols of Subsec.~\ref{sec:orthogonal-mixing}, including optimization over orthogonal mixing; they do not optimize over arbitrary quantum measurements or estimators.

The reduction in Eq.~\eqref{eq:orthogonal-mixing-cost-reduction} requires the mixed observables $B_p$ themselves to be measured.
Applying the same orthogonal matrix only to classical outcomes previously collected from the original list $\bm A$ does not implement measurements of $O\bm A$ and therefore does not realize the spectral minimum.
To show the effect of measuring orthogonal combinations at fixed state, take the spin-up state along the $z$ axis,
\begin{align}
\rho_z\coloneqq\frac12(\1_2+\sigma_z),
\end{align}
and use the nonorthogonal qubit observable list $\bm A$ defined in Eq.~\eqref{eq:angular-qubit-frame}.
Using Eq.~\eqref{eq:spin-half-pauli}, the covariance matrix at this fixed state is
\begin{align}
\Gamma_{\rho_z}(\bm A)=\frac14\begin{pmatrix}1&\cos\theta&0\\ \cos\theta&1&0\\ 0&0&0\end{pmatrix}.
\label{eq:fixed-qubit-frame-covariance}
\end{align}
Its diagonal entries and eigenvalues give the fixed state values
\begin{align}
\Mfun_{\rho_z}(\bm A)=\frac14,\quad \Mfun_{\rho_z}^{\mathrm{sp}}(\bm A)=\frac{\sin\theta}{4}.
\end{align}
These quantities are evaluated at the prescribed state $\rho_z$, whereas the minima in Thm.~\ref{thm:nonorthogonal_qubit_minima} are obtained by optimizing over the state.

Equation~\eqref{eq:main-hybrid-cost} and Thm.~\ref{thm:spectral_hybrid_sampling_excess} therefore give
\begin{align}
R_{\mathrm{dir}}(\rho_z;\bm A)=\frac12,\quad R_{\mathrm{sep}}(\rho_z;\bm A)=1,\quad \min_{O\in\mathrm O(3)}R_{\mathrm{sep}}^{O}(\rho_z;\bm A)=\frac{1+\sin\theta}{2}.
\end{align}
The reduction in the separate sampling coefficient is
\begin{align}
R_{\mathrm{sep}}(\rho_z;\bm A)-\min_{O\in\mathrm O(3)}R_{\mathrm{sep}}^{O}(\rho_z;\bm A)=\frac{1-\sin\theta}{2}=2\left[\Mfun_{\rho_z}(\bm A)-\Mfun_{\rho_z}^{\mathrm{sp}}(\bm A)\right].
\end{align}

An orthogonal transformation attaining the minimum in Thm.~\ref{thm:spectral_hybrid_sampling_excess} is
\begin{align}
O_*=\frac{1}{\sqrt2}\begin{pmatrix}1&-1&0\\ 1&1&0\\ 0&0&\sqrt2\end{pmatrix},
\end{align}
which gives observables along the principal axes of the covariance matrix
\begin{align}
B_1=\frac{A_1-A_2}{\sqrt2},\quad B_2=\frac{A_1+A_2}{\sqrt2},\quad B_3=A_3.
\end{align}
For the fixed state $\rho_z$, their covariance matrix is
\begin{align}
\Gamma_{\rho_z}(\bm B)=O_*\Gamma_{\rho_z}(\bm A)O_*^{\T}=\operatorname{diag}\left(\frac{1-\cos\theta}{4},\frac{1+\cos\theta}{4},0\right).
\end{align}
Thus $O_*$ satisfies the attainment condition in Thm.~\ref{thm:spectral_hybrid_sampling_excess} and realizes the spectral sampling minimum.

The corresponding standard deviations are
\begin{align}
\Delta_{\rho_z}B_1=\frac12\sqrt{1-\cos\theta},\quad \Delta_{\rho_z}B_2=\frac12\sqrt{1+\cos\theta},\quad \Delta_{\rho_z}B_3=0.
\end{align}
By Eq.~\eqref{eq:optimal-allocation-fractions}, the asymptotically optimal batch fractions satisfy
\begin{align}
\frac{n_1}{n}\longrightarrow\frac{\sqrt{1-\cos\theta}}{\sqrt{1-\cos\theta}+\sqrt{1+\cos\theta}},\quad \frac{n_2}{n}\longrightarrow\frac{\sqrt{1+\cos\theta}}{\sqrt{1-\cos\theta}+\sqrt{1+\cos\theta}},\quad \frac{n_3}{n}\longrightarrow0.
\end{align}
Since $B_3$ has zero variance in $\rho_z$, assigning it any fixed admissible number of copies does not affect the asymptotic coefficient, and one copy suffices.

At $\theta=\pi/2$, Eq.~\eqref{eq:fixed-qubit-frame-covariance} is already diagonal, and therefore the original observables already minimize the separate measurement protocol cost.
Accordingly, $\Mfun_{\rho_z}(\bm A)=\Mfun_{\rho_z}^{\mathrm{sp}}(\bm A)=1/4$ and orthogonal mixing produces no reduction in the separate sampling coefficient.
In the limit $\theta\to0^+$, the observables $A_1$ and $A_2$ become identical and the coefficient matrix loses rank.
In this limit, $\Mfun_{\rho_z}^{\mathrm{sp}}(\bm A)\to0$ and $\min_{O\in\mathrm O(3)}R_{\mathrm{sep}}^{O}(\rho_z;\bm A)\to R_{\mathrm{dir}}(\rho_z;\bm A)$, so the entire separate measurement protocol excess can be eliminated by measuring orthogonal combinations of the observables.
The endpoint $\theta=0$ is excluded from the full rank family and is shown only as a continuous rank-deficient limit.

Optimizing over states and orthogonal combinations gives exact excess sampling costs from the spectral minima established in Appendices~\ref{app:su2-proofs} and~\ref{app:su11-proofs}.
For the spin case, let $M\in\mathbb R^{N\times3}$ have rank three, with singular values $0<\sigma_1(M)\leqslant\sigma_2(M)\leqslant\sigma_3(M)$.

\begin{mycor}{Exact Spin Sampling Excess}{exact_spin_sampling_excess}
Let $M\in\mathbb R^{N\times3}$ have rank three.
For the irreducible spin-$j$ representation with $j>0$,
\begin{align}
\min_{\rho}\min_{O\in\mathrm O(N)}
\left[
R_{\mathrm{sep}}^{O}(\rho;M\bm J)
-
R_{\mathrm{dir}}(\rho;M\bm J)
\right]
=
j \sigma_1(M)\sigma_2(M).
\label{eq:spin-optimal-spectral-excess}
\end{align}
The minimum is attained by a spin coherent state polarized along a unit right singular vector of $M$ associated with $\sigma_3(M)$, together with any orthogonal transformation that diagonalizes the corresponding covariance matrix.
\end{mycor}

\begin{proof}
For each fixed state $\rho$, Thm.~\ref{thm:spectral_hybrid_sampling_excess} gives an excess of $2\Mfun_\rho^{\mathrm{sp}}(M\bm J)$ after optimization over orthogonal mixtures.
Theorem~\ref{thm:spin-frame-minimum} gives the minimum $j\sigma_1(M)\sigma_2(M)/2$ of this spectral uncertainty, attained by a spin coherent state polarized along a unit right singular vector of $M$ associated with $\sigma_3(M)$.
For this state, the equality condition in Thm.~\ref{thm:spectral_hybrid_sampling_excess} is satisfied by any orthogonal transformation that diagonalizes the corresponding covariance matrix.
This proves Eq.~\eqref{eq:spin-optimal-spectral-excess} and the attainability statement.
\end{proof}

For the positive discrete series with Bargmann index $k>0$, consider coefficient matrices $M=DL$ of the form specified in Eq.~\eqref{eq:su11-weighted-lorentz-frame}.

\begin{mycor}{Exact Lorentz Sampling Excess}{exact_lorentz_sampling_excess}
For the positive discrete series with Bargmann index $k>0$ and any coefficient matrix $M=DL$ in Eq.~\eqref{eq:su11-weighted-lorentz-frame},
\begin{align}
\min_{\rho\in\mathcal S_2(\mathcal D_k^+)}
\min_{O\in\mathrm O(3)}
\left[
R_{\mathrm{sep}}^{O}(\rho;M\bm K)
-
R_{\mathrm{dir}}(\rho;M\bm K)
\right]
=
abk.
\label{eq:su11-optimal-spectral-excess}
\end{align}
The minimum is attained by the state $\rho_L$ in Eq.~\eqref{eq:su11-lorentz-minimizer}, together with any orthogonal transformation that diagonalizes $\Gamma_{\rho_L}(M\bm K)$.
\end{mycor}

\begin{proof}
For $M=DL$, Cor.~\ref{cor:su11_lorentz_frame_minima} establishes the spectral minimum $abk/2$, attained by $\rho_L$.
Minimizing the spectral sampling identity of Thm.~\ref{thm:spectral_hybrid_sampling_excess} over admissible states gives Eq.~\eqref{eq:su11-optimal-spectral-excess}.
At $\rho_L$, any orthogonal transformation that diagonalizes $\Gamma_{\rho_L}(M\bm K)$ attains this excess.
\end{proof}

\end{document}